\documentclass[reprint,prx]{revtex4-2}
\usepackage{blindtext}
\usepackage{amsmath, amssymb, graphicx, mathtools, multirow,makecell,subfigure,color,float,stmaryrd}

\renewcommand{\thesection}{\Roman{section}}
\renewcommand{\thesubsection}{\thesection.\arabic{subsection}}
\makeatletter
\renewcommand{\p@subsection}{}
\renewcommand{\p@subsubsection}{}
\makeatother

\usepackage[normalem]{ulem}
\usepackage{hyperref}

\usepackage{cleveref}

\hypersetup{
	colorlinks=false,       % false: boxed links; true: colored links
	linkcolor=blue,          % color of internal links (change box color with linkbordercolor)
	citecolor=magenta,        % color of links to bibliography
	urlcolor=magenta           % color of external links
}

\newcommand{\blue}{\color{blue}}

\begin{document}
	%%%%%%%%%%%%%%%%%%%%%%%%%%%%%%
	\title{Segregation,
 Finite Time Elastic Singularities and Coarsening
% in a Two-component 
 in Renewable Active Matter}
	
	%%%%%%%%%%%%%%%%%%%%%%%%%%%%%%%%%%%
	\author{Ayan Roychowdhury}
	\thanks{Joint first author}
	\affiliation{Simons Centre for the Study of Living Machines, National Centre for Biological Sciences-TIFR, Bengaluru, India  560065.}
	%%%%%%%%%%%%%%%%%%%%%%%%%%%%%%%%%%%
	\author{Saptarshi Dasgupta}
        \thanks{Joint first author}
	\affiliation{Simons Centre for the Study of Living Machines, National Centre for Biological Sciences-TIFR, Bengaluru, India  560065.}
	%%%%%%%%%%%%%%%%%%%%%%%%%%%%%%%%%%%
	\author{Madan Rao}
	\thanks{rao.madan@gmail.com}
	\affiliation{Simons Centre for the Study of Living Machines, National Centre for Biological Sciences-TIFR, Bengaluru, India  560065.}
%%%%%%%%%%%%%%%%%%%%%%%%%%%%%%%%%%%	
	
%%%%%%%%%%%%%%%%%%%%%%%%%%%%%%%%%%%%%%%%%%%%%%%%%%%%%%%%%%%%%%%%%%%%%%%%%%%%%%%%%%%%%%%%%%%%%%%%%%%%%%%%%%%%%%%%%%%%%%%%%%%%%%%%%%%%%%%%%%%%%%%%%%%%%%%%%%%%%%%%%%%%%%%%%%%%%%%%%%%%%%%%%%%%%%%%%%%%%%%%%%%%%%%%%%%%%%%%%%%%%%%%%%%%%%%%%%%%%%%%%%%%%%%%%%%%%%%%	

\begin{abstract}
%Viewed under a fluorescence microscope, the actomyosin cytoskeleton presents vivid streaks of lines together with persistent oscillatory waves.
Material renewability in active living systems, such as in cells and tissues,  can drive the large-scale patterning of forces, with distinctive phenotypic consequences. This is especially significant in the cell cytoskeleton, where multiple species of myosin bound to actin, apply differential contractile stresses and undergo differential turnover, giving rise to patterned force channeling. 
Here we study the dynamical patterning of stresses that emerge in a hydrodynamic description of a renewable active actomyosin elastomer comprising two myosin species. Our analytical framework also holds for an actomyosin elastomer with a single myosin species.
%The time reversal symmetry breaking of the underlying renewable active material leads to a dynamic elastic nonreciprocity and the active extraction of work when subjected to a cyclic load.
We find that a uniform active contractile elastomer spontaneously segregates
into spinodal stress patterns, 
%in the absence of attractive interactions. This is
followed by a finite-time collapse into  tension carrying singular structures that display self-similar scaling and caustics.  Our numerical analysis carried out in 1D, shows that  these singular structures move and merge, and gradually result in a slow coarsening dynamics.
% \sout{In addition, the nonreciprocal nature of  the underlying dynamics gives rise to exceptional points that are  associated with a variety of travelling states -- from peristalsis to swap and trains of regular and singular stress patterns, that may coexist with each other. 
% Both the novel segregation and excitability are consequences of time reversal symmetry breaking of the underlying active dynamics. }
We discuss the implications of our findings to the emergence of stress fibers and the spatial patterning of actomyosin. Our study suggests, that with state-dependent turnover of crosslinkers and myosin, the {\it in vivo} cytoskeleton can navigate through the space of material parameters to achieve a variety of functional phenotypes.
%We explore the implications  of these coexisting
%nonequilibrium phases for a larger class of active mixtures with non-reciprocal couplings.
%A supercritical Hopf bifurcation lying on the other side of these last two exceptional transitions leads to regular travelling and/or swapping patterns.
%These elastic singularities move and  merge to become stronger. {\red (or repel each other)}. 
%In addition, the renewable active elastomer exhibits travelling waves and swapping, a consequence of its nonreciprocal dynamics. 
%\sout{Our work has implications for the emergence of tension carrying actomyosin  stress fibers coexisting with mechanical excitability that is observed in substrate-attached cells.}
%In the finite geometry of the cell, the collection of active tension chains can form an active web held together by specific anchoring at the cell boundary.
%On the other hand, preferential wetting at the cell boundary can reinforce active segregation in a mixture of stresslets leading to stratification.
\end{abstract}

%%%%%%%%%%%%%%%%%%%%%%%%%%%%%%%%%%%%%%%%%%%%%%%%%%%%%%%%%%%%%%%%%%%%%%%%%%%%%%%%%%%%%%%%%%%%%%%%%%%%%%%%%%%%%%%%%%%%%%%%%%%%%%%%%%%%%%%%%%%%%%%%%%%%%%%%%%%%%%%%%%%%%%%%%%%%%%%%%%%%%%%%%%%%%%%%%%%%%%%%%%%%%%%%%%%%%%
\maketitle
%%%%%%%%%%%%%%%%%%%%%%%%%%%%%%%%%%%%%%%%%%%%%%%%%%%%%%%%%%%%%%%%%%%%%%%%%%%%%%%%%%%%%%%%%%%%%%%%%%%%%%%%%%%%%%%%%%%%%%%%%%%%%%%%%%%%%%%%%%%%%%%%%%%%%%%%%%%%%%%%%%%%%%%%%%%%%%%%%%%%%%%%%%%%%%%%%%%%%%%%%%%%%%%%%%%%%%%%%%%%%%%%%%%%%%%%%%%%%%%%%%%%%%%%%%%%%%%%%%%%%%%%%%%%%%%%%%%%%%%%%%%%%%%%%%%%%%%%%%%%%%%%%%%%%%%%%%%%%%%%%%%%%%%%%%%%%%%%%%%%%%%%%%%%%%%%%%%%%%%%%%%%%%%%%%%%%%%%%%%%%%%%%%%%%%%%%%%%%%%%%%%%%%%%%%%%%%%%%%%%%%%%%%%%%%%%%%%%%%%%%%%%%%%%%%%%%%%%%%%%%%%%%%%%%%%%%%%%%%%%%%%%%%%%%%%%%%%%%%%%%%%%%%%%%%%%

\section{Introduction}

Since actomyosin is the primary agency of cell and tissue scale forces, 
 its patterning along system spanning stress fibers~\cite{PollardGoldman2016}, cables or arcs~\cite{HotulainenLappalainen2006,Lehtimakietal2021} reveals the patterning of forces along tension chains.
The cytoskeleton represent  a class of active matter called {\it renewable active matter}; renewability is a distinctive feature of living materials that allows it 
to navigate through the space of material parameters, resulting in
unusual mechanical responses~\cite{RRT2024} and system spanning force channeling~\cite{Vignaudetal2021}.
Here we study the nature of dynamical force patterning using a hydrodynamic description of a renewable active material.

%The 
%molecular agencies of biological force are
%distribution of  forces across the scale of the cell is dynamically templated by
%dependent on motor-filament complexes of 
%the active actin cytoskeleton, 
%most commonly as 
%assemblies of a variety of myosins together with actin filaments and their crosslinkers, that appear as 
% `stress fibers' that carry
 %cellular
% tension~\cite{PollardGoldman2016}.
%Cytoskeletal organisation and remodelling give the cell its dynamical shape and form~\cite{Lecuitetal2011,Tanejaetal2020}, as well as its adaptive mechanical response~\cite{Matthewsetal2006,banerjeeetal2020,Kuceraetal2022}. In addition, it sets up a global scaffold for the patterning and localisation of mesoscale condensates~\cite{Brangwynne2021} and the positioning of subcellular organelles~\cite{Marshall2020}, such as the centrosome~\cite{jimenezetal2021} and nucleus~\cite{Makhija2015,Sunetal2020}.
The  actomyosin cytoskeleton is
put together by the nonequilibrium self-assembly of a variety of crosslinkers and myosin species, that both exert and sense forces,  the latter via their strain dependent turnover~\cite{Kovacsetal2007, Fernandez-Gonzalezetal2009,Mullaetal2022,HawkinsLiverpool2014,HiraiwaSalbreux2016,MunroGardel2021,Greenbergetal2016}.
We will refer to these units of mechanotransduction as {\it stresslets}; the spatiotemporal patterning of these stresslets mark the patterning of forces. High resolution microscopy reveals that
distinct actomyosin stress fibers~\cite{Vignaudetal2021, HotulainenLappalainen2006,Lehtimakietal2021,Bershadsky2017,Luoetal2013,Weissenbruchetal2021},
%and web-like structures~\cite{Vignaudetal2021, Bershadsky2017,Weissenbruchetal2021,Luoetal2013}, 
are associated with different stresslet species displaying different cellular localisations~\cite{Beachetal2014,Weissenbruchetal2021}.
%More recently, there have been systematic studies of spatial patterning of cytoskeletal structures within cells and tissues~\cite{Vicente-Manzanaresetal2008},
%{\it in vitro} reconstitutions~\cite{silvaetal2011,koenderink-paluch2018}, and in cell extracts on micropatterned substrates~\cite{Vignaudetal2021,jimenezetal2021,zhangetal2021}.
% {\it In vivo}, these structures often coexist with persistent waves of oscillation of actomyosin, reminiscent of an excitable system~\cite{zallen2011,munjal2015,debsankar2017,Lehtimakietal2021,Dierkesetal2014}.

%A generic physical basis for these pervasive spatiotemporal force patterning is lacking.

%pervasive phenomena is lacking, as is a dynamical theory for the establishment of the cellular framework, the emergent force patterning and their homeostatic response~\cite{Weissenbruchetal2021}.

%%%%%%%%%%%%%%%%%%%%%%%%%%%%%%%%%%%%%%%%
Indeed, the human genome   
encodes around $40$ myosin genes, grouped into $18$ classes, with class II being the most prominent in muscle and non-muscle cells \cite{Bergetal2001,HartmanSpudich2012}. 
Non-muscle myosin II (NM II), found in all non-muscle eukaryotic cells, assembles into bipolar filaments of up to 30 hexamers
% typically about 300 nm in length 
\cite{Beachetal2014,Billingtonetal2013,Bershadsky2017}.
Mammalian cells express three NM II isoforms---NM IIA, NM IIB, and NM IIC---that share identical light chains but differ in their heavy chains, conferring distinct mechanical properties \cite{Shutovaetal2017,Billingtonetal2013,Weissenbruchetal2021,Weissenbruchetal2022}. These myosin isoforms differ in their contractile action \cite{Kovacsetal2003,Hippleretal2020}, duty ratio \cite{Wangetal2003} and binding-unbinding rates
\cite{Billingtonetal2013,Weissenbruchetal2021}.
Their coordinated interplay is crucial for cellular tension generation and maintenance \cite{Beachetal2014,Weissenbruchetal2021}, with the spatially regulated relative abundance leading to strong subcellular localization. In particular, NM IIA and NM IIB are known to undergo self-sorting \cite{Weissenbruchetal2021,Winkleretal2025}, leading to spatially organized and functionally distinct contractile zones within the cell.

Several studies have explored pattern formation in actomyosin systems, treating them either as active fluids \cite{Kruseetal2004,Boisetal2011,Rechoetal2013,Kumaretal2014,Gowrishankaretal2012,GowrishankarRao2016,HusainRao2017,Winkleretal2025,BarberiKruse2023,BarberiKruse2024} or as active elastomers with a single myosin species \cite{BanerjeeLiverpoolMarchetti2011,BanerjeeMarchetti2011,debsankar2017}.
In contrast to the aster-like patterns and oscillations typical of active fluids, the long-range marginalized elastic interactions in active solids can generate system-spanning stress singularities \cite{RRT2024}. 
The present work goes beyond a routine extension of these studies to multi-species myosin systems; it focuses on how differential stresslet activity and strain dependent turnover can efficiently remodel the effective elastic energy landscape---offering a new way of looking at evolving materials navigating the space of material moduli and driving towards and maintaining itself in mechanically adaptable states \cite{banerjeeetal2020}.

%{\red Renewable matter, such as Cytoskeleton offers a new way of looking materials - navigating the space of moduli. elaborate in the discussion}

%An important aspect of these molecular mechanotransducers, such as Myosin-II minifilaments, are that they are both generators and sensors of force, the latter via via their strain dependent unbinding rates from a substrate, exhibiting either catch or slip bond response. Activated myosin aggregates are predominantly generators of contractile stresses. In addition, they behave as stress sensors, via their strain dependent unbinding rates from a substrate, and can exhibit either catch or slip bond response. We will refer to these units of mechanotransduction as {\it stresslets}; the spatiotemporal patterning of these stresslets will then mark the patterning of forces.

In this paper, we explore the unusual features of such renewable active matter, that exhibits  dynamic material renormalization, leading to segregation and singular stress  patterns.
% \sout{can be ultimately traced to the violation of time reversal symmetry (TRS) of
% the underlying active dynamics.}
%we make an attempt at such an understanding, 
% \sout{To see this,}
We construct
active hydrodynamic equations~\cite{Marchettietal2013} for a mixture of contractile stresslets, such as myosin IIA and IIB~\cite{Bershadsky2017,Shutovaetal2017,Weissenbruchetal2021, Weissenbruchetal2022}, on a permanently crosslinked actin elastomer, 
while allowing turnover of myosin
stresslets~\cite{BanerjeeLiverpoolMarchetti2011,BanerjeeMarchetti2011,debsankar2017}. 
% This is the natural setting to understand the dynamics of emergence of actomyosin force patterns {\it in vivo}, such as dorsal-ventral stress fibers and transverse actomyosin arcs~\cite{Beachetal2014,HotulainenLappalainen2006,Bershadsky2017,Weissenbruchetal2021,Weissenbruchetal2022,Lehtimakietal2021,jimenezetal2021,Vignaudetal2021,Luoetal2013}. This analysis can be easily extended to the case where the cell background is a fluid at long time scales. Irrespective, 
Since the action (stress generation) and reaction (stress sensing dependent turnover) are regulated by independent biochemical cycles, the underlying hydrodynamics is nonreciprocal \cite{Fruchartetal2021}.
%The fact that the turnover and the contractile force generation of Myo-II are regulated by independent biochemical cycles,  
% mechanical nonreciprocity is the source of the spatially extended mechanical excitability reported here.
A straightforward material consequence of this nonreciprocity is a renormalization of the elastic moduli that can in principle drive the material to elastic marginality~\cite{RRT2024}.
%(ii) a dynamic elastic nonreciprocity that manifests as a violation of Maxwell-Betti relation, and (iii) the active extraction of work when subjected to a cyclic load. 
 We find that a slight difference between the contractile activities or turnover rates of the  stresslets, can drive spontaneous segregation, {\it even in the absence of any attractive interaction} -- the myosin species with higher contractile activity tend to cluster together (a similar analysis for a single myosin species,  leads to a segregation into myosin-rich and myosin-poor regions~\cite{mabhishek2024}). 
%This results in distinctive force  patterns at macroscopic scales, much larger than the scale of the stresslets. 
However, unlike conventional segregation driven by gradients in chemical potential~\cite{Bray2002}, the spontaneous segregation instability of the myosin stresslets is driven by an effective elastic stress relaxation. At later times, the  growing segregated domains collapse into well separated, singular structures of enhanced contractility (tensile ``stress fibers''), in striking departure from conventional coarsening. We derive scaling forms for these finite time singular structures~\cite{eggersfontelos-book2015} and verify them with careful numerics in 1D. The amplification of 
contractile stresses in the singular tension structures recalls the study in~\cite{Roncerayetal2016}. %We find that 
These singular tension structures  can be static or moving~\cite{Vignaudetal2021} and we determine the conditions for these. The moving singular tension structures merge over time and exhibit a slow coarsening dynamics at late times controlled by the differential myosin turnover.
%we derive equations for their mass and force balance and analyse conditions for their merger and repulsion.
%In the finite geometry of the cell, more complex active webs of these tension lines can be shaped and stabilised by cell geometry and cell surface anchors. This is reminiscent of the patterning of actomyosin networks stabilised by cadherin-mediated adherens junctions (AJ) and integrin-mediated focal adhesion (FA), and exhibit fragile mechanical behaviour that sensitively depend on the geometry of anchoring~\cite{GuptonWaterman-Storer2006, GeigerSpatzBershadsky2009,Vignaudetal2021, Kassianidouetal2017}.

Recent experiments done on cells plated on micro-patterned substrates~\cite{Vignaudetal2021} provide the ideal setup to  test  our predictions and explore new possibilities on the disassembly of stress fibers  upon treatment with the Rho kinase inhibitor Y27632, and their reassembly dynamics following washout of the inhibitor \cite{Bershadsky2017,Vignaudetal2021,Hippleretal2020}. This provides a perfect experimental platform to carry out systematic quantitative studies on how myosin activation can drive the segregation of an initially homogeneous actomyosin network into domains with differential contractility, and how the highly contractile domains subsequently collapse to form cell-spanning actomyosin cables or stress fibers. In the present setting of continuum elasticity, such taut fibers or cables manifest as singularities or localized concentrations in the tensile stress field along lines (or as ``force punctae'' in one dimension, as illustrated in this work).

Details of the derivation of the hydrodynamic equations, linear stability analysis and dispersion curves, finite time scaling and convergence analysis, and Movie captions are presented in the Appendix.

% \sout{In addition, renewable active materials exhibit spontaneous mechanical excitability --
% %in the context of dynamic pattern formation, 
% we find that the TRS violating  hydrodynamic equations lead to a non-Hermitian dynamical matrix which exhibits nonreciprocal features~\cite{Fruchartetal2021} such as exceptional points~\cite{Kato1984} which presage the  appearance of excitable travelling wave~\cite{Youetal2020} and standing wave (swap) phases~\cite{Fruchartetal2021}. These excitable phases can form long-lived transients that eventually resolve into segregated singular tension structures. 
% %This suggests that the segregation of singular tension structures can coexist with excitability. 
% We leave a detailed analysis of this novel form of coexistence of  nonequilibrium phases for a future study.} 

%{\red shall we talk about the evolutionary perspective here briefly, on the myosin turnover rates/renewability w.r.t. ''functional'' phenotypes such as force chains?}

%Shiqiong Hu, Kinjal Dasbiswas, Zhenhuan Guo, Yee-Han Tee, Visalatchi Thiagarajan, Pascal Hersen, Teng-Leong Chew, Samuel A. Safran, Ronen Zaidel-Bar & Alexander D. Bershadsky, Long-range self-organization of cytoskeletal myosin II filament stacks, Nature Cell Biology volume 19, pages 133–141 (2017).

%%%%%%%%%%%%%%%%%%%%%%%%%%%%%%%%%%%%%%%%%%%%%%%%%%%%%%%%%%%%%%%

%{\red MR: Compare the spinodal  stress patterns observed here with Golubovic's workon dynamics of buckling in 2d plates \cite{Golubovicetal1998}.} 

\section{Hydrodynamic description of actomyosin cytoskeleton}

    \begin{figure}[t]
        \includegraphics[scale=0.25]{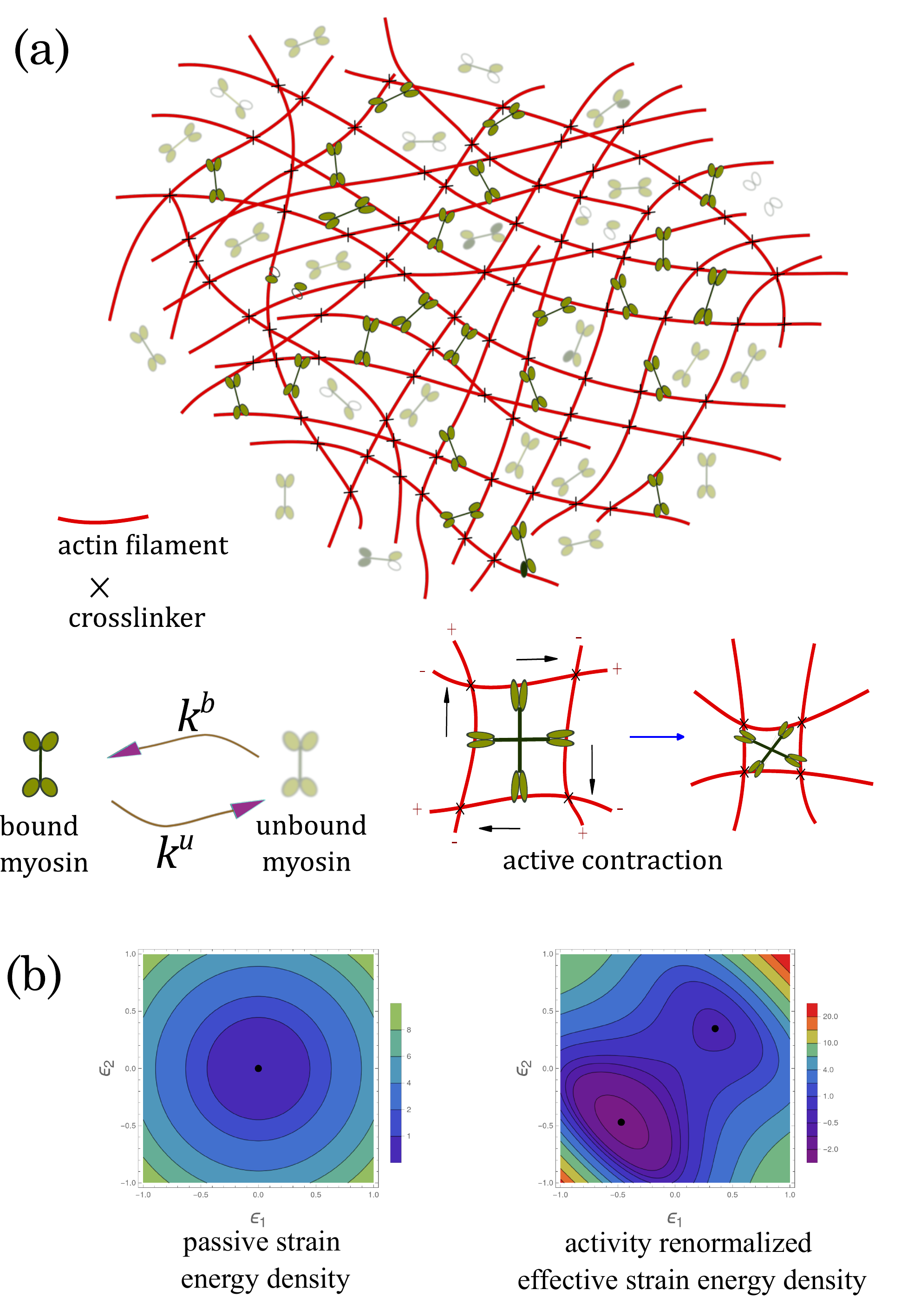}
        \caption{{\bf Actomyosin cytoskeleton as renewable active matter.} (a) ({\it top}) Schematic of the actin elastomer with crosslinkers that give it rigidity, together with bound/unbound Myo-II minifilaments. ({\it bottom left}) Transitions from bound to unbound Myo-II and vice versa occur with rates $k_u$ and $k_b$, respectively. ({\it bottom right}) Bound Myo-II applies contractile stresses on the actin meshwork as shown. (b) ({\it left}) Elastic energy contour in the space of principal strains $\epsilon_{1,2}$ for the passive elastomer showing linear elastic behaviour with a single energy minimum (black point). ({\it right}) Contours of the effective elastic energy $w(\boldsymbol{\epsilon})=\int_0^{\boldsymbol{\epsilon}}\boldsymbol{\sigma}(\boldsymbol{\epsilon}')\cdot d\boldsymbol{\epsilon}'$ renormalized by activity. Activity generates additional  minima
     %   for the effective elastic energy 
        and nonlinear stabilization~\cite{debsankar2017,RRT2024}.
        %(c) dynamic tension-compression non-reciprocity driven by strain dependent turnover.
        }
    \end{figure}

%To explore these ideas, %it suffices to 
We work with a %scalar 
%elasticity model of the 
$2$-dimensional
permanently crosslinked actin elastomer of mass density $\rho_a$,
 embedded in the cytosol, described by a 
 linearized strain tensor $\boldsymbol{\epsilon}:=(\nabla\boldsymbol{u}+\nabla\boldsymbol{u}^T)/2$, where $\boldsymbol{u}$ represents the
 displacement relative to an unstrained reference state.
The hydrodynamic linear momentum balance of the overdamped elastomer is 
%\rho_a\,\ddot{\boldsymbol{u}}+
$\Gamma\,\dot{\boldsymbol{u}}=\nabla\cdot\boldsymbol{\sigma}$, 
%where overdot represents ordinary time derivative, 
where $\Gamma$ is the friction of the elastomer with respect to the fluidic cytosol (whose dynamics we neglect since the volume fraction of the mesh is high), and $\boldsymbol{\sigma}(\boldsymbol{\epsilon},\rho_a,\{\rho_i\})$ is the total stress in the elastomer, which depends on the strain $\boldsymbol{\epsilon}$, the density of the actin mesh $\rho_a$, and the densities $\rho_i$ of the $i^{th}$ species of active bound myosin (stresslets).

The 
%constitutive equation for the 
total stress $\boldsymbol{\sigma}$ %, in the long time limit {\red ?}, 
is the summation of elastic stress $\boldsymbol{\sigma}^e$, the viscous stress $\boldsymbol{\sigma}^d$, and active stress $\boldsymbol{\sigma}^a$:  $\boldsymbol{\sigma}=\boldsymbol{\sigma}^e+\boldsymbol{\sigma}^d+\boldsymbol{\sigma}^a$~\cite{BanerjeeLiverpoolMarchetti2011,BanerjeeMarchetti2011,debsankar2017}. The elastic stress $\boldsymbol{\sigma}^e = \frac{\delta F}{\delta \,\boldsymbol{\epsilon}}$ is computed   
 from a  free-energy functional  $F[\boldsymbol{\epsilon},\rho_a]
=\int d{\bf r} f_B$, describing an isotropic elastomer,
%an isotropic, linear elastic material, 
$f_B=\frac{B}{2}\epsilon^2+\mu|\tilde{\boldsymbol{\epsilon}}|^2+C\,\delta\rho_a\,\epsilon+\frac{A}{2}\delta\rho_a^2$,
where $\epsilon:=\text{tr}\boldsymbol{\epsilon}$ is the dilational strain and $\tilde{\boldsymbol{\epsilon}}:=\boldsymbol{\epsilon}-(\epsilon/d)\mathbf{I}$ is the deviatoric strain, $B>0$ and $\mu>0$ 
are the passive elastic bulk and shear moduli, respectively, 
$\delta\rho_a:=\rho_a-\rho_a^0$ is the local deviation of $\rho_a$ from its state value $\rho_a^0$, $A^{-1}>0$ is the isothermal compressibility at constant strain, and $C>0$ couples dilation with local density variation.
The passive viscous stress of the elastomer is 
$\boldsymbol{\sigma}^d
=\eta_b\,\dot{\epsilon}\,\mathbf{I}+2\eta_s\,\dot{\tilde{\boldsymbol{\epsilon}}}$, where $\eta_{b,s}$ are the bulk and shear viscosities, respectively.
We take the active stress  $\boldsymbol{\sigma}^a$ to be isotropic and
of the form:
$\boldsymbol{\sigma}^a=\chi(\rho_a)\sum_i \zeta_i\,\rho_i\,\mathbf{I}$, 
where  $\zeta_i>0$ are activity coefficients representing contractile stress,
and $\chi(\rho_a)$ is a sigmoidal function which captures the dependence of the active stress on the local actin mesh density. %and hence on the local mesh strain. 
% \sout{We will further assume that the turnover of the actin mesh is fast compared to the its stress relaxation time scale}.
For this permanently crosslinked actin mesh,   the local density fluctuation $\delta\rho_a$ of actin is slaved  to the mesh strain, $\delta\rho_a=-\frac{C}{A}\epsilon$  (Appendix \ref{sect:a2}).
%obtained by setting $\frac{\delta F}{\delta\, (\delta\rho_a)}=0$.
%(as $\rho_a$ is enslaved to $\epsilon$: $\delta\rho_a=-\frac{C}{A}\epsilon$, {\red see Appendix Sec.\,?}).

The bound active stresslets exert contractile stresses and undergo  turnover, which can in general be
%	the advection-diffusion dynamics is given by, $\dot\rho_i+\nabla\cdot(\rho_i\,\dot{\boldsymbol{u}})=D\,\nabla^2\rho_i+k_i^{b}\,\rho_a - k_i^{u}(\boldsymbol{\epsilon})\,\rho_i$.
strain dependent; here we take the unbinding rates to be of the Bell-type form \cite{Bell1978}:
$k_i^{u}(\boldsymbol{\epsilon})=k_{i0}^{u}\, e^{\alpha_i\,\epsilon}$,
where $k_{i0}^{u}>0$,  are the strain independent parts of the respective rates, and $\alpha_i$ are dimensionless numbers; $\alpha_i>0$ represent {\it catch bonds} where local contraction (extension) will decrease (increase) the unbinding rate of the stresslets, while $\alpha_i<0$ represent {\it slip bonds} where local extension (contraction) will decrease (increase) the unbinding rate of the stresslets~\cite{Kovacsetal2007,Fernandez-Gonzalezetal2009,Mullaetal2022,debsankar2017}.

For a binary mixture of stresslets, the physics of segregation and the subsequent force patterning is explored by casting the hydrodynamic equations in terms of the average density
$\rho:=(\rho_1+\rho_2)/2$, and relative density $\phi:=(\rho_1-\rho_2)/2$ ($\rho_1$ is the more contractile species): 
%which in the overdamped limit reduces to, 
\begin{subequations}
\begin{align}
	& \dot{\boldsymbol{u}} = \nabla\cdot \boldsymbol{\sigma}, \label{eqn:maineqn:main:a}\\
	& \dot{\rho}+\nabla\cdot(\rho\,\dot{\boldsymbol{u}})=D\nabla^2 \rho +1 -\frac{C\,\epsilon}{A} -k^{u}_{\text{avg}}\,\rho-k^{u}_{\text{rel}}\,\phi, \label{eqn:maineqn:main:b}\\
	& \dot{\phi}+\nabla\cdot(\phi\,\dot{\boldsymbol{u}})=D\nabla^2 \phi + k^{b}_{\text{rel}}\bigg(1-\frac{C\,\epsilon}{A}\bigg) -k^{u}_{\text{avg}}\,\phi-k^{u}_{\text{rel}}\,\rho; \label{eqn:maineqn:main:c}
\end{align}
\label{eqn:main}
\end{subequations}
\hspace{-0.15cm}made dimensionless by setting the units of time, length and density,
as $t^\star:=1/k^{b}_{\text{avg}}$, $l^\star:=\sqrt{\eta_b/\Gamma}$ and $\rho_a^0$, where 
$k^{b,u}_{\text{avg}}:=(k^{b,u}_1+k^{b,u}_2)/2$, $k^{b,u}_{\text{rel}}:=(k^{b,u}_1-k^{b,u}_2)/2$ are the average and relative binding (unbinding) rates of the stresslets, respectively. The derivation of these equations together with the underlying assumptions
%and parameter ranges 
are presented in Appendix\,\ref{sect:goveqn}.

Note that despite writing the hydrodynamic equations explicitly for a binary mixture of stresslets, we may trivially use this to study the dynamics of a single myosin species by simply setting the contractility of the second species to zero ($\zeta_2=0$). If in addition, we set the unbinding rate $k^u_2$  to have a slip bond nature, then species 2 can act as a {\it probe} molecule, highlighting regions that are depleted in myosin (species 1). \footnote{Even a single component active isotropic fluid medium (e.g., small actomyosin patches/bundles in the cytosol), can  segregate into
high and low density regions, provided the active stress  depends on myosin density non-monotonically.}

%{\red For the actin mesh, the unit of length ($l=\sqrt{\eta_b/\Gamma}$) is $0.5 \mu \text{m}$ \cite{rauzi_lecuit}. For myosin II, the unit of time, $k_b^{-1}$ , can be estimated from the myosin FRAP data. From experiments we find $k_u = 0.2 \pm 0.08 s^{-1}$ \cite{munjal2015}
%The viscosity of the mesh is 50 Pa.s\cite{munro1}.} 

%The value of the Bulk's Modulus of the acto-myosin cytoskeleton can be taken to be $B= 42$ Pa \cite{munro2}.

%For the case of a single species of myosin stresslets, a similar set of dynamical equations in terms of $\boldsymbol{u}$ and $\rho$ can be written down (see SISec.\,4).

%the equations non-dimensionalised by setting the units of time 	Let  $k^{b}_{\text{avg}}:=(k^{b}_1+k^{b}_2)/2>0$, $k^{b}_{\text{rel}}:=(k^{b}_1-k^{b}_2)/2$ be the average and relative binding rates, respectively. We set the characteristic time scale by $t^\star:=1/k^{b}_{\text{avg}}$, the characteristic length scale by  $l^\star:=\sqrt{\eta/\Gamma}$ ($\eta$ is either  $\eta_b$ or $\eta_s$), and the characteristic density  $\rho_a^0$.
%We non-dimensionalize all the variables with these scales (details in the Supplementary). 

%	We will further assume the overdamped limit of the elastomer, i.e., $|\rho_a\,\ddot{\boldsymbol{u}}|\ll |\Gamma\,\dot{\boldsymbol{u}}|$.

By Taylor expanding the function $\chi(\rho_a)$ about its state value  $\rho_a^0$, and with $\delta\rho_a$ enslaved to $\epsilon$, the total stress can be recast
to cubic order in strain as, 
\begin{equation}
\boldsymbol{\sigma}=\Big(\sigma_0+\tilde{B}\,\epsilon
+B_2\,\epsilon^2+B_3\,\epsilon^3+\eta_b\,\dot{\epsilon}\Big)\mathbf{I}+2\mu\tilde{\boldsymbol{\epsilon}}+2\eta_s\,\dot{\tilde{\boldsymbol{\epsilon}}}
%(\nabla\epsilon,\ldots,,\rho,\phi,\nabla\rho,\nabla\phi,\ldots),
%2\mu\tilde{\boldsymbol{\epsilon}}+\dot{\boldsymbol{\epsilon}},
\label{constitutive-rel}
\end{equation}
with a purely active prestress $\sigma_0(\rho,\phi):=2\chi(\rho^0_a)(\zeta_{\text{avg}}\,\rho+\zeta_{\text{rel}}\,\phi)$,
where 	$\zeta_{\text{avg}}:=(\zeta_1+\zeta_2)/2>0$, $\zeta_{\text{rel}}:=(\zeta_1-\zeta_2)/2$ are the average and relative contractility, respectively, 
an activity renormalized linear elastic bulk 
modulus $\tilde{B}(\rho,\phi):=B-\frac{C^2}{A}-2\chi'(\rho^0_a)\,\frac{C}{A}(\zeta_{\text{avg}}\,\rho+\zeta_{\text{rel}}\,\phi)$ and  activity generated  nonlinear elastic bulk 
moduli $B_2(\rho,\phi), B_3(\rho,\phi)$,  linearly dependent on $\rho$ and $\phi$ (see Appendix\,\ref{sect:a3}). 

The material model, described by Eqs.\,\eqref{eqn:main},\,\eqref{constitutive-rel}, is dynamic, renewable and active, and breaks time reversal symmetry (TRS) in that it contains terms that are not derivable from a free energy. %As a consequence it exhibits unusual %material behaviour including elastic marginality and nonreciprocity.
The spatiotemporally inhomogeneous material moduli $\tilde{B}(\rho,\phi)$ and $B_{2,3}(\rho,\phi)$ lead to a rich effective elastic energy landscape with multiple wells, tuned dynamically by activity and material renewal. 
As we explore in detail later, accumulation of the stronger contractile motors (i.e., $\rho\gg 1$, $\phi>0$) can make $\tilde{B}(\rho,\phi)\to 0$ locally and even drive it to negative values, thus, triggering localized linear elastic instability, in which case the purely active higher order moduli $B_{2,3}$ would stabilize the nonlinear response~\cite{RRT2024}. 
%that generates additional minima in the effective elastic energy and provides nonlinear stabilisation of the material.
%$\tilde{B}:=B-\frac{C^2}{A}-2\chi'(\rho^0_a)\,\frac{C}{A}(\zeta_{\text{avg}}\,\rho+\zeta_{\text{rel}}\,\phi)$ is the renormalized/effective bulk modulus; $B_2:=\chi''(\rho^0_a)\bigg(\frac{C}{A}\bigg)^2(\zeta_{\text{avg}}\,\rho+\zeta_{\text{rel}}\,\phi)$ and $B_3:=-\frac{\chi'''(\rho^0_a)}{3}\bigg(\frac{C}{A}\bigg)^3(\zeta_{\text{avg}}\,\rho+\zeta_{\text{rel}}\,\phi)$ are purely active nonlinear elastic moduli;  
%	$k^{u}_{\text{avg}}:=(k^{u}_1+k^{u}_2)/2>0$, $k^{u}_{\text{rel}}:=(k^{u}_1-k^{u}_2)/2$ are the average and relative unbinding rates, respectively;  and 

%	(we set	$\Delta\mu=1$ for convenience). 
The renormalization of elastic moduli show up in the unusual mechanical behaviour of renewable active materials~\cite{RRT2024}.
Here we focus on how these dynamic material properties drive %dynamic 
stress patterning, 
%that are of 
directly relevant to the actomyosin cytoskeleton, leading to the emergence of singular tension structures
%(``stress fibers'')
and excitability.

%%%%%%%%%%%%%%%%%%%%%%%%%%%%%%%%%%%%%%%%%%%%%%%%%%%%%%%%%%%%%%%%%%%%%%%%%%%%%%%%%%%%%%%%%%%%%%%%%%%%%%%%%%%%%%%%%%%%%%%%%%%%%%%%%%%%%
%%%%%%%%%%%%%%%%%%%%%%%%%%%%%%%%%%%%%%%%%%%%%%%%%%%%%%%%%%%%%%%%%%%%%%%
%%%%%%%%%%%%%%%%%%%%%%%%%%%%%%%%

    \begin{figure*}[t]
        \includegraphics[scale=1.85]{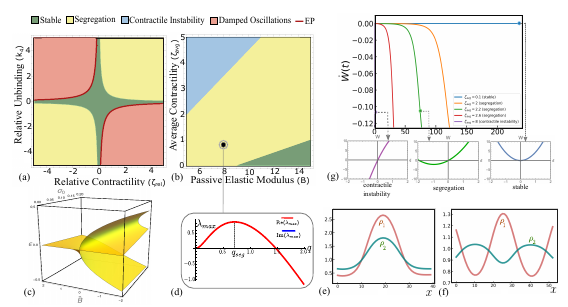}
        \caption{{\bf Linear stability phase diagrams and segregation instability in a binary mixture of stresslets.} (a) Phase diagram in relative unbinding versus contractility, with $B=8$, $\zeta_{\text{avg}}=1$, shows segregation instability (yellow) for $\zeta_{\text{rel}}k_4>0$. 
   %     (and damped oscillations for $\zeta_{\text{rel}}k_4<0$). 
        The red line corresponds to a line of exceptional points (EP), where the eigenvectors of the dynamical matrix ${\mathbf M}$ co-align. (b) Phase diagram in bare elastic modulus versus average contractility for  $\zeta_{\text{rel}}k_4>0$,  shows  transitions from the stable phase to segregation followed by elastic contractile  instability, with increasing $\zeta_{\text{avg}}$.
        (c) Graph of the minimum of the effective elastic energy at a fixed value of $\rho$ and $\phi$, i.e. $\sigma_0+\tilde{B}\epsilon+B_2\epsilon^2+B_3\epsilon^3=0$, showing the elastic instability $\tilde{B}\le 0$ as a pitchfork bifurcation of the unstrained state into a spinodal pattern of expansion and contraction, with the active prestress $\sigma_0$ acting as an ``external field''. Tensile prestress $\sigma_0>0$ biases the stable branch towards contraction ($\epsilon<0$).
	(d) Typical dispersion curve obtained from a linear analysis of the segregation instability, where  $q_{seg}$ is the fastest growing mode.
	(e,f) Snapshots of segregation %$\phi$-profile showing their time-evolution starting from a uniform configuration of stresslets (sequence of evolving profiles denoted by dashed line with progressively dark hues), %the tendency towards
 obtained from a numerical solution of the scalar version of Eq.\,\eqref{eqn:main} with periodic boundary condition in 1D (see \href{https://drive.google.com/file/d/14CVxT7SFx-76mTdeD2PDLCU-ONjg75jI/view?usp=sharing}{Movie 1}, \href{https://drive.google.com/file/d/16a1KWvMt95U8zEvvgcskpv9YLXXqApWs/view?usp=sharing}{Movie 2}). Co-localisation and separation of density peaks of the individual segregating stresselets, when (e) the two stresslets have catch bonds with $k_3=2$ (see Eq.\,\eqref{strainunbinding}), and when (f) stresslet 1(2) has catch(slip) bond with $k_3=-0.5$. The ratio $\rho_1/\rho_2$ within the domain depends on activity, turnover and the stress jump across the domain.
  We have set $C=A=D=1$, 
	$k_1=1$, $k_2=0$, $\chi(\rho_a^0)=1$ and $\chi'(\rho_a^0)=1$, $\zeta_{rel}=1$ and $\zeta_{avg}=2.8$ throughout. 
 %In panel (e), $k_3=2$ and in panel (f), $k_3=-0.5$. 
 (g) Power density $\dot{W}(t)$ (see text) monotonically decreases in the linear segregation regime; the rate of decrease goes to $-\infty$ as the average activity $\zeta_{\text{avg}}$ increases and the elastic instability is approached. Outsets show the minimum $\epsilon_{min}$ of the effective strain energy density $w_{lin}$ at fixed $(\rho,\phi)$ moving to $-\infty$ as $\zeta_{\text{avg}}$ increases (see Eq.\,\eqref{wlin}).}
\label{fig:phasedia-disp}
    \end{figure*}

%\section*{Positive reinforcement between activity and turnover leads to segregation of stresslets and force patterning}

\section{Positive reinforcement between activity and turnover leads to segregation of stresslets}

As the first step towards force patterning, we discuss a novel segregation into domains of high and low contractility. This is readily seen from a linear stability analysis 
%Experimental evidence for excitable singular force patterns in cytoskeleton (cite Thomas). 
%For a dynamic force pattern to emerge, a homogeneous and uniform mixture of stresslets must segregate first into regimes containing individual species. This active segregation happens spontaneously through a long wavelength linear instability in the homogeneous and symmetric ($\phi=0$), unstrained steady state of the actomyosin elastomer described by \eqref{eqn:main}. 
%Stability %about
about the undeformed symmetric mixture ($\boldsymbol{u}=\mathbf{0}$, $\phi=0$)
%the homogeneous unstrained steady state of the system
%considering perturbation along the strain direction to be purely isotropic,
%starting from a symmetric mixture of stresslets $\phi=0$,
%is $u=u_0=0$, $\rho=\rho_0:=(k_1-k_2\,k^{b}_{\text{rel}})/(k_1^2-k_2^2)$, $\phi=\phi_0:=(k_1\,k^{b}_{\text{rel}}-k_2)/(k_1^2-k_2^2)$.
%	focussing on 
%the special case  
described by the linear dynamical system, $\dot{\mathbf{w}}=\mathbf{M}\mathbf{w}$ (see Appendix\,\ref{sect:b2}), which governs the evolution of the perturbation  $\mathbf{w}=\Big(\hat{\delta \epsilon}(t,{\bf q})\,\,\hat{\delta \rho}(t,{\bf q})\,\,\hat{\delta \phi}(t,{\bf q})\Big)^T$ that depends on the Fourier wave vector ${\bf q}$. Here, $\hat{\delta \epsilon}\equiv\delta\hat{u}_{\parallel}:=\hat{\delta\boldsymbol{u}}\cdot\mathbf{q}$ is the Fourier transform of the dilation strain field $\epsilon$, %is the `longitudinal' deformation mode in the Fourier space, while the `transverse' deformation mode $\hat{\delta u}_{\perp}:=\hat{\delta\boldsymbol{u}}\times\mathbf{q}$, the Fourier transform of $\text{curl}\,\boldsymbol{u}$, is uncoupled from the rest of the dynamics as shear elasticity remains unaffected by the activity and turnover.
%for a one dimensional system, i.e.
%    The neccesary condition to maintain this is $k^{b}_{\text{rel}}=k_2/k_1$. We further see that      $\rho_0=1/k_1$, and $\tilde{B}_0=B-\frac{C^2}{A}-2\,\chi'(\rho_a^0)\,\frac{C}{A}\,\frac{\zeta_{\text{avg}}}{k_1}$. For the linearized dynamics about this steady state, we consider the scalar version of the equations in the Fourier space: 
%	\begin{equation}
%		\dot{\mathbf{w}}=\mathbf{M}\math%bf{w}
%	\end{equation}
%	where $q$ is the 1D wave `vector',  
 and the dynamical matrix is 
\begin{widetext}
\begin{equation}
	\mathbf{M}=\left[\begin{array}{ccc}
		-\frac{(\tilde{B}+\mu)\,q^2}{1+q^2} & -\frac{2\chi(\rho_a^0)\,\zeta_{\text{avg}}\,q^2}{1+q^2} & -\frac{2\chi(\rho_a^0)\,\zeta_{\text{rel}}\,q^2}{1+q^2} \\
		-\bigg(\frac{C}{A}+\frac{k_3}{k_1}-\frac{(\tilde{B}+\mu)\,q^2}{(1+q^2)k_1}\bigg) & -D q^2-k_1+\frac{2\chi(\rho_a^0)\,\zeta_{\text{avg}}\,q^2}{(1+q^2)k_1} & -k_2+\frac{2\chi(\rho_a^0)\,\zeta_{\text{rel}}\,q^2}{(1+q^2)k_1} \\
		-\bigg(\frac{k_2}{k_1}\frac{C}{A}+\frac{k_4}{k_1}\bigg) & -k_2 & -D q^2- k_1
	\end{array}\right],
	\label{lindynamicalmatrix}
\end{equation}
\end{widetext}
 where $q:=|\mathbf{q}|$, and $\tilde{B}:=B-\frac{C^2}{A}-2\chi'(\rho^0_a)\,\frac{C}{A}\frac{\zeta_{\text{avg}}}{k_1}$ is the activity renormalized bulk modulus of the homogeneous symmetric mixture. Here the turnover rates appearing in  Eq.\,\eqref{eqn:main} have been expanded to linear order in $\epsilon$,
\begin{subequations}
	\begin{align}
		& k^{u}_{\text{avg}}(\boldsymbol{\epsilon})=k_1+k_3\,\epsilon+o(|\boldsymbol{\epsilon}|),\\
		& k^{u}_{\text{rel}}(\boldsymbol{\epsilon})=k_2+k_4\,\epsilon+o(|\boldsymbol{\epsilon}|);
	\end{align}
\end{subequations}
where the parameters that appear in Eq.\,\eqref{lindynamicalmatrix}, namely
\begin{equation}
	k_1:=\frac{k^{u}_{10}+k^{u}_{20}}{2}>0,~~\text{and}~~k_2:=\frac{k^{u}_{10}-k^{u}_{20}}{2}\, ,
\end{equation}
represent the  average and relative bare unbinding rates (i.e., strain independent), respectively, and
\begin{equation}
\label{strainunbinding}
	k_3:=\frac{k^{u}_{10}\alpha_1+k^{u}_{20}\alpha_2}{2},~~\text{and}~~k_4:=\frac{k^{u}_{10}\alpha_1-k^{u}_{20}\alpha_2}{2},
\end{equation}
represent the coefficients of the strain dependent parts of the  relative and average unbinding rates, respectively. We see that the linear dynamics of the transverse Fourier displacement mode $\delta\hat{u}_{\perp}:=\hat{\delta\boldsymbol{u}}\cdot\mathbf{q}_{\perp}$ uncouples from the rest of the dynamics.

Note that $\mathbf{M}$ is
non-Hermitian  due to TRS breaking; as a consequence, the eigenvectors along which the perturbations propagate are no longer orthogonal to each other and may even co-align for some parameter values, as we will see later.

With the simplifying assumption that the bare rates of binding and unbinding are identical in the two myosin species, the instabilities are determined solely by the maximum eigenvalue of $\mathbf{M}$,
%: $k_2=0$, hence, $k^{b}_{\text{rel}}=0$, meaning that the bare unbinding rates (strain independent) and binding rates are identical. 
\iffalse
(i.e., $k^{u}_{10}=k^{u}_{20}$), 
and the binding rates are identical as well (i.e., $k^{b}_{1}=k^{b}_{2}$). It follows that $k_4=\frac{k^{u}_{10}}{2}(\alpha_1-\alpha_2)$. 
\fi	
%	The three eigenvalues of $\mathbf{M}$ are
\begin{equation}
%		\lambda_1=-k_1-D\,q^2,~~~~
%		\lambda_2=-\frac{\lambda_a+\sqrt{\lambda_b}}{2(1+q^2)},~~~~
\lambda_{max}=-\frac{\lambda_a-\sqrt{\lambda_b}}{2(1+q^2)}
\label{egvalues}
\end{equation}
 where 
\begin{subequations}
\begin{align}
	&\lambda_a:=k_1+\bigg( \tilde{B}+\mu-\frac{2\chi(\rho_a^0)\zeta_{\text{avg}}}{k_1}+D+k_1\bigg)\,q^2+D\,q^4,\label{lam_a}\\
	&\lambda_b:= \lambda_a^2-4\,q^2(1+q^2)k_1^2\bigg[(\tilde{B}+\mu)\,D\,q^2+\nonumber\\
	&\hspace{3mm}k_1\Bigg(\tilde{B}+\mu-2\chi(\rho_a^0)\bigg(\frac{\zeta_{\text{avg}}}{k_1}\Big(\frac{C}{A}+\frac{k_3}{k_1}\Big)+\frac{\zeta_{\text{rel}}}{k_1}\frac{k_4}{k_1}\bigg)\Bigg)\bigg].\label{lam_b}
\end{align}
\end{subequations}
%$q:=|\mathbf{q}|$, $\tilde{B}_0:=B-\frac{C^2}{A}-2\chi'(\rho^0_a)\,\frac{C}{A}\frac{\zeta_{\text{avg}}}{k_1}$ is the activity renormalized bulk modulus of the homogeneous symmetric mixture,  $k_1$ is the average bare unbinding rate, and $k_3$ and $k_4$ are the average and relative (strain dependent) unbinding rates, respectively.

%%%%%%%%%%%%%%%%%%%%%%%%%%%%%%%%%%%%

With this in hand, we are in a position to describe both the elastic and segregation instabilities of the uniform elastomer driven by active stresses and turnover (Fig.\,\ref{fig:phasedia-disp}).

%\subsection{Contractile activity and turnover driven elastic instability} 
{\bf Elastic (contractile) instability.} Starting from a stable unstrained elastomer, we see that large enough average activity $\frac{\zeta_{\text{avg}}}{k_1}$
drives the renormalized  longitudinal elastic modulus $\tilde{B}+\mu$ of the symmetric mixture to negative values, leading to a linear elastic
instability of the underlying elastomer~\cite{debsankar2017,RRT2024} that affects all modes $q\in[0,\infty)$ (see Fig.\,\ref{fig:phasedia-disp}(b) and dispersion curve in Fig.\,\ref{dispersionsi}). 
 This elastic spinodal decomposition into patterns of positive and negative $\epsilon$ is a pithfork bifurcation, with prestress as an `external field' controlling it (shown in Fig.\,\ref{fig:phasedia-disp}(f)).
%This 
The contracted ($\epsilon<0$) domains eventually show up as self-penetration and subsequent collapse (halted by steric effects) of the 
%\redout{uniform} 
contractile  mixture,
%	to a single point 
unless constrained by appropriate boundary conditions~\cite{RRT2024}.
%, as we will see later.  
%Since the elastomer is built from cross-linked stiff actin filaments, 
The microscopic origin of this elastic instability can be attributed to the microbuckling \cite{Roncerayetal2016},  the loss of crosslinkers, resulting in relative sliding of the semiflexible actin filaments  \cite{MurrellGardel2012}, to micro-wrinkling \cite{MullerKierfeld2014} and 
 %can also be linked to the  
 stretching-to-bending transition \cite{Salman2019}.
%\cite{debsankar2017}.   %	Material instability of the underlying elastomer (unbounded growth of $\delta u$) is related to the change of sign of the renormalized elastic modulus $\tilde{B}_0$ from being positive to negative. Since $B-\frac{C^2}{A}>0$ (since we started with a stable passive elastomer), a large positive value of the normalized average activity $\frac{\zeta_{\text{avg}}}{k_1}$ leads to $\tilde{B}_0<0$.
%	In this linear approximation, contractile instability means material self-penetration.  

%	In reality, steric hindrance of the actomyosin complexes provide an ultraviolet cut-off for these contracted configurations. The cutoff length scale is around $300$\,nm, that corresponds to the dimension of myosin-II bipolar filament \cite{Bershadsky2017}.

%%%%%%%%%%%%%%%%%%%%%%%%%%%%%%%%%%%%%%%%%
We will see that the force patterning that we discussed earlier is achieved through entrapping this 
%(system spanning) 
contractile instability into segregated domains.
%within the cell body. 
In the linear theory, this segregation, and other nonequilibrium phases, show up in the mechanically stable regime of the active elastomer,
$\tilde{B}+\mu>0$.

%; the nature of the instability being governed by the value of $\lambda_{max}$.

%%%%%%%%%%%%%%%%%%%%%%%%%%%%%%%%%%%%%%
{\bf Segregation instability.}
As $\lambda_b$ increases beyond $0$,  $\lambda_{max}$ goes from being negative (stable) to positive, leading to a long-wavelength instability in $\phi$. 
The fastest growing wave-vector $q_{seg}$
%\begin{equation}
%	q_{seg}:=\sqrt{-\frac{k_1}{D {\tilde{B}_0}}\left(\tilde{B}_0-2\frac{\zeta_{\text{avg}}}{k_1}\bigg(\frac{C}{A}+\frac{k_3}{k_1}\bigg)-2\frac{\zeta_{\text{rel}}}{k_1}\frac{k_4}{k_1}\right)}
%	\label{qseg}
%\end{equation}
sets the characteristic width $q_{seg}^{-1}$ of the segregated pattern (Fig.\,\ref{fig:phasedia-disp}(c), Eq.\,\eqref{eqnqseg}), provided $\tilde{B}$ is bounded between,
\begin{equation}
0	<\tilde{B}+\mu< 2\chi(\rho_a^0)\Bigg(\frac{\zeta_{\text{avg}}}{k_1}\bigg(\frac{C}{A}+\frac{k_3}{k_1}\bigg) + \frac{\zeta_{\text{rel}}}{k_1}\frac{k_4}{k_1}\Bigg).
\label{segrecond}
\end{equation}
%	{\red All modes $q$ in the interval $[0,q_{seg}]$ undergo segregation (see the dispersion curve Fig.\,\ref{fig:phasedia-disp}(d))}. 
This linear segregation regime is typically realised 	%	Note that for fixed values of normalized average activity $\zeta_{\text{avg}}/k_1$ and average strain dependent unbinding $k_3/k_1$ such that $\tilde{B}_0>0$ and $\tilde{B}_0-2\frac{\zeta_{\text{avg}}}{k_1}\bigg(\frac{C}{A}+\frac{k_3}{k_1}\bigg)>0$, segregation instability is generic 
when the relative activity $\zeta_{\text{rel}}$ and strain dependent unbinding $k_4$ have the same sign, which implies  $k^u_1(\epsilon)<k^u_2(\epsilon)$, since the stresslets are contractile.  To drive segregation, the stresslet with stronger contractile activity must have a lower strain-dependent unbinding rate.  Note that the density peaks of the individual stresslets colocalise (Fig.\,\ref{fig:phasedia-disp}(e), \href{https://drive.google.com/file/d/14CVxT7SFx-76mTdeD2PDLCU-ONjg75jI/view?usp=sharing}{Movie 1}) unlike in conventional phase separation. %giving rise to {\it granular}~\cite{SahaGolest} or {\it speckled} domains. 
This is reminiscent of the co-assembly of Myosin IIA and IIB during the formation of stress fibers~\cite{Beachetal2014, Weissenbruchetal2021,Weissenbruchetal2022}, and occurs when both stresslets exhibit catch-bond behaviour ($\alpha_1, \alpha_2>0$). For slip-bond response of the weaker species ($\alpha_1>0$, $\alpha_2<0$), the individual density peaks separate (Fig.\,\ref{fig:phasedia-disp}(f), \href{https://drive.google.com/file/d/16a1KWvMt95U8zEvvgcskpv9YLXXqApWs/view?usp=sharing}{Movie 2}). 
% {\blue This segregation peak-separation can be observed in single myosin species systems (with $\zeta_2\equiv0$). If in addition, we set the unbinding rate $k^u_2$  to have a slip bond nature, then species 2 can act as a {\it probe} molecule, highlighting regions that are depleted in myosin (species 1).}

%	Setting $\zeta_{\text{rel}}=0$ (identical activity) or $k_4=0$ (identical strain dependent  unbinding) leads to the usual clumping instability~\cite{kabir,banerjeeetal2017}.

%	observed in single component stresslets

%%%%%%%%%%%%%%%%%%%%%%%%%%%%%%%%%%%%%%%%%%%%%%%%%%%%%%%%%%%%%%%%%%%%%%%%%%
	
 {\bf Driving force for linear segregation.}
Since the stresslets do not directly interact with each other, the driving force for this segregation in the linear theory must come from their indirect interaction through the elastomer strain. 
%We find from a 1D numerical solution in a system of size $L$, that the power density $\dot{W}=(1/L)\int \dot{w}_{lin}\,dx$ associated with 
This is best seen by defining the activity renormalized ``elastic energy density'' in the 
%linearized regime
linear elastic regime, as $w_{lin}=\int_0^{\epsilon} \sigma(\epsilon') d{\epsilon}':=\sigma_0(\rho,\phi)\,\epsilon+\frac{1}{2}\tilde{B}(\rho,\phi)\,\epsilon^2$.
%is negative 
%in the segregated phase
%(Fig.\,\ref{fig:phasedia-disp}(g)), i.e., $W$ is a Lyapunov functional driving segregation of the stresslets. 	
%%The appearance of $\rho_2$ micro-domains within the $\rho_1$ (more contractile) domain, is a consequence of the interplay between the strain dependent catch-bond turnover and this driving force.
%Note that the value of strain in the linearly segregated domains of high contractility is set by the minima of $w$,   $\epsilon_{\text{min}}=-\frac{\sigma_0}{B_{\textbf{pass}}-\frac{\chi'(\rho_a^0)}{\chi(\rho_a^0)}\frac{C}{A}\sigma_0}$, and depends directly on the active prestress $\sigma_0$. The active prestress is significant in keeping the segregated domains of stronger contractile %%stresslets well-separated, preventing them from forming one single clump.
Now for any fixed value of $\rho$ and $\phi$, the minimum of $w_{lin}$ occurs at the strain
\begin{equation}
\label{wlin}
	\epsilon_{min}=-\frac{\sigma_0}{\tilde{B}}=-\frac{2\,\chi(\rho^0_a)\big(\zeta_{\text{avg}}\,\rho+\zeta_{\text{rel}}\,\phi\big)}{B-\frac{C^2}{A}-2\,\chi'(\rho^0_a)\,\frac{C}{A}\big(\zeta_{\text{avg}}\,\rho+\zeta_{\text{rel}}\,\phi\big)}.
\end{equation}
%the energy at this minima is
%\begin{eqnarray}
	%&&w_{min}=w(\epsilon_{min})=-\frac{\sigma^2_0}{2\,\tilde{B}}\nonumber\\
% &&=-\frac{1}{2}\Bigg(\frac{\Big(2\,\chi(\rho^0_a)\big(\zeta_{\text{avg}}\,\rho+\zeta_{\text{rel}}\,\phi\big)\Big)^2}{B-\frac{C^2}{A}-2\,\chi'(\rho^0_a)\,\frac{C}{A}\big(\zeta_{\text{avg}}\,\rho+\zeta_{\text{rel}}\,\phi\big)}\Bigg).
%\end{eqnarray}
Without  activity, $\epsilon_{min}=0$. With activity and  large $\zeta_{\text{rel}}$ (implying segregation), $\sigma_0$ is large and $\tilde{B}$ is a small positive number; hence, $\epsilon_{min}$ becomes highly contractile. As the elastic instability is approached with further increase in activity, $\epsilon_{min}\to-\infty$ (see Fig.\,\ref{fig:phasedia-disp}(g) insets). 

Though we have not found a completely analytical proof, we see from a 1D numerical solution in a system of size $L$, that the power density $\dot{W}=(1/L)\int \dot{w}_{lin}\,dx$ associated with 
linear segregation regime
is negative 
(Fig.\,\ref{fig:phasedia-disp}(g)), i.e., $W$ is a Lyapunov functional driving segregation of the stresslets.

\begin{figure}[t]
	\centering
	\includegraphics[scale=0.7]{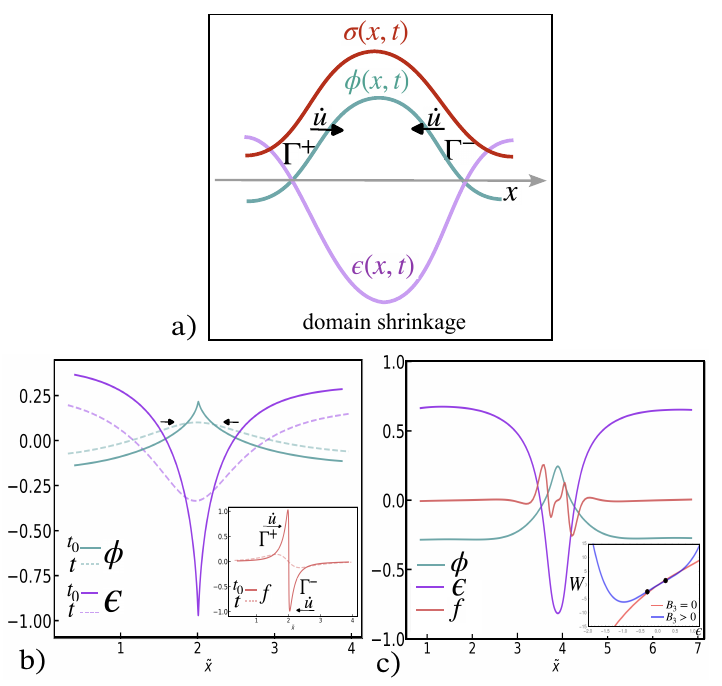}
	\caption{%Profiles of $\phi$, $\epsilon$ and the stress $\sigma$ at the end of linear segregation instability. At this stage nonlinear effects become significant and drive the domain fronts to move towards each other with speed $\dot u$, shrinking the domain to  eventually form a singular tension field in a finite time.
 {\bf Mechanism of singularity formation.} 
 (a) Schematic of a single domain showing the profiles of the segregation field $\phi$, strain $\epsilon$ and stress $\sigma$. The magnitude of the velocities of
the interfaces $\Gamma^\pm$ bounding the domain are $\dot{u}$, causing them to move towards each other as the singularity develops.
 (b) Snapshots from 1D numerical simulation of a single segregated domain showing the segregation parameter $\phi$ and strain $\epsilon$ profiles at an early time point $t$ (dashed line) and later time point $t_0$ (solid line) at which the singularity develops when the steric term $B_3=0$. Inset shows the corresponding force $f=\partial_x\sigma$ ($\equiv\dot{u}$) that drives the velocity of the interfaces $\Gamma^\pm$ towards each other,
 %near the transition from linear to nonlinear regime, with the steric term $B_3=0$. Dotted lines mark states away from singularity and solid lines, right at the singularity. %Displacement $u$ tends to become discontinuous at the sites of high contractility $\epsilon\ll 0$.
%Jumps in the stress $\sigma$ of opposite signs across the interfaces 
%Forces of opposite signs at the interfaces $\Gamma^\pm$ bounding the domain cause them to move towards each other with speed $\dot{u}$, 
leading to growing amplitudes of $\phi$ and $\epsilon$ and an eventual singularity. 
%\sout{The ensuing displacement discontinuity at and beyond the singularity marks self-penetration of the elastomer.} 
All plots are shown in the deformed coordinate $\tilde{x}=x+u(x,t)$. Parameters 
%used for the numerics in panel (b) 
are $B=6, \chi(\rho_a^0)=1, \chi(\rho_a^0)'=1, \chi(\rho_a^0)''=0.5, \chi(\rho_a^0)'''=0, \zeta_{avg}=2, \zeta_{rel}=2$. 
(c) Physical resolution of the singularities in $\phi$, $\epsilon$ and $f$,  by introducing a large steric term $B_3=10$. ({\it Inset}) Comparison of the effective elastic energy landscapes for $B_3=0$ and $B_3=10$. The two black points $(-B_2\pm\sqrt{B_2^2-3\tilde{B}B_3})/3B_3$, for fixed $(\rho,\phi)$, mark the elastic spinodals bounding the unstable regime of the energy landscape for a finite $B_3>0$; the spinodals move to infinity as $B_3\to 0$. See \href{https://drive.google.com/file/d/1nAkaJR_G6udob4C0n7vA6DdXK_b3O7D4/view?usp=sharing}{Movie 3} for (b) and (c). } 
	\label{frontsmove}
\end{figure}

%%%%%%%%%%%%%%%%%%%%%%%%%%%%%%%%%%%%%%%%%%%%%%%%%%%%%%%%%%%%%%%%%%%%%%%%%%%%%%%%%%%%%%%%%%%%%%%%%%%%%%%%%%%%%%%%%%%%%%%%%%%%%%%%%%%%%%%%%%%%%%%%%%%%%%%%%%%%%%%%%%%%%%%%%%%%%%%%%%%%%%%%%%%%%%%%%%%%%%%%

\section{Finite-time Elastic Singularities} 
As time passes, nonlinear terms in the dynamical equations become prominent;
the growing spinodal patterns predicted by the linear analysis, start to shrink into singular  structures that carry tension.
%, such as punctae,  1-D singular tension chains, or 2-D sheets,
%depending on the form of the  active nonlinear anisotropic stress $\boldsymbol{\Sigma}^a$.
%The exponential growth of the linear segregation instability quickly leads to a stage where nonlinear effects become significant. However, 
This behaviour is quite unlike the usual segregating mixture, where nonlinearities  temper the exponential growth of the domains to a slower power-law growth~\cite{Bray2002}.

To see this, let us focus on a %1-D section of a 
single domain of the evolving spinodal pattern in 1D of width $\sim q^{-1}_{seg}$ and bounded by two interfaces $\Gamma^{\pm}$, 
Fig.\,\ref{frontsmove}(a). 
The segregation field $\phi >0$ inside the domain 
and $\phi<0$ outside.
The forces $\partial_x\sigma$ acting across this domain  cause it to gradually shrink,
%Within the strip, the stress is highly tensile ($\epsilon <0$), whereas outside the stress is relatively of low magnitude  ($\epsilon <0$). 
%The result of this is that, as a result of stress jump across the interfaces $\Gamma^\pm$, the interfaces move {\it towards} each other, 
concentrating the more contractile stresslet within. Eventually, the center of this shrinking domain enters the elastic instability regime $\partial^2_{\epsilon\epsilon}w(\epsilon;\rho,\phi)\le0$
when %$\phi$
the stresslet density is large enough (Fig.\,\ref{frontsmove}), following which there is no escape from collapse, leading to the formation of singular force carrying structures in finite time. These singular tensile structures which harbour the more contractile stresslet,
%(lying along the curve $\mathcal{S}$, the `axis' of the collapsing strip, Fig.~\ref{fig:singularformation}(a)),
%made predominantly of the stronger stresslets (such as myosin IIA), 
remain well separated owing to the active prestress in the surrounding 
elastic medium which predominantly harbours the less contractile stresslet.

In an actual physical situation, the elastic singularity is never reached and one ends up with highly concentrated tensile regions of finite width set by the constraint of steric hindrance, represented by the $\epsilon^3$ term in the effective elastic stress (Eq.\,\eqref{constitutive-rel}), see Fig.\,\ref{frontsmove}(c). This is consistent with the observations~\cite{Beachetal2014,Weissenbruchetal2021}  on the segregation of the more contractile non-muscle Myosin IIA from the less contractile Myosin IIB,C.

\iffalse
If in the dynamical equations, we  were to prevent %now
the turnover of the stresslets immediately after the initial linear segregation regime, then their densities would be conserved. In this situation, the domain shrinkage %of the central bulk region 
would ultimately stop due to the $\epsilon^4$ term in the effective elastic energy density, arising from steric effects that are inevitably present (Fig.\,\ref{frontsmove}). 
%by the pseudo wells in the effective strain energy landscape %strain minima 
%introduced by the activity induced 
As a result, the singularity is never reached and one ends up with highly concentrated tensile regions of finite width set by the physical constraint of steric hindrance. In presence of turnover of catch bond type, with the catch bond for $\rho_1$ stronger than $\rho_2$ (i.e., $k_4>0$, same sign as $\zeta_{\text{rel}}$), this width will be modified due to violation of local detailed balance.
\fi

{\bf Self-similar solution for finite-time singularity.}
Following the above discussion, we analyse
the approach to the finite-time singularity starting from 
the linear segregation pattern,  by dropping the steric stabilising cubic term in elastic stress and the turnover of stresslets. We also neglect the  viscous stresses for convenience, though this does not pose any difficulties. The
dynamical equations in 1D then reduce to the conservative system
\begin{subequations}
	\begin{align}
		& \dot{\rho}=\partial_{x}\big(D\partial_{x}\rho-\partial_x\sigma\,\rho\big)\label{rho-1d-simple}\\
		& \dot{\phi}=\partial_{x}\big(D\partial_{x}\phi-\partial_x\sigma\,\phi\big)\label{phi-1d-simple}\\
		&\dot{\epsilon}=\partial^2_{xx}\sigma,\label{epsilon-1d-simple}
	\end{align}
	\label{dyn-1d-simple}
\end{subequations} 
where
\begin{eqnarray}
	\sigma &=& 2\,\chi(\rho^0_a)\,(\zeta_{\text{avg}}\,\rho+\zeta_{\text{rel}}\,\phi)\nonumber\\
	&&+\bigg(B-C^2-2\,\chi'(\rho^0_a)\,C\big(\zeta_{\text{avg}}\,\rho+\zeta_{\text{rel}}\,\phi\big)\bigg)\,\epsilon \nonumber\\
 &&+\chi''(\rho^0_a)\,C^2\,\big(\zeta_{\text{avg}}\,\rho +\zeta_{\text{rel}}\,\phi\big)\,\epsilon^2.
	\label{stress:eqn}
\end{eqnarray}

%Starting from a white noise perturbation of the homogeneous unstrained symmetric mixture, the approach to singularity happens through stages.
During the initial stages of linear segregation when $\rho$, $\phi$ and $\epsilon$ are relatively small, the first two terms (active prestress and linear elasticity) in the stress field Eq.\,\eqref{stress:eqn} dominate. However, as $\rho$, $\phi$ and $\epsilon$ grow exponentially due to linear instability, they become large inside the segregated domains where local density is high. When this happens, the  lowest order nonlinearity, the $\epsilon^2$ term in Eq.\,\eqref{stress:eqn} starts dominating over other terms. 
%in regions where the density is high. 
%this $\epsilon^2$ nonlinearity in the stress is the dominant driving mechanism,
This dominance of a particular power of the driving field in the final approach to the singularity, 
is at the origin of the scale invariance of the equation, and, hence, of the solution near the singularity 
\cite{eggersfontelos-book2015}. Note that in the relatively low density regions between the high density domains, the dominant term is the active prestress, as the effective bulk modulus in the second term in Eq.\,\eqref{stress:eqn} is small. Thus the interface between the high and low density regions, is {\it pushed} by the active prestress and {\it pulled} by the contractile instability.

%From our discussion above, 
We expect that there exists a fixed space-time location $(x_0,t_0)$ at which the the first singularity appears. The scale invariance of the solution near this singularity implies that,
starting from a typical segregated domain at time $t<t_0$ as shown in Fig.\,\ref{frontsmove}, the height of the profiles $\rho$, $\phi$ and $\epsilon$ must grow and the domain width must shrink at $t\to t_0$, as power laws with appropriate {\it universal} scaling exponents. 
%This universality of the scaling exponents is 
This a consequence of the fact that they arise solely from the structure of the partial differential equations, and are independent of initial conditions \cite{eggersfontelos-book2015}.
This motivates us to express the solution as a self-similar form,
\begin{subequations}
	\begin{align}
		&\rho(x,t)=\frac{C_0}{(t_0-t)^{s}}R(\xi),\label{simtrans-rho}\\
		&\phi(x,t)=\frac{A_0}{(t_0-t)^{p}}\Phi(\xi),\label{simtrans-phi}\\
		&\epsilon(x,t)=\frac{B_0}{(t_0-t)^{q}}E(\xi),\label{simtrans-e}
	\end{align}
	\label{simtrans}
\end{subequations}
		with the scaling variable, $\xi:=(x-x_0)/(t_0-t)^r$.
In the above,		
	$A_0$, $B_0$ and $C_0$ are constants; $s>0$, $p>0$, $q>0$ and $r>0$ are the scaling exponents; and $R$, $\Phi$ and $E$ are analytic scaling functions.

With these scaling forms, we are left with the following  unknowns -- the blowup time and location $t_0, x_0$, the exponents $s$, $p$, $q$ and  $r$, and the  similarity profiles $R(\xi)$, $\Phi(\xi)$ and $E(\xi)$. These are obtained by substituting the similarity forms Eqs.\,\eqref{simtrans} into the Eqs.\,\eqref{dyn-1d-simple}. Since diffusion cannot produce any finite-time singularity, we will ignore the diffusion term in the $\rho$ and $\phi$-equations.

%%%%%%%%%%%%%%%%%%%%%%%%%%%%%%%%%%%%%%%%%%%%%%%%%%%

\begin{figure}[t]
\centering
\includegraphics[scale=.52]{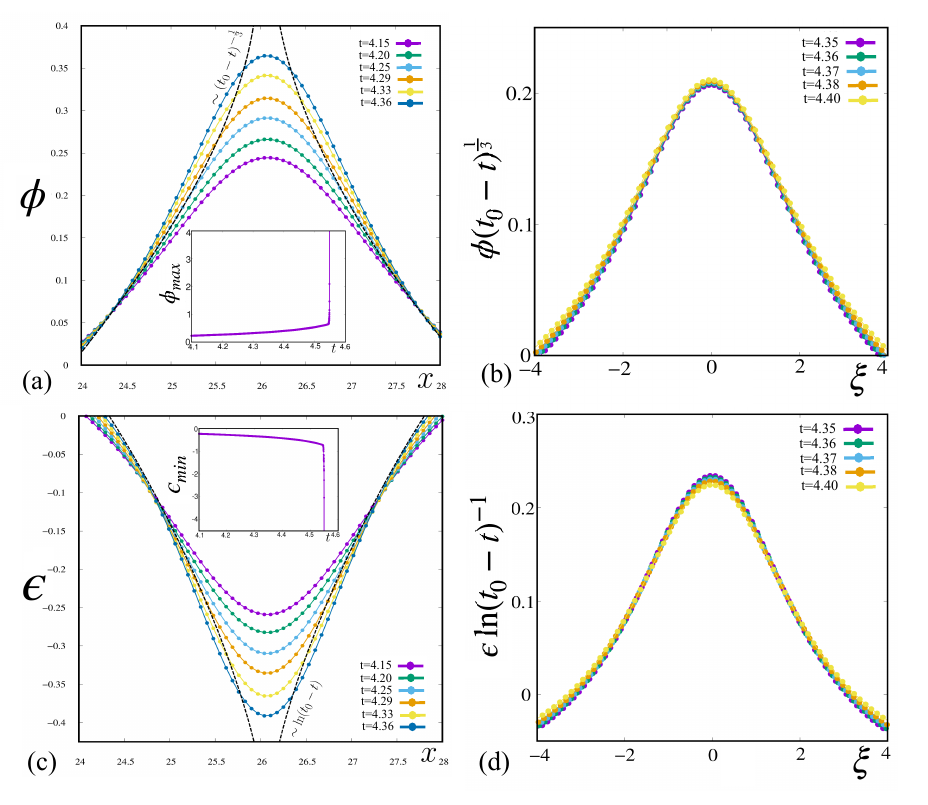}
\caption{{\bf %Emergence 
Numerical analysis of finite time singularities.} 
%Inset (a): Snapshot from 1D numerical simulation of a single segregated domain showing the segregation order parameter $\phi$, strain $\epsilon$, stress $\sigma$ {\red and displacement $u$} near the transition from linear to nonlinear regime. Displacement $u$ tends to become discontinuous at the sites of high contractility $\epsilon\ll 0$. Jumps in the stress $\sigma$ of opposite signs across the interfaces $\Gamma^\pm$ cause them to move towards each other with speed $\dot{u}$, leading to ever growing amplitudes of $\phi$ and $\epsilon$ and an eventual singularity in the middle. The ensuing displacement discontinuity at these sites marks self-penetration of the elastomer. 
(a)-(d) Numerical results verifying the scaling profiles of $\phi$ and $\epsilon$ and the formation of a singularity in finite time. (b,d) show scaling collapse of the $\phi$ and $\epsilon$ profiles near the singularity, as predicted from theory; $\xi=(x-x_0)/(t_0-t)^{\frac{1}{3}}$ is the scaling variable. Starting from a homogeneous unstrained symmetric state, a 1D numerical solution of Eq.\,\eqref{eqn:main} gives the value of  the finite time blowup $t_0=4.55$ 
%(in the unit of time unit $t^{\star}$), 
and 
blowup location $x_0=26.06$ (see insets in (a) and (c)).
	%(f) Cauchy convergence of $\phi(x,t)$ in time $t$ as the singularity is approached; points closer to the blow-up location $x_0$ diverge faster.
}
\label{fig:singularformation}
\end{figure}

%%%%%%%%%%%%%%%%%%%%%%%%%%%%%%%%%%%%%%%%%%%%%%%%%%%%%%%%%%%%%%%

Since we have ignored binding-unbinding,  
$\rho(x,t)$ and $\phi(x,t)$ are conserved, thus
	\begin{subequations}
		\begin{align}
			&\int_{-L/2}^{L/2}\rho(x,t)\,dx=\int\frac{C_0}{(t_0-t)^{s-r}}R\,d\xi=const.,~~\text{and}\label{consvrho}\\
			&\int_{-L/2}^{L/2}\phi(x,t)\,dx=\int\frac{A_0}{(t_0-t)^{p-r}}\Phi\,d\xi=const.\label{consvphi}
		\end{align}
	\label{consvcod}
	\end{subequations}
%	Equation\,\eqref{consvcod} 
 which immediately implies $s=p=r$.
	To get the asymptotic form of the singularity, we analyse the  coupled nonlinear ODEs 
written in terms of the scaling variable $\xi$ (see Eqs.\,\eqref{epsilon-ode},\,\eqref{rho-ode}\,and\,\eqref{phi-ode} in Appendix\,\ref{sect:nonlin}) 
using the method of dominant balance~\cite{eggersfontelos-book2015}. The largest order nonlinearity (coming from the $\epsilon^2$ term in Eq.\,\eqref{stress:eqn}) should dominate near the singularity, and must balance the time derivative terms. Thus, 
\begin{subequations}
	\begin{align}
	    &1+q=s+r+2q+r,\\
     &1+s=s+s+r+2q+r,\\
     &1+p=p+s+r+2q+r\\
     &\Rightarrow\,s=p=r=\frac{1}{3},\,q=0,
	\end{align}
 \label{dombalance}
 \end{subequations}
which immediately leads to the following scaling forms,
\begin{subequations}
	\begin{align}
	& 	\rho\sim\frac{1}{(t_0-t)^{\frac{1}{3}}} R\Big(\frac{x-x_0}{(t_0-t)^{\frac{1}{3}}}\Big),~\dot{\rho}\sim\frac{1}{(t_0-t)^{\frac{4}{3}}} R_1\Big(\frac{x-x_0}{(t_0-t)^{\frac{1}{3}}}\Big),\\	
	&	\phi\sim\frac{1}{(t_0-t)^{\frac{1}{3}}} \Phi\Big(\frac{x-x_0}{(t_0-t)^{\frac{1}{3}}}\Big),~\dot{\phi}\sim\frac{1}{(t_0-t)^{\frac{4}{3}}} \Phi_1\Big(\frac{x-x_0}{(t_0-t)^{\frac{1}{3}}}\Big),\\
	&	\dot{\epsilon} \sim \frac{1}{t_0-t} E_1\Big(\frac{x-x_0}{(t_0-t)^{\frac{1}{3}}}\Big)~\Rightarrow~\epsilon\sim\ln(t_0-t)E\Big(\frac{x-x_0}{(t_0-t)^{\frac{1}{3}}}\Big).
	\end{align}
	\label{scalingforms}
\end{subequations}

	\iffalse
	The dominant balance gives the following leading order nonlinear advection equation as the singularity is approached:
	\begin{subequations}
		\begin{align}
			&\dot{\rho}+\partial_{x}\big(\rho\,\partial_{x}\sigma_2\big)=0,\label{rho-1d-simple-lo}\\
			& \dot{\phi}+\partial_{x}\big(\phi\,\partial_{x}\sigma_2\big)=0,\label{phi-1d-simple-lo}\\
			&\dot{\epsilon}=\partial^2_{xx}\sigma_2,\label{epsilon-1d-simple-lo}\\
			&\sigma_2:=\chi''(\rho^0_a)\,C^2\,\big(\zeta_{\text{avg}}\rho+\zeta_{\text{rel}}\phi\big)\,\epsilon^2.
		\end{align}
		\label{dyn-1d-simple-lo}
	\end{subequations} 
 \fi

The initial data in $\phi$ is of the form $\phi(x,0)\sim\phi_0\Big(\frac{x-x_0}{l}\Big)$, parameterized by an initial width $l$. We obtain the blowup time $t_0$ using dimensional analysis: $t_0\sim \tau (l/\ell)^3$, where $\tau:=\Big(\frac{\Gamma}{{B_2}^2}\Big)^{\frac{1}{3}}$,  $\ell:=\Big(\frac{B_2}{\Gamma^2}\Big)^{\frac{1}{3}}$, and  $B_2:=\chi''(\rho^0_a)\,C^2\,\zeta_{\text{rel}}$. Thus the space-time location of the first blowup $(x_0, t_0)$ will depend on the initial data that we determine from the numerical solution.

To obtain the scaling functions, we solve the nonlinear ODEs near the singularity 
\begin{subequations}
		\begin{align}
			&\big(\xi\,R\big)'=-3\chi''(\rho^0_a)\,B_0^2\,\bigg(R\,\bigg(\big(C_0\zeta_{\text{avg}}R+A_0\zeta_{\text{rel}}\Phi\big)\,E^2\bigg)'\bigg)'\\
			& \big(\xi\,\Phi\big)'=-3\chi''(\rho^0_a)\,B_0^2\,\bigg(\Phi\,\bigg(\big(C_0\zeta_{\text{avg}}R+A_0\zeta_{\text{rel}}\Phi\big)\,E^2\bigg)'\bigg)',\\
			&\xi\,E'=3\chi''(\rho^0_a)\,B_0\,\bigg(\big(C_0\zeta_{\text{avg}}R+A_0\zeta_{\text{rel}}\Phi\big)\,E^2\bigg)''.
		\end{align}
		\label{REPodes}
	\end{subequations} 
with appropriate boundary conditions. This second order system of ODEs requires two boundary conditions each. One natural choice for the 
boundary conditions comes from symmetry:
\begin{equation}
    R'(0)=\Phi'(0)=E'(0)=0.\label{bc1}
\end{equation}
The other boundary conditions come from the asymptotics of the profiles at $\xi\to \pm\infty$ that corresponds to the limit $t\to t_0$. In this  limit, for $x- x_0\ne 0$ (away from the singularity location), the singular solution must match the time independent outer solution \cite{eggersfontelos-book2015}. Hence, the boundary conditions for the above ODE system is obtained by setting the time derivative terms on the left hand sides of Eq.\,\eqref{REPodes} to zero:
\begin{equation}
\big(\xi\,R\big)'=0,~~\big(\xi\,\Phi\big)'=0,~~\xi\,E'=0,~~~~\xi\to\pm\infty
\end{equation}
implying
\begin{equation}
R\propto\xi^{-1},~~\Phi\propto\xi^{-1},~~E\propto\xi^{-a},~~~~\xi\to\pm\infty\label{bc2}
\end{equation}
for any choice of $a\ge 1$. The precise similarity profiles $R$, $\Phi$ and $E$ can in principle be obtained by solving the nonlinear ODE system Eq.\,\eqref{REPodes}, subject to boundary conditions Eq.\,\eqref{bc1} and Eq.\,\eqref{bc2}. Note that, as the singularity is approached at $t\to t_0$, the strain dependent turnover term, $e^{\alpha \epsilon}\sim e^{\alpha\ln(t_0-t)E(\xi)}\to 0$. 
This provides an aposteriori justification for our neglect of turnover in our scaling analysis.

%%%%%%%%%%%%%%%%%%%%%%%%%%%%%%%%%%%%%%%%%%%%%%%%%%%%%%%%%%%%%%%%%%%%%%%%

{\bf Numerical solutions of the blowup.} 
We use Dedalus pseudospectral solver \cite{dedalus2020} to solve the system of PDEs Eqs.\,\eqref{eqn:main} in 1D,
%While analysing the sharp singular structures formed in the segregation regime, the finite difference solver written in the Euler scheme has severe limitations. Solving the set of pde's using Spectral schemes help with obtaining sharper singular profiles without the numerical scheme breaking down. 
%To study segregation, the 
with variables represented on a periodic Fourier basis. Periodic basis functions like the Fourier basis provide exponentially converging approximations to smooth functions. The fast Fourier transform (implemented using sciPy and FFTW libraries \cite{dedalus2020}) 
%can compute the series coefficients in $\mathcal{O}(N \text{log}_{2} N)$ time (where $N$ is of the input in bits), 
enables computations requiring both the series coefficients and grid values to be performed efficiently. For time evolving the system of equations, we use a SBDF2 time-stepper, a 2nd-order semi-implicit Backward Differentiation formula scheme~\cite{wang2008}. Our  initial conditions are a homogeneous unstrained symmetric mixture $\phi=0$, to which we add noise sampled from a uniform distribution between $[-0.1, 0.1]$. We use $500$ modes ($N=500$) in a domain length of $L=35$ and a time-step ($dt$) of $10^{-4}$ to solve Eqs.\,\eqref{eqn:main}.
We use the following set of parameters: $B=10$, $C=1$, $D=1$, $k_1=1$, $k_2=0$, $k^b_{\text{rel}}=0$, $k_3=1$, $\zeta_{\text{avg}}=3$, $\zeta_{\text{rel}}=2$, $k_4=2$, $\chi(\rho_a^0)=1$, $\chi'(\rho_a^0)=0.01$,  $\chi''(\rho_a^0)=0.001$, $\chi'''(\rho_a^0)=-0.001$.

%and obtain numerical solutions of Eqs.\,\eqref{eqn:main} in the segregation regime with the above initial conditions. %For this part of the analysis, 
%The PDEs were solved using the spectral methods based scheme Dedalus~\cite{dedalus2020}. 
%We use  SBDF2 time-stepper with a time step size of $dt=10^{-4}$, taking 500 modes (Nx=500) in a domain length $Lx=35$.
%The initial condition is set at a white noise perturbation of magnitude $0.2$ about the homogeneous unstrained symmetric mixture $\phi=0$.
%Then, we evolve the system of equations in time and plot the profiles of $\phi$ and $\epsilon$ at different time points. 
On solving we find that after the initial linear stability regime, singularities in $\rho$, $\phi$ and $\epsilon$ develop in regions of high myosin density. The profiles of $\phi$ and $\epsilon$ near the singularity, at different time points, are shown in Fig.\,\ref{fig:singularformation}(a) and Fig.\,\ref{fig:singularformation}(c), respectively.
We numerically determine the first blowup time $t_0$ and location $x_0$, from the divergence of the maximum (minimum) value of $\phi$ ($\epsilon$), as seen in the insets of  Fig.\,\ref{fig:singularformation}(a),\,(c). In agreement with our scaling analysis, we see an excellent collapse of the data for the scaled profiles of $\phi$ and $\epsilon$, see  Figs.\,\ref{fig:singularformation}(b),\,(d).
This agreement between the numerical solutions of Eqs.\,\eqref{eqn:main}
and the scaling analysis of Eqs.\,\eqref{dyn-1d-simple} reaffirms the irrelevance of myosin turnover in the neighbourhood of the singularity.

%We find that at the numerical blowup location $x_0=26.06$ and  blowup time $t_0=4.55$, 

	%	\begin{figure}[H]
	%	\centering
	%	\includegraphics[scale=0.65]{convergence_phi.pdf}
	%	\caption{Cauchy convergence of scaled values of $\phi(x, t_{n+1})- \phi(x, t_n)$ as we approach singularity at $x=x_0$. }
	%	\label{fig:convergence}
	%\end{figure}

	Since our numerical method, does not allow us to approach the singular point  arbitrarily closely, we use our numerics to test the asymptotic convergence of the profiles approaching the singularity. We employ the Cauchy-convergence criterion for the sequence $\phi(x,t_n)$, for  a fixed value of $x$ near $x_0$ and where the sequence $\{t_n\}$ steadily approaches $t_0$. In Fig.\,\ref{fig:cauchyconvergence}, we have plotted 
	$\phi(x,t_{n+1})-\phi(x,t_{n})$ as a function of $t_n$, at three spatial locations $x-x_0=0$, $x-x_0=0.6$, and $x-x_0=1$. We see that $\phi(x,t_{n+1})-\phi(x,t_{n})\to 0$ faster for points closer to the centre of the profile; the convergence rate is maximum at $\phi_{max}$.

These singularities are  {\it physical} in that their resolution involves incorporation of additional
physical effects such as steric hindrance (represented by the $\epsilon^4$ term in $w$).
%, and the gradient nonlinearities in $\boldsymbol{\Sigma}^a$ which provides a length scale. 
%However, the shrinking $\phi>0$ domain stops at the largest negative strain set by the $\epsilon^4$ term in $w$ due to steric hindrance (see Figure \ref{frontmove}(a) inset).
With this, the singularity is never reached, resulting in highly concentrated tensile regions of finite width, 
%$\sim 100$\,nm, 
which could be taken to be the size of non-muscle myosin-II bipolar filaments~\cite{Bershadsky2017}.

%\sout{If in the dynamical equations, we  were to prevent %now
%the turnover of the stresslets immediately after the initial linear segregation regime, then their densities would be conserved. In this situation, the domain shrinkage %of the central bulk region 
%would ultimately stop due to the $\epsilon^4$ term in the effective elastic energy density, arising from steric effects that are inevitably present (Fig.\,\ref{frontsmove}). 
%by the pseudo wells in the effective strain energy landscape %strain minima 
%introduced by the activity induced 
%As a result, the singularity is never reached and one ends up with highly concentrated tensile regions of finite width set by the physical constraint of steric hindrance. In presence of turnover of catch bond type, with the catch bond for $\rho_1$ stronger than $\rho_2$ (i.e., $k_4>0$, same sign as $\zeta_{\text{rel}}$), this width will be modified due to violation of local detailed balance.}

The appearance of finite time elastic singularities that lead to zones of high concentration of both the actin meshwork and myosin stresslets is reminiscent of  caustics discussed in the context of the Zel'dovich model for the formation of large scale self-gravitating structures of the universe~\cite{Arnoldetal1982} and verified in extensive numerical simulations, such as in~\cite{Feldbruggeetal2018}. We observe that at the blowup location $x=x_0$ where $\epsilon\approx-1$ (Fig.\,\ref{frontsmove}(c)), actin meshwork density in the deformed or Eulerian configuration $\tilde{x}_t(x)=x+u(x,t)$, given by $\tilde{\rho}_a(\tilde{x}_t(x))=\frac{\rho_a}{1+\epsilon}=\frac{\rho_a^0-(C/A)\epsilon}{1+\epsilon}$, goes to $+\infty$ as $t\to t_0$. Moreover, this meshwork density blowup occurs before the stresslet density blowup, as seen both from the similarity forms in Eq.\,\eqref{scalingforms} which shows that the  strain rate $\dot{\epsilon}\to-\infty$ faster than $\rho_{i}\to\infty$, at the same spatial blowup location $x_0$, and from the numerical solution (see Fig.\,\ref{caustic}). These {\it elastic caustics} as singularities of Lagrange maps $\tilde{x}_t\colon\mathbb{R}^d\to\mathbb{R}^d$ in higher dimensions will appear as more general Lagrangian catastrophes \cite{Arnoldetal1982,Arnoldbook1985}, of a variety of co-dimensions (points, lines, surfaces etc.) and topology, where the associated deformation map $\partial_x \tilde{x}_t$ degenerates. 
%\sout{Note that, while in a gravitational fluid context, the `particles' can overtake each other and the material can go past the singularity, such self-penetrations (``pancakes'') are prevented in compatible elastic deformations, as seen here, through steric interactions. }

We note from our 1D numerical analysis, that the maximum strain magnitude within singular structures is of order $0.8$ while outside it can reach a value of $2.5$, suggesting that geometrical nonlinearities are significant. This will be particularly relevant in 2D.

%dynamics of inertia-less active particles driven by a background vortical flow~\cite{Arnoldetal1982}, although in a continuum hydrodynamic setup.
%In turbulent aerosol flows (heavy particles in a background incompressible flow), particles tend to cluster in very thin regions, known as {\it caustics}, away from the vortices in the fluid. Caustics occur when, at a finite time, a fast moving particle overtakes an adjacent slow moving particle, i.e., when their relative separation $\Delta\mathbf{r}$ becomes zero but their relative velocity $\Delta\mathbf{v}$ is non-zero \cite{Chajwaetal2024}. There are two singularities at play here: (1) the fluid vortex at the center where the {\it curl of the fluid velocity diverges}, and (2) the caustics encircling the fluid vortex where the {\it divergence of the particle velocity diverges}.  {\it ``Note in particular that particles which asymptotically approach each other at long times do not give rise to caustics, although they do contribute to clustering''} \cite{Ravichandran2015}.
%Consistent with this interpretation, we find that the  strain rate $\dot{\epsilon}_{min}\to-\infty$ faster than $\phi_{max}\to\infty$, at the same blowup spatial location $x_0$. This is also clear from the similarity forms in Eq.\,\eqref{scalingforms}.

%%%%%%%%%%%%%%%%%%%%%%%%%%%%%%%%%%%%%%%%%%%%%%%%%%%%%%%%%%%%%%%%%%%%%%%%%%%%%%%%%%%
 \begin{figure*}[t]
	\centering
	\includegraphics[scale=.09]{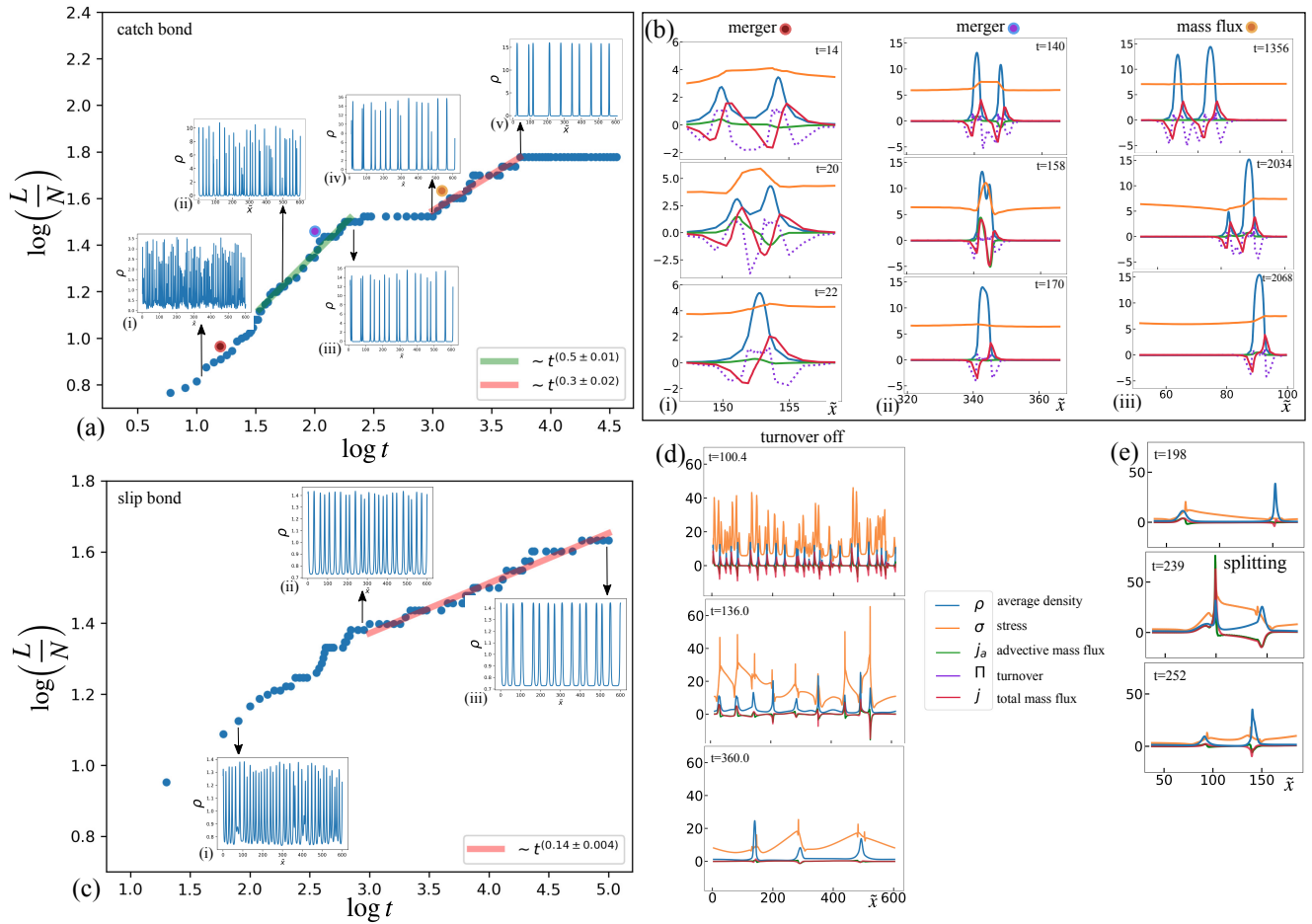}
	\caption{{\bf Merging of singular structures and coarsening dynamics.} In figure (a) we see coarsening dynamics for a system with catch bond. The system of equations are numerically solved for the following set of parameters :  $B=5, \chi(\rho_a^0)=1, \chi(\rho_a^0)'=1, \chi(\rho_a^0)''=0.1, \chi(\rho_a^0)'''=15, \zeta_{avg}=1.4, \zeta_{rel}=1$, $k_4=1$, $k_3=2$. Panel (b) describes the dynamics of merger events driven by stress gradients and mass flux via snapshots from the numerical simulation described in (a). Panel (c) captures the coarsening dynamics for a system with slip bond. All the parameters remain unchanged apart from $k_3=-0.5$. Panel (d) shows how the configurations generated in the numerical simulation in (a) at $t=100$, evolve after the turnover is switched off. The jump in total flux $\llbracket j\rrbracket$ and stress $\llbracket \sigma \rrbracket$ is much higher in the absence of turnover and the merger events proceed faster as seen from the snapshots. (e) Sequence of snapshots showing splitting of singular structures after the turnover is switched off (see \href{https://drive.google.com/file/d/1XYkDBGoiNu1z5DrmqNfOgcsHBNfQXiD2/view?usp=sharing}{Movie 5}).}	\label{coarsening2}
\end{figure*}
%%%%%%%%%%%%%%%%%%%%%%%%%%%%%%%%%%%%%%%%%%%%%%%%%%%%%%%%%%%%%%%%%%%%%%%%%%%%%%%%%%%%%%%%%%%%%%%%%%%%%%%%%%%%%%%%%%%%%%%%%%%

\section{Moving, Merging and Coarsening of singularities}

In the above scaling analysis,  we have assumed that the singular structures are described by symmetric profiles of the fields, which lead to static finite-time singularities.
However, our numerical analysis shows that the singular structures, tempered by steric hindrance, can move. Such {\it moving} singularities must be associated with 
an asymmetry in its density profile and discontinuities in bulk stress and mass flux across it. 
%give rise to {\it moving} singularities.
Indeed moving singularities show up as moving stress fibers whose interactions and coalescence  have been reported in {\it in vivo} studies~\cite{Vignaudetal2021}. 

%For our discussion henceforth, we will reintroduce the turnover of the myosin stresslets.

Before proceeding, we draw attention to 
the study of the {\it nucleation} instability of clusters of bound myosin~\cite{mabhishek2024}, as opposed to the {\it spinodal} instability studied here.
We show that on nucleating a symmetric myosin pulse of finite width, the pulse grows if the amplitude is beyond a threshold, else it relaxes to a uniform background. Further, on nucleating an asymmetric pulse, a large enough asymmetry drives a translational instability. Using a boundary layer analysis, we derive analytical forms for the asymmetric myosin density and strain profile and its velocity~\cite{mabhishek2024}. This information will be useful in the analysis presented below.

{\bf Isolated moving singularity.}
  The moving singularity at $x_0(t)$ carries with it the singular component of all the bulk fields, accompanied by jumps across $x_0(t)$.
We denote the jump  of the limiting values (i.e., difference between right and left)  of bulk fields at $x_0(t)$ by $\llbracket \rrbracket$ 
and their average by $\langle \rangle$. 
Continuity of the bulk displacement field ${u}$ (i.e., $\llbracket u \rrbracket=0$) implies %well-known Hadamard compatibility condition for the deformation gradient $\mathbf{F}=\mathbf{I}+\nabla\boldsymbol{u}$, and the
the velocity compatibility condition~\cite{AbeyaratneKnowles1991}, a kinematic requirement:
\begin{equation}
\llbracket\dot{{u}}\rrbracket +V\,\llbracket\epsilon\rrbracket ={0},\label{velcomp2}
\end{equation}
where $V:=\dot{x}_0$ denotes material velocity of the singular point (see Appendix \ref{sect:singfield}). 
The spatial velocity of the singularity is then defined by  ${v}_S:=\langle\dot{{u}}\rangle+V\,\langle\epsilon\rangle$. The force and mass balance equations at the singularity can be obtained, using the divergence and transport theorems for discontinuous fields, as
\begin{subequations}
    \begin{align}
        &  \Gamma_S v_S=\llbracket\sigma\rrbracket,\\
        & \dot{\rho}_S-V\llbracket\rho\rrbracket=-\llbracket j\rrbracket +\Pi_S,
    \end{align}
    \label{jumpconds}
\end{subequations}
where $\Gamma_S$ is the friction coefficient at the singularity, $\rho_S$ is the stresslet density at the singularity,  $j:=-D\partial_x\rho+\rho\dot{u}$ is the bulk mass flux and $\Pi_S$ is the (strain dependent) singular turnover term  %\sout{(see \cite{RDR2022} for derivation in 2D)} 
(see Appendix \ref{sect:singfield}). Hence, the jumps in stress, mass density and mass flux  across each singularity act as the driving force for movement of the singularity and its height variation.

{\bf Interacting moving singularities.} When the number density of singularities is high, they will interact with each other via the long-range elastic strain. Since the moving singularities are associated with an asymmetric density profile, there could be additional non-reciprocal interactions between them, possibly 
leading to dynamical phenomena discussed in 
~\cite{JPetal2022, Guptaetal2022}.

\begin{figure}[h]
\centering
\includegraphics[scale=.2]{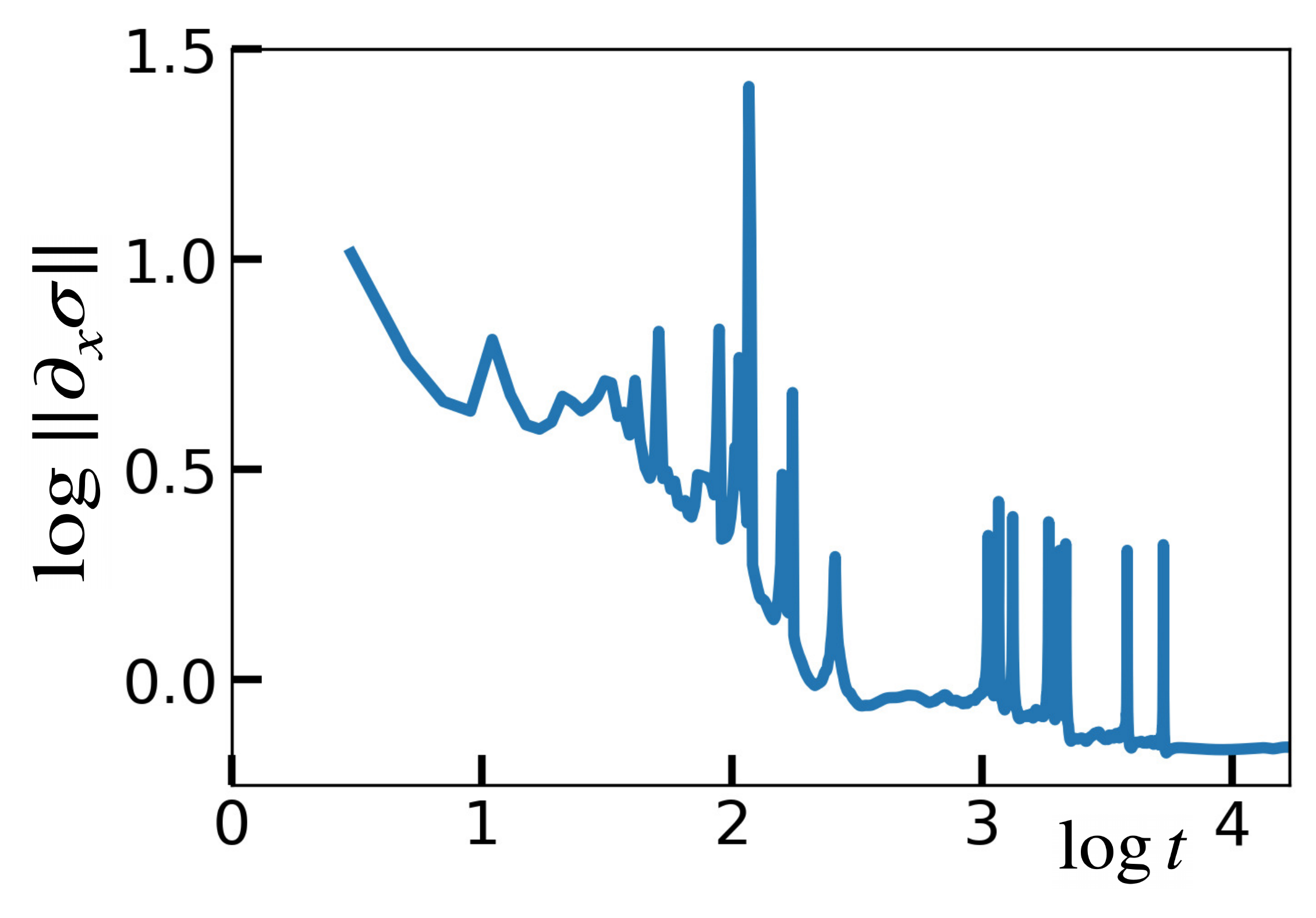}
\caption{Log of $L^2$ norm of $\partial_{x} \sigma$, $\lVert\partial_x\sigma\rVert:=\sqrt{\int|\partial_x\sigma|^2dx}$, is plotted as a function of log time, in the case where the two myosin species have catch bonds (with parameters same as Fig.\,\ref{coarsening2}(a), \href{https://drive.google.com/file/d/1ci_D42aP20kXC45w0JwtzV5Nisizz1hS/view?usp=sharing}{Movie S4}). The stress gradient
shows an overall decrease in time, except at isolated time points associated with the onset of singularity mergers. This could be interpreted as a  Lyapunov functional, if smoothened with a suitable mollifier.
}
\label{lyapunov_merge}
\end{figure}

% This leads to a nonreciprocal
%sensing wherein a trailing particle senses and catches up
%with a leading particle moving ahead of it.
%%%%%%%%%%%%%%%%%%%%%%%%%%%%%%%%%%%%%%%%%%%%%%%
{\bf Merging and Coarsening.} 
Thus far we have established that following the initial segregation and finite time collapse, we are generically left with a distribution of moving and interacting singularities. Subsequent evolution involves the merger and subsequent coarsening of singularities. There are two important features of this coarsening dynamics that emerge from our 1D analysis -- (i) because of the long range strain, and velocity compatibility Eq.\,\eqref{velcomp2},
the analysis of the dynamics of merger needs to be global, and (ii) coarsening proceeds so as to make the stress homogeneous, i.e. $\sigma= const.$, except at isolated points associated with the singularities. This is reflected in the time dependence of 
$\int (\partial_x \sigma)^2 \,dx$ (see Fig.\,\ref{lyapunov_merge}),
which shows an overall decrease in time, except at isolated time points associated with the onset of singularity mergers, a kind of Lyapunov functional.

 We see that the inverse density of singularities $L/N$ which measures the typical distance between singularities, grows initially and then saturates, see Fig.\,\ref{coarsening2}(a) and (c), for the cases where both the species are of  catch and slip type, respectively. The power law growth associated with early time coarsening is indicated in the figures. The saturation at late times, either indicate arrested coarsening or at most a  slow logarithmic growth indicative of activated dynamics.

 To understand the dynamics of coarsening, we resort to the singular balance equations \eqref{jumpconds} for isolated singularities, and monitor the time dependence of the local  density fields, stresses and currents. Soon after the establishment of the distribution of singularities, the number density of singular structures is high and the variation in the heights of the singular density profiles 
$\rho_{S_i}$ 
is large; here, the subscript $i$ labels the singular peaks (Fig.\,\ref{coarsening2}(a),\,inset\,(i), and \href{https://drive.google.com/file/d/1ci_D42aP20kXC45w0JwtzV5Nisizz1hS/view?usp=sharing}{Movie 4}). As a result all the driving jump terms $\llbracket\rho\rrbracket_i$,  $\llbracket j\rrbracket_i$ and $\llbracket\sigma\rrbracket_i$, across each singularity $i$, are large, leading to a large speed $V_i$ and large mass flux  $\dot{\rho}_{S_i}$ out of/into the singularities (Fig.\,\ref{coarsening2}(b)(i)).
This drives the early mergers, since the elastic energy barrier between
neighbouring singularities are easily overcome, and results in fast coarsening with a rate $t^{0.5}$ (Fig.\,\ref{coarsening2}(a), green line). After several  merger events, the singularities get sparser, the variation in the heights of the singularities gets smaller (Fig.\,\ref{coarsening2}(a),\,insets\,(ii,iii,iv)), the jumps  in the density fields and the mass flux across the singularities become smaller -- and as a result, the 
elastic stress gets more homogeneous (Fig.\,\ref{coarsening2}(b)(ii)).
%{\red We have to discuss, with representation in Fig.\,\eqref{coarsening2}, the advective fluxes from the shorter singularities to the higher ones.}
Together, this leads to a slower coarsening with a rate $t^{0.3}$ (Fig.\,\ref{coarsening2}(a), pink line), associated with a driving force that mainly comes from
the advective current flowing from slightly smaller singularities to higher ones,
(and to a lesser extent through the turnover of myosin), see Fig.\,\ref{coarsening2}(b)(iii). We note that the fitted growth exponents reported here represent intermediate time dynamics, and are not universal. For $k_3<0$, i.e., slip bond response, the stress becomes homogeneous relatively quickly, and the coarsening gets much slower with a rate $t^{0.14}$ (Fig.\,\ref{coarsening2}(c) and \href{https://drive.google.com/file/d/1ezNY2PXOxmdJnOaqTryy-zk9IpxBE7jo/view?usp=sharing}{Movie 6}).

Eventually the variation in the heights of the singularity profiles vanishes, and the dynamics reaches a steady state, thus, arresting the coarsening, with a distinct distribution of singular structures, see the plateau in Fig.\,\ref{coarsening2}(a) and inset (v).

Interestingly,  we find that myosin turnover, is crucial in sustaining this dynamics of singular structures over a long time. Thus, if we switch off the turnover during this coarsening stage, say at $t=100$ in Fig.\,\ref{coarsening2}(a), we find a large increase in the mass flux from diffusion and advection, causing the singular structures to spread and overlap, leading to merger and faster coarsening, see Fig.\,\ref{coarsening2}(d) and
\href{https://drive.google.com/file/d/1XYkDBGoiNu1z5DrmqNfOgcsHBNfQXiD2/view?usp=sharing}{Movie 5}. We also observe the splitting of singularities, see Fig.\,\ref{coarsening2}(e) and 
\href{https://drive.google.com/file/d/1XYkDBGoiNu1z5DrmqNfOgcsHBNfQXiD2/view?usp=sharing}{Movie 5}, which is reminiscent of stress fiber splitting reported in  \cite{Vignaudetal2021,Beachetal2017a}. This quickly leads to a single domain with high myosin density, a contractile collapse, as predicted by the phase diagram when the turnover rates are set to zero. In vivo, this eventuality can be interrupted by anchoring on other extended structures such as on microtubules or on the plasma membrane or internal membranes. 

This suggests that material renewability acts as a control for the coarsening dynamics, maintaining the distribution and organisation of the singular force structures over long time scales.

\section{Discussion}

Renewable active matter is 
a new %kind
class of active %matter 
materials %with 
that exhibit novel mechanical  properties %displaying 
such as elastic marginality, force chains, mechanical fragility and spontaneous excitability~\cite{debsankar2017,mabhishek2024,RRT2024}.
In this work, we have studied 
how the interplay between activity and material renewability can give rise to unusual segregation and nonequilibrium force patterning in living materials. 
A striking example is the active cytoskeleton, where a variety of myosin species---as discussed in the Introduction---act as molecular force generators and sensors, 
giving rise to %striking 
singular extended structures carrying tension  (the stress fibers).

We derive hydrodynamic equations in terms of the local actin mesh
displacement and the densities of two types of myosin species that differ in their contractility and turnover rates, but which otherwise do not interact with each other. Despite this, the two myosin species undergo a spinodal segregation instability, with myosin density patterns that depend on the nature of the strain-dependent unbinding, namely, whether they have catch or slip bond response. 

Subsequently the spinodal patterns become unstable 
due to a nonlinear coupling between the linear segregation and contractile instability. This drives domain shrinkage,  leading to a finite-time singularity where the density and elastic fields display scaling behaviour. We formally associate the distribution of these singular structures with caustics. These singular structures or caustics move, merge and coarsen over time. The rate of coarsening gets slower as time progresses and eventually gets arrested. This segregation of myosin species into singular structures immediately implies spatiotemporal force patterning that spans across the system.
Importantly, {\it material renewability in the form of myosin turnover, is crucial in maintaining this dynamics and eventual arrest of singular structures, and thus force patterning}. 

We will, in another publication, extend our analysis to a fully tensorial elasticity in two dimensions; while this extension is nontrivial, the basic message of segregation followed by a finite time collapse survives. 
However, we note that we would need to include active boundary anchoring at the cell surface in the form of focal adhesions and adherent junctions, whose strength and positioning will select specific bulk force patterns~\cite{Theryetal2006,Kassianidouetal2019}.
% In addition, force patterning within the cell may also be affected by other spatially extended structures, such as
% microtubules and internal membranes.
This will allow us to make direct quantitative comparisons to the dynamical patterning of stress fibers, such as those observed in experiments 
in cells~\cite{Vicente-Manzanaresetal2008}, 
in cell extracts on
micropatterned substrates~\cite{Vignaudetal2021,jimenezetal2021}, and in vitro reconstituted systems with controlled mechanical constraints \cite{Krishnaetal2024}.

From an evolutionary materials perspective, our work suggests that
neutral variations in the space of myosin binding/unbinding affinities and contractility, may facilitate 
favourable phenotypes \cite{KirschnerGerhart2005},
%{\blue Refer to Kirschner and Gerhardt's book, Plausibility of Life}, 
such as,
differences in cellular localisation and the patterning of a network of forces across the scale of the cell. Further, in the context of {\it physical learning} \cite{SternMurugan2023}, our work provides insights on active materials synthesis   whose components learn parameter values that drive it towards specific material response, such as mechanical adaptation \cite{banerjeeetal2020} and tensegrity \cite{ingberetal2014}.

 We conclude by reflecting on the broader implications of material renewability in active systems—particularly in relation to material evolution and physical learning \cite{SternMurugan2023}.
While active elastomers composed of a single myosin species can display pattern formation and contractile segregation, introducing multiple myosin species with distinct contractility and strain dependent turnover rates not only greatly enhances the richness of the emergent dynamics, but also provides a new conceptual framework for understanding how self-organizing materials can evolve toward mechanical adaptation. The wide range of contractility and binding-unbinding kinetics found in living systems  equips the material to explore an extended landscape of mechanical responses (i.e., material moduli). This enables the system to self-tune \cite{Goodrichetal2015} and evolve toward mechanically adaptive states with different functional phenotypes \cite{RRT2024}.

\section{Acknowledgements}

We thank members of the Simons Centre at NCBS
%, especially Abhishek M. 
%Alkesh Yadav, Amit Kumar, Archishman Raju, Kabir Husain, Mukund Thattai and Sandeep Krishna, 
for clarifying discussions. 
We acknowledge support from the Department of Atomic Energy (India) under project no.\,RTI4006, the Simons Foundation (Grant No.\,287975) and the computational facilities at NCBS.
MR acknowledges Department of Science and Technology (India), for a JC Bose Fellowship
(SERB File No.JCB/2018/00030).

\appendix
%\renewcommand\section{Appendix}
%\section{The first appendix} 

%%%%%%%%%%%%%%%%%%%%%%%%%%%%%%%%%%%%%%%%%%%%%%%%%%%%%%%%%%%%%%%%%%%%%%%%%%%%%%%%%%%%%%%%%%%%%%%%%%%%%%%%%%%%%%%%

%\setcounter{figure}{0}
%\setcounter{table}{0}
%\setcounter{equation}{0}
%\setcounter{page}{1}
%\setcounter{section}{0}

%\renewcommand{\theequation}{A\arabic{equation}} 
%\renewcommand{\thepage}{A\arabic{page}} 
%\renewcommand{\thesubsection{\arabic{section}}  
%\renewcommand{\thetable}{A\arabic{table}}
\renewcommand{\thesubsection}{\thesection\arabic{subsection}}
\renewcommand{\thefigure}{\thesection\arabic{figure}}
%\section*{Appendix}\label{Appendix}

\section{Hydrodynamic Equations for a Mixture of Contractile Stresslets on an Elastomer}\label{sect:goveqn}

We describe the dynamics of active stress propagation in the active medium of the cell using hydrodynamic equations for the crosslinked actin mesh and the density of different species of myosin 
filaments, embedded in the viscous cytosol.

\subsection{Equations for the Crosslinked Actin Meshwork embedded in the Cytosol}\label{sect:a1}

We start with a  passive $2$-dimensional elastomeric meshwork of mass density $\rho_a$, whose displacement with respect to an unstrained reference state is $\boldsymbol{u}$. The meshwork moves in the fluidic cytosol whose velocity is $\boldsymbol{v}$. The hydrodynamic equations for its linear momentum balance and mass balance are, therefore,
\begin{subequations}
	\begin{align}
		&\rho_a\ddot{\boldsymbol{u}}+\Gamma\,(\dot{\boldsymbol{u}}-\boldsymbol{v})=\nabla\cdot\boldsymbol{\sigma},~~\text{and}\label{actin-dyn1}\\
		& \dot{\rho_a}+\nabla\cdot(\rho_a\,\dot{\boldsymbol{u}})=M\nabla^2\frac{\delta F}{\delta \rho_a}.\label{actin-dyn2}
	\end{align}
	\label{actin-dyn}
\end{subequations}
Here, $\Gamma>0$ is the friction coefficient of the elastomer with respect to the fluidic cytosol, and $\boldsymbol{\sigma}$ is the total stress in the elastomer; $M$ represents the mobility of permeation of the meshwork. 
%and $\mathcal{S}_a$ represents turnover of the actin meshwork. 
$F$ is the free energy functional for the passive meshwork: 
\begin{equation}
	F(\boldsymbol{\epsilon},\rho_a)=\int_{\Omega}f_B\,d^{d}r;
\end{equation}
where the free energy density $f_B(\boldsymbol{\epsilon},\rho_a)$ depends on the linearized strain $\boldsymbol{\epsilon}:=(\nabla\boldsymbol{u}+\nabla\boldsymbol{u}^T)/2$ of the elastomer, and the mass density $\rho_a$.

The hydrodynamic equations of the fluidic cytosol are given by

\begin{equation}
	\rho_f\Big(\dot{\boldsymbol{v}}+\boldsymbol{v}\cdot\nabla\boldsymbol{v}\Big)=\eta^s_f\nabla^2\boldsymbol{v}+\eta^b_f\nabla(\nabla\cdot\boldsymbol{v})-\nabla p+\Gamma\,(\dot{\boldsymbol{u}}-\boldsymbol{v});
\end{equation}
here, $\rho_f$ is the density of the fluid, and $\eta^s_f$ and $\eta^b_f$ are the shear and bulk viscosities of the fluid. The fluid pressure $p$ appears due to the total incompressibility of the meshwork-fluid system:
\begin{equation}
	\nabla\cdot\Big(c_a\,\dot{\boldsymbol{u}}+(1-c_a)\boldsymbol{v}\Big)=0;
\end{equation} 
here, $c_a$ is volume fraction of the meshwork. 
Here, for convenience, we ignore the hydrodynamics of the fluid,  
permissible in the limit when $c_a \approx 1$.

\subsection{Equations for the Stresslets}\label{sect:a2}

Consider a mixture of active contractile stresslets with different contractilities undergoing turnover onto this elastomer. We assume that the stresslets binding onto the elastomer are recruited from an infinite pool of stresslets unbound to the elastomer, and, hence, disregard the dynamics of these unbound stresslets.  Restricting to a binary mixture, let $\rho_i$, $i=1,2$,  be the density fields of the two species of bound stresslets. The bound stresslets get advected by the local velocity $\dot{\boldsymbol{u}}$ of the elastomer, and diffuse on it with the same diffusion coefficient $D$. Let the stresslets bind onto the elastomer with rates $k_i^{b}>0$, and unbind from the elastomer with rates $k_i^{u}(\boldsymbol{\epsilon})>0$ that in principle depends on the strain $\boldsymbol{\epsilon}$ of the elastomer. The dynamics of these bound stresslets is, hence, governed by
\begin{equation}
	\dot\rho_i+\nabla\cdot(\rho_i\,\dot{\boldsymbol{u}})=\nabla\cdot(D\,\nabla\rho_i)+k_i^{b}\,\rho_a - k_i^{u}(\boldsymbol{\epsilon})\,\rho_i.
	\label{stresslets-dyn}
\end{equation}
%here, $\mathcal{S}_i:=k_i^{b}\,\rho_a - k_i^{u}(\boldsymbol{\epsilon})\,\rho_i$ represent turnover of the bound stresslets.
We will assume the Hill  form for the  unbinding rates:
$k_i^{u}(\boldsymbol{\epsilon})=k_{i0}^{u}\, e^{\alpha_i\,\epsilon}$, $i=1,2$ where $k_{i0}^{u}>0$, $i=1,2$,  are the strain independent parts of the respective rates, $\epsilon:=\text{tr}\,\boldsymbol{\epsilon}$ is the dilational strain, and the  dimensionless numbers $\alpha_i$ capture whether the bond is catch or slip type: $\alpha_i>0$  ensures  that local  contraction (extension) will decrease (increase) the unbinding of the stresslets, i.e., the bond is of catch type, while $\alpha_i<0$  ensures  that local  contraction (extension) will increase (decrease) the unbinding of the stresslets, i.e., the bond is of slip type.

% \sout{In the small timescale of the turnover processes of the stresslets comparatively larger than the turnover of the crosslinked actin meshwork, the elastomer can be considered as long lived, i.e.,}
We consider the actin elastomer to be  permanently crosslinked and that the elastomer density is slaved to the strain, i.e. 
% thus setting the}
% right hand side of \eqref{actin-dyn2} to zero:  
$\frac{\delta F}{\delta (\delta\rho_a)}=0$. 
% This implies that elastomer density $\rho_a$ is enslaved to the isotropic strain of the elastomer: 
Thus, $\delta\rho_a\propto-\epsilon$ (obtained from a variation of \eqref{enrg} below), where  $\delta\rho_a:=\rho_a-\rho_a^0$ is the deviation of the elastomer density from its state value  $\rho_a^0$.

%%%%%%%%%%%%%%%%%%%%%%%%%%%%%%%%%%%%%%%%%%%%%%%%%%%%%%%%%%%%%%%%%%%%%%%%
\subsection{Constitutive Equations}
\label{sect:a3}

%The total isotropic stress $\boldsymbol{\sigma}=\boldsymbol{\sigma}^e+\boldsymbol{\sigma}^a$ is the summation of passive elastic stress $\boldsymbol{\sigma}^e$ and the active stress $\boldsymbol{\sigma}^a$.

%\subsubsection{Passive elastic stress}
  The elastic stress $\boldsymbol{\sigma}^e := \frac{\delta F}{\delta \,\boldsymbol{\epsilon}}$ comes from the  free-energy functional   $F(\boldsymbol{\epsilon},\rho_a)
=\int d^2r f_B$ of an isotropic linear elastic medium, where 
\begin{eqnarray}
	f_B&=&\frac{1}{2}B\epsilon^2+\mu|\tilde{\boldsymbol{\epsilon}}|^2+C\,\delta\rho_a\,\epsilon+\frac{A}{2}\delta\rho_a^2+\frac{K}{2}|\nabla{\epsilon}|^2
	\label{enrg}
\end{eqnarray}
is the free energy density; 
with the elastic moduli  $B,\,\mu>0$, and $C>0$, $A>0$ from thermodynamic stability. The isotropic elastic stress is
\begin{eqnarray}
	\boldsymbol{\sigma}^e&=&\frac{\delta F}{\delta \boldsymbol{\epsilon}}
	=\Big(B\,\epsilon+C\,\delta\rho_a-K\nabla^2\epsilon\Big)\,\mathbf{I}+2\mu\tilde{\boldsymbol{\epsilon}}\nonumber\\
	&=&\Big(B-\frac{C^2}{A}\Big)\,\epsilon\,\mathbf{I}-K\nabla^2\epsilon\,\mathbf{I}+2\mu\tilde{\boldsymbol{\epsilon}},
\end{eqnarray}
noting that $\delta\rho_a=-\frac{C}{A}\epsilon$. The elastic modulus $B$ of the actin mesh is set by the crosslinker density. The passive viscous stress of the elastomer is 
$\boldsymbol{\sigma}^d
=\eta_b\,\dot{\epsilon}\,\mathbf{I}+2\eta_s\,\dot{\tilde{\boldsymbol{\epsilon}}}$, where $\eta_{b,s}$ are the bulk and shear viscosities, respectively.

%\subsubsection{Active stress}

At the macroscopic/coarse-grained scale, the isotropic active stress  $\boldsymbol{\sigma}^a$ is of the form  $\boldsymbol{\sigma}^a=\Delta\mu\,\chi(\rho_a)\zeta(\{\rho_i\})\mathbf{I}$, where $\Delta\mu$ is the chemical potential change due to ATP hydrolysis, and $\chi(\rho_a)$ is a sigmoidal function that encodes the dependence of the active stress on the meshwork density. We will take $\Delta\mu=1$ for simplicity.

%For a single stresslet system, Taylor expanding the function $\chi(\rho_a)$ about the state value $\rho_a=\rho_a^0$ upto cubic order leads to

%\begin{widetext}
%\begin{eqnarray}
%	{\sigma}^a&=&\chi(\rho_a)\, \zeta(\rho)\nonumber\\
	%&=&\Bigg(\chi(\rho_a^0)+\chi'(\rho_a^0)\,\delta\rho_a+\frac{1}{2!}\chi''(\rho_a^0)\,\delta\rho_a^2+\frac{1}{3!}\chi''(\rho_a^0)\,\delta\rho_a^3+o(\delta\rho_a^3)\Bigg)\,\zeta(\rho)\nonumber\\
	%&=&\Bigg(\chi(\rho_a^0)-\chi'(\rho_a^0)\frac{C}{A}\,\epsilon+\frac{1}{2!}\chi''(\rho_a^0)\,\Big(\frac{C}{A}\,\epsilon\Big)^2-\frac{1}{3!}\chi''(\rho_a^0)\,\Big(\frac{C}{A}\,\epsilon\Big)^3+o(\epsilon^3)\Bigg)\,\zeta(\rho).
	%\label{activestress-sigle}
%\end{eqnarray}
%\end{widetext}

We assume that the stresslets do not interact directly but only via the strain of the elastomer. Hence, the function $\zeta(\{\rho_i\})$ can be additively decomposed into individual contributions coming from each stresslet species, i.e., $\zeta(\{\rho_i\})=\sum_i\zeta_i(\rho_i)$. We assume the functions $\zeta_i(\rho_i)$ to be linear in $\rho_i$, i.e., $\zeta_i(\rho_i)=\zeta_i\rho_i$ (no sum over $i$) where the contractilities $\zeta_i>0$ are different for different species $i$. 
For a binary mixture, Taylor expanding the function $\chi(\rho_a)$ about the state value $\rho_a=\rho_a^0$ upto cubic order leads to
\begin{widetext}
\begin{eqnarray}
	\boldsymbol{\sigma}^a&=&\chi(\rho_a) \Big(\zeta_{1}\,\rho_1+\zeta_{2}\,\rho_2\Big)\mathbf{I}\nonumber\\
	&=&\Bigg(\chi(\rho_a^0)-\chi'(\rho_a^0)\frac{C}{A}\,\epsilon+\frac{1}{2!}\chi''(\rho_a^0)\,\Big(\frac{C}{A}\,\epsilon\Big)^2-\frac{1}{3!}\chi''(\rho_a^0)\,\Big(\frac{C}{A}\,\epsilon\Big)^3+O(\epsilon^4)\Bigg)\times2\Big(\zeta_{\text{avg}}\,\rho+\zeta_{\text{rel}}\,\phi\Big)\mathbf{I},
	\label{activestress3}
\end{eqnarray}
\end{widetext}
where we have introduced the the average and the relative densities of the bound stresslets:
\begin{equation}
	\rho:=\frac{\rho_1+\rho_2}{2}~~\text{and}~~ \phi:=\frac{\rho_1-\rho_2}{2};
\end{equation}
and
\begin{equation}
\zeta_{\text{avg}}:=\frac{\zeta_{\text{avg}}+\zeta_{\text{rel}}}{2}>0,~\zeta_{\text{rel}}:=\frac{\zeta_{\text{avg}}-\zeta_{\text{rel}}}{2}
\end{equation}
as the average and relative contractility, respectively. Note that $\phi$ is the order parameter for segregation.

%%%%%%%%%%%%%%%%%%%%%%%%%%%%%%%%%%%%%%%%%%%%%%%%%%%%%%%%%%%%%%%%%%%%%%%%%%%%%%%%%%

\subsection{Non-dimensional Governing Equations}\label{sect:a4}

We will further assume the overdamped limit of the elastomer, i.e., $|\rho_a\,\ddot{\boldsymbol{u}}|\ll |\Gamma\,\dot{\boldsymbol{u}}|$, 
%With this, \eqref{actin-dyn1} and \eqref{stresslets-dyn} reduce to the following system of equations
%\begin{widetext}
%\begin{subequations}
%	\begin{align}
%		&\Gamma\,\dot{\boldsymbol{u}}= \nabla{\sigma}, \label{dyn-u-f12}\\
%		& \dot{\rho} +\nabla\cdot(\rho\,\dot{\boldsymbol{u}}) = {D}\,\nabla^2\rho+k^b_{\text{avg}}\,\bigg(\rho_a^0-\frac{C}{A}\nabla\cdot\boldsymbol{u}\bigg)-k^{u}_{\text{avg}}({\epsilon})\,\rho-k^{u}_{\text{rel}}({\epsilon})\,\phi,\label{dyn-e-f12}\\
%		&\dot{\phi} +\nabla\cdot(\phi\,\dot{\boldsymbol{u}}) = {D}\,\nabla^2\phi+k^b_{\text{rel}}\,\bigg(\rho_a^0-\frac{C}{A}\nabla\cdot\boldsymbol{u}\bigg)-k^{u}_{\text{avg}}({\epsilon})\,\phi-k^{u}_{\text{rel}}({\epsilon})\,\rho;\label{dyn-c-f12}
	%\end{align}
	%\label{dyn-f12}
%\end{subequations}
%\end{widetext}
%where $\boldsymbol{\sigma}$ is defined in \eqref{totalstress};
and introduce
\begin{equation}
	k^{u}_{\text{avg}}:=\frac{k^u_1+k^u_2}{2},~~~k^{u}_{\text{rel}}:=\frac{k^u_1-k^u_2}{2},
\end{equation}
the average and relative unbinding rates, respectively;
\begin{equation}
	k^{b}_{\text{avg}}:=\frac{k^{b}_1+k^{b}_2}{2}>0~~\text{and}~~k^{b}_{\text{rel}}:=\frac{k^{b}_1-k^{b}_2}{2}
\end{equation}
the average and relative binding rates, respectively. Taylor expanding the Hill form of unbinding, we write
\begin{subequations}
	\begin{align}
		& k^{u}_{\text{avg}}({\epsilon})=k_1+k_3\,\epsilon+o(|{\epsilon}|),\\
		& k^{u}_{\text{rel}}({\epsilon})=k_2+k_4\,\epsilon+o(|{\epsilon}|);
	\end{align}
\end{subequations}
where
\begin{equation}
	k_1:=\frac{k^{u}_{10}+k^{u}_{20}}{2}>0,~~\text{and}~~k_2:=\frac{k^{u}_{10}-k^{u}_{20}}{2}
\end{equation}
are the bare (i.e., strain independent) average and relative unbinding rates, respectively, and
\begin{equation}
	k_3:=\frac{k^{u}_{10}\alpha_1+k^{u}_{20}\alpha_2}{2},~~\text{and}~~k_4:=\frac{k^{u}_{10}\alpha_1-k^{u}_{20}\alpha_2}{2}
\end{equation}
are the coefficients of the linear strain dependent parts of the  relative and average unbinding rates, respectively. 

Note that if $\zeta_{\text{rel}}=0$, $k_2=0$ and $k_4=0$, the distinction between the two contractile species disappears and the system becomes effectively one species. Hence, these three parameters in our model cannot be made zero simultaneously.

Let the characteristic time scale be $t^\star:=1/k^{b}_{\text{avg}}$, the characteristic length scale $l^\star:=\sqrt{\eta_b/\Gamma}$, and the characteristic density  $\rho_a^0$. We non-dimensionalize all the variables with the following redefinitions:
\begin{subequations}
	\begin{align}
		&\frac{t}{t^\star}\to t,~~\frac{{x}}{l^\star}\to{x},~~l^\star\partial_{x}\to \partial_{x},~~{l^\star}^2\partial^2_{xx}\to \partial^2_{xx}\\
		&\frac{{u}}{l^\star}\to {u},~~\frac{\rho}{\rho_a^0}\to \rho,~\frac{\phi}{\rho_a^0}\to\phi,~~\frac{D\,t^\star}{{l^{\star}}^2}\to D,\\
		&t^\star\,k^{b}_{\text{rel}}\to k^{b}_{\text{rel}},~~t^\star\,k^{u}_{\text{avg}}\to k^{u}_{\text{avg}},~~t^\star\,k^{u}_{\text{rel}}\to k^{u}_{\text{rel}},\\
		&\frac{B\, t^\star}{\Gamma {l^\star}^2}\to B,~~\frac{C \,t^\star \rho_a^0}{\Gamma {l^\star}^2}\to C,~~A\to A,\\
		&\frac{\zeta_{\text{avg}} \,t^\star \rho_a^0}{\Gamma {l^\star}^2}\to\zeta_{\text{avg}},~~\frac{\zeta_{\text{rel}}\, t^\star \rho_a^0}{\Gamma {l^\star}^2}\to \zeta_{\text{rel}}.
	\end{align}
\end{subequations}

With this, the non-dimensional form of the governing equations become
%\begin{widetext}
\begin{subequations}
	\begin{align}
		&\dot{\boldsymbol{u}}= \nabla\cdot\boldsymbol{\sigma}, \label{dyn-d-u-f12}\\
		& \dot{\rho} +\nabla\cdot(\rho\,\dot{\boldsymbol{u}}) = {D}\,\nabla^2\rho+\bigg(\rho_a^0-\frac{C}{A}\epsilon\bigg)-k^{u}_{\text{avg}}({\epsilon})\,\rho\nonumber\\
        &\hspace{52mm}-k^{u}_{\text{rel}}({\epsilon})\,\phi,\label{dyn-nd-e-f12}\\
		&\dot{\phi} +\nabla\cdot(\phi\,\dot{\boldsymbol{u}}) = {D}\,\nabla^2\phi+k^b_{\text{rel}}\,\bigg(\rho_a^0-\frac{C}{A}\epsilon\bigg)-k^{u}_{\text{avg}}({\epsilon})\,\phi\nonumber\\
        &\hspace{55mm}-k^{u}_{\text{rel}}({\epsilon})\,\rho.\label{dyn-nd-c-f12}
	\end{align}
	\label{dyn-nd-f12}
\end{subequations}
%\end{widetext}
with
\begin{equation}
	\boldsymbol{\sigma}
	=\Big(\sigma_0
	+\tilde{B}\,\epsilon
	+B_2\,\epsilon^2
	+B_3\,\epsilon^3+\dot{{\epsilon}}-K\nabla^2\epsilon\Big)\mathbf{I}+2\mu\tilde{\boldsymbol{\epsilon}},
	\label{totalstress-nondim}
\end{equation}
as the non-dimensional stress, where
\begin{eqnarray}
	\sigma_0&:=&2\,\chi(\rho^0_a)\big(\zeta_{\text{avg}}\,\rho+\zeta_{\text{rel}}\,\phi\big),\\
	\tilde{B}&:=&B-\frac{C^2}{A}-2\,\chi'(\rho^0_a)\,\frac{C}{A}\big(\zeta_{\text{avg}}\,\rho+\zeta_{\text{rel}}\,\phi\big),\\
	B_2&:=&\chi''(\rho^0_a)\,\bigg(\frac{C}{A}\bigg)^2\,\big(\zeta_{\text{avg}}\,\rho +\zeta_{\text{rel}}\,\phi\big),\\
	B_3&:=&-\frac{\chi'''(\rho^0_a)}{3}\,\bigg(\frac{C}{A}\bigg)^3\,\big(\zeta_{\text{avg}}\,\rho +\zeta_{\text{rel}}\,\phi\big).
\end{eqnarray}
Here, $\sigma_0$ is the purely active back pressure, $\tilde{B}$ is the activity renormalized bulk modulus of linear elasticity,  $B_2$ and $B_3$ are purely active nonlinear bulk moduli, which depend on
$\rho_1$ and $\rho_2$. For the effective material to show contractile response, $\chi''(\rho_a^0)$ and hence $B_2$ needs to be positive. For stability of the nonlinear elastic material,
$\chi'''(\rho_a^0)$  must be negative (rendering $B_3$ positive).

\section{Linear Stability Analysis}\label{sect:b}

\subsection{Homogeneous unstrained steady state}\label{sect:b1}

For the homogeneous unstrained steady states of the system \eqref{dyn-nd-f12}, we have $\boldsymbol{u}=\mathbf{0}$, and $\nabla\rho=\nabla\phi=\mathbf{0}$. The dynamical equations reduce to
\begin{subequations}
	\begin{align}
		&k_1\,\rho+k_2\,\phi = 1,\\
		& k_2\rho + k_1\,\phi= k^{b}_{\text{rel}},
	\end{align}
\end{subequations}
which yields
\begin{equation}
	\rho=\rho_0:=\frac{k_1-k_2\,k^{b}_{\text{rel}}}{(k_1)^2-(k_2)^2},~~\text{and}~~	\phi=\phi_0:=\frac{k_1\,k^{b}_{\text{rel}}-k_2}{(k_1)^2-(k_2)^2}.
	\label{hom-unstrained-ss}
\end{equation}
For $\rho_0 >0$, we require either  $k_1>k_2 k^{b}_{\text{rel}}$ and $(k_1)^2>(k_2)^2$, or $k_1<k_2 k^{b}_{\text{rel}}$ and $(k_1)^2<(k_2)^2$.
%%%%%%%%%%%%%%%%%%%%%%%%%%%%%%%%%%%%%%%%%%%%%%%%%%%%%%%%%%%%%%%%%%%%%%%%%%%%%%%%%%%%%%%%%%%%%%%%%%%%%%%%%%%%%%%%%%%%

\subsection{Linear stability of the homogeneous unstrained  steady state}\label{sect:b2}

To study the linearized dynamics of \eqref{dyn-nd-f12} around the homogeneous unstrained steady state, we substitute $\boldsymbol{u}=\delta\boldsymbol{u}$, $\rho=\rho_0+\delta\rho$ and $\phi=\phi_0+\delta\phi$ in \eqref{dyn-nd-f12}, neglect all the terms containing higher powers of ${\delta\boldsymbol{u}}$, ${\delta \rho}$ and ${\delta \phi}$, and obtain
\begin{widetext}
\begin{subequations}
	\begin{align}
		& \dot{\delta u}_{i} = \tilde{B}\, \delta u_{j,ji}+\mu\,\delta u_{i,jj}+2\,\chi(\rho_a^0)\big(\,\zeta_{\text{avg}}\, \delta\rho_{,i}+\zeta_{\text{rel}}\, \delta\phi_{,i} \big) 
		+ \dot{\delta u}_{j,ji}, \label{u-lin1-1d}\\
		& \dot{\delta\rho}+\rho_0\,\dot{\delta u}_{i,i}=D\,\delta\rho_{,ii} -k_1\,\delta\rho  -\bigg(\frac{C}{A} +k_3\,\rho_0 + k_4\,\phi_0\bigg) \delta u_{i,i} -k_2\,\delta\phi, \label{rho-lin1-1d}\\
		& \dot{\delta\phi}+\phi_0\,\dot{\delta u}_{i,i}=D\,\delta\phi_{,ii} -k_1\,\delta\phi  - \bigg(k^{b}_{\text{rel}}\frac{C}{A}  +k_3\,\phi_0 + k_4\,\rho_0\bigg)\,\delta u_{,ii}- k_2\,\delta\rho, \label{phi-lin1-1d}
	\end{align}
	\label{dyn-lin1-1d}
\end{subequations}
\end{widetext}
where $\tilde{B}:=B-\frac{C^2}{A}-2\chi'(\rho_a^0)\frac{C}{A}\big(\zeta_{\text{avg}}\,\rho_0
	+\zeta_{\text{rel}}\,\phi_0\big)$
is now the (spatiotermporally constant) renormalized linear elastic modulus at the homogeneous steady state.

The Fourier transform of \eqref{dyn-lin1-1d} with respect to $\mathbf{x}$ is obtained by substituting the anstatz $\delta A(\mathbf{x},t)=\frac{1}{(2\pi)^d}\int{\delta \hat{A}(t)}e^{i\mathbf{q}\cdot\mathbf{x}}\,d\mathbf{q}$, where $A$ stands for $\boldsymbol{u}$, $\rho$, $\phi$, and ${\bf q}$ is the wave vector, with $q:=|{\bf q}|$. We rewrite the resulting equations in terms of dilation mode $\hat{u}_\parallel\equiv\hat{\delta \epsilon}:=\hat{\delta u}_i q_i$, 
the Fourier transform of $\epsilon=\text{div}\,\boldsymbol{u}$. We do not write the dynamics of the shear mode $\hat{\delta u}_{\perp}:=e_{ij}\hat{\delta u}_i q_j$ here as it uncouples from the rest of the dynamics ($e_{ij}$ denotes the 2D Levi-Civita symbol, with $e_{11}=e_{22}=0$ and $e_{12}=-e_{21}=1$). The final linear equations (assuming the interface term $K=0$) are
\begin{subequations}
	\begin{align}
		&\dot{\hat{\delta \epsilon}}= -\frac{(\tilde{B}+\mu)\,q^2\,\hat{\delta \epsilon}+2\,\chi(\rho_a^0)\big(\,\zeta_{\text{avg}}\, \hat{\delta\rho}+\zeta_{\text{rel}}\, \hat{\delta\phi} \big)q^2}{1+q^2}, \label{u-lin1-ft}\\
		& \dot{\hat{\delta\rho}}+\rho_0\,\dot{\hat{\delta \epsilon}}=-\big(D\,q^2 +k_1\big)\hat{\delta\rho}  -\bigg(\frac{C}{A} +k_3\,\rho_0 + k_4\,\phi_0\bigg) \hat{\delta \epsilon}\nonumber\\ 
        &\hspace{50mm}-k_2\,\hat{\delta\phi}, \label{rho-lin1-ft}\\
		& \dot{\hat{\delta\phi}}+\phi_0\,\dot{\hat{\delta \epsilon}}=-\big(D\,q^2  + k_1\big)\hat{\delta\phi} - \bigg(k^{b}_{\text{rel}}\frac{C}{A}  +k_3\,\phi_0 + k_4\,\rho_0\bigg)\,\hat{\hat{\delta \epsilon}}\nonumber\\
        &\hspace{50mm}- k_2\,\hat{\delta\rho}.
  \label{phi-lin1-ft}
	\end{align}
	\label{dyn-lin1-ft}
\end{subequations}
%and $\dot{\hat{\delta u }}_{\perp}=-\frac{\mu\,q^2}{1+\eta_s q^2}\hat{\delta u}_{\perp}$, i.e., the transverse mode is decoupled from the rest of the system.
%%%%%%%%%%%%%%%%%%%%%%%%%%%%%%%%%%%%%%%%%%%%%%%%%%%%%%%%%%%%%%%%%%%%%%%%%%%%%%%%%%%%%%%%%%
%\subsection{Symmetric mixture of stresslets} 
%\label{phi0case}
We focus on the special case  $\phi_0=0$, i.e., symmetric binary mixture. From \eqref{hom-unstrained-ss}, we obtain the necessary condition to maintain this, namely,  $k^{b}_{\text{rel}}=k_2/k_1$. As a consequence we see that      $\rho_0=1/k_1$, and $\tilde{B}=B-\frac{C^2}{A}-2\,\chi'(\rho_a^0)\,\frac{C}{A}\,\frac{\zeta_{\text{avg}}}{k_1}$. 
The resulting dynamical system in the Fourier space can be written as $\dot{\mathbf{w}}=\mathbf{M}\mathbf{w}$,
%\begin{equation}
	%\dot{\mathbf{w}}=\mathbf{M}\mathbf{w}
%\end{equation}
where $\mathbf{w}({\bf q},t)=\Big(\hat{\delta \epsilon}({\bf q},t)\,\,\hat{\delta \rho}({\bf q},t)\,\,\hat{\delta \phi}({\bf q},t)\Big)^T$, and
\begin{widetext}
\begin{equation}
	\mathbf{M}=\left[\begin{array}{ccc}
		-\frac{(\tilde{B}+\mu)\,q^2}{1+q^2} & -\frac{2\chi(\rho_a^0)\,\zeta_{\text{avg}}\,q^2}{1+q^2} & -\frac{2\chi(\rho_a^0)\,\zeta_{\text{rel}}\,q^2}{1+q^2} \\
		-\bigg(\frac{C}{A}+\frac{k_3}{k_1}-\frac{\tilde{B}\,q^2}{(1+q^2)k_1}\bigg) & -D q^2-k_1+\frac{2\chi(\rho_a^0)\,\zeta_{\text{avg}}\,q^2}{(1+q^2)k_1} & -k_2+\frac{2\chi(\rho_a^0)\,\zeta_{\text{rel}}\,q^2}{(1+q^2)k_1} \\
		-\bigg(\frac{k_2}{k_1}\frac{C}{A}+\frac{k_4}{k_1}\bigg) & -k_2 & -D q^2- k_1
	\end{array}\right].
	\label{simplified-lin-system}
\end{equation}
\end{widetext}

If $\lambda_i(q)$ are distinct eigenvalues of  $\mathbf{M}(q)$ and $\mathbf{v}_i(q)$ are the corresponding (linearly independent) eigenvectors, then we can write the general solution as
\begin{equation}
	\mathbf{w}(q,t)=\sum_{i=1}^{3}c_i(q)e^{\lambda_i(q)\,t}\,\mathbf{v}_i(q),
\end{equation}
where the coefficients $c_i(q)$ are the projections of the initial data $\mathbf{w}(q,0)$ along the respective eigenvectors:\\ $\Big(c_1(q)\,c_2(q)\,c_3(q)\Big)^T={\mathbf{V}(q)}^{-1}\mathbf{w}(q,0)$, where  $\mathbf{V}(q):=\Big[\mathbf{v}_1(q)\,\mathbf{v}_2(q)\,\mathbf{v}_3(q)\Big]$ is the matrix containing the eigenvectors as columns.

The three eigenvalues, assuming $A=1$, %and $\chi(\rho_a^0)=\chi'(\rho_a^0)=1$,  
are $\lambda_1=-k_1-D\,q^2$ and $\lambda_{2,3}=-\frac{\lambda_a\pm\sqrt{\lambda_b}}{2(1+q^2)}$,
%\begin{eqnarray}
	%\lambda_1&=&-k_1-D\,q^2,\\
	%\lambda_{2,3}&=&-\frac{\lambda_a\pm\sqrt{\lambda_b}}{2(1+q^2)};
	%\label{eigenvalues}
%\end{eqnarray}
where 
\begin{subequations}
	\begin{align}
		&\lambda_a:=k_1+\Big(\tilde{B}+\mu-\frac{2\chi(\rho_a^0)\zeta_{\text{avg}}}{k_1}+D+k_1\Big)\,q^2+D\,q^4,\label{lambda_a}\\
		&\lambda_b:= \lambda_a^2-4\,q^2(1+q^2)\bigg[(\tilde{B}+\mu)D\,q^2\nonumber\\
    &+k_1\bigg(\tilde{B}+\mu-2\chi(\rho_a^0)\frac{\zeta_{\text{avg}}}{k_1}\bigg(\frac{C}{A}+\frac{k_3}{k_1}\bigg)-2\chi(\rho_a^0)\frac{\zeta_{\text{rel}}}{k_1}\frac{k_4}{k_1}\bigg)\bigg];\label{lambda_b}
	\end{align}
\end{subequations}
and the corresponding eigenvectors are
\begin{equation}
	\mathbf{v}_1=\left[\begin{array}{c}
		0\\
		-\frac{\zeta_{\text{rel}}}{\zeta_{\text{avg}}}\\
		1
	\end{array}
	\right],~\mathbf{v}_{2,3}=\left[\begin{array}{c}
		\frac{1}{2(1+q^2)}\frac{k_1}{k_4}\Big(v_a\mp\sqrt{v_b}\Big)\\
		\frac{1}{2(1+q^2)}\frac{1}{k_4}\Big(v_c\mp\frac{1}{k_1}\sqrt{v_b}\Big)
		
	\end{array}
	\right],
\end{equation}
where $v_{a,b,c,d}$ have rather long expressions that we do not write here.
\iffalse
\begin{widetext}
    \begin{eqnarray}
	v_a&:=&-k_1+\Big(\tilde{B}_0-\frac{2\zeta_{\text{avg}}}{k_1}-D-k_1\Big)q^2-Dq^4;\\
	v_b&:=&2q^2(1+q^2)(-B+C^2+D+Dq^2)k_1^3+(1+q^2)^2k_1^4+4\zeta^2_{\text{avg}}(1+C)^2q^4    \nonumber   \\
	&&+q^2k_1^2\Bigg[q^2\Big(-B+C^2+D+Dq^2\Big)^2+4(3C-1)(1+q^2)\zeta_{\text{avg}}\Bigg]\nonumber\\
	&&+4q^2k_1\Big[\zeta_{\text{avg}}\Big(\Big(-B(1+C)-D+C(C+C^2+D)+(C-1)Dq^2\Big)+2(1+q^2)k_3\Big)+2(1+q^2)k_4\zeta_{\text{rel}}\Big];\\
	v_c&:=&-(1-2C)k_1+2k_3-\Bigg(\tilde{B}_0-\frac{2\zeta_\text{avg}}{k_1}+D+(1-2C)k_1-2k_3\Bigg)q^2-Dq^4.
\end{eqnarray}
\end{widetext}
\fi
Since the matrix $\mathbf{M}$ is non-Hermitian (due to presence of activity and turnover), its eigenvalues $\lambda_i$ are not real and the eigenvectors $\mathbf{v}_i$ are not orthogonal in general.
We make one further simplifying assumption: $k_2=0$, hence, $k^{b}_{\text{rel}}=0$, meaning that the bare (strain independent) part of the unbinding rates are idential (i.e., $k^{u}_{10}=k^{u}_{20}$), and the binding rates are identical as well (i.e., $k^{b}_{1}=k^{b}_{2}$). It follows that $k_4=\frac{k^{u}_{10}}{2}(\alpha_1-\alpha_2)$.

Note that, (a) the coupling $M_{12}$ between $\hat{\delta \epsilon}$ and $\hat{\delta\rho}$ in the $\hat{\delta \epsilon}$-equation  and the coupling $M_{21}$ between $\hat{\delta \epsilon}$ and $\hat{\delta\rho}$ in the $\hat{\delta\rho}$-equation can be  of opposite signs depending upon the relative signs and magnitudes of $k_3$ and $\tilde{B}$; (b)  the coupling $M_{13}$ between $\hat{\delta \epsilon}$ and $\hat{\delta \phi}$ in the $\hat{\delta \epsilon}$-equation  and the coupling $M_{31}$ between $\hat{\delta \epsilon}$ and $\hat{\delta\phi}$ in the $\hat{\delta\phi}$-equation are  of opposite signs when $\zeta_{\text{rel}}$ and $k_4$ have the opposite signs; and (c) the coupling $M_{23}$ between $\hat{\delta\rho}$ and $\hat{\delta\phi}$ in the $\hat{\delta\rho}$-equation can have any sign while the coupling $M_{32}$ between $\hat{\delta\phi}$ and $\hat{\delta\rho}$ in the $\hat{\delta\phi}$-equation is zero. Hence, $\hat{\delta\rho}$-$\hat{\delta\phi}$ interaction is always non-reciprocal, while the $\hat{\delta \epsilon}$-$\hat{\delta\phi}$ interaction is  non-reciprocal when the stronger stresslet  unbinds faster; $\hat{\delta \epsilon}$-$\hat{\delta\rho}$ interaction can be  non-reciprocal depending upon the relative signs and magnitudes of $k_3$ and $\tilde{B}$.

If $\lambda_i(q)$ are distinct eigenvalues of  $\mathbf{M}(q)$ and $\mathbf{v}_i(q)$ are the corresponding (linearly independent) eigenvectors, then we can write the general solution as
\begin{equation}
	\mathbf{w}(q,t)=\sum_{i=1}^{3}c_i(q)e^{\lambda_i(q)\,t}\,\mathbf{v}_i(q),
\end{equation}
where the coefficients $c_i(q)$ are the projections of the initial data $\mathbf{w}(q,0)$ along the respective eigenvectors: $\Big(c_1(q)\,c_2(q)\,c_3(q)\Big)^T={\mathbf{V}(q)}^{-1}\mathbf{w}(q,0)$, where  $\mathbf{V}(q):=\Big[\mathbf{v}_1(q)\,\mathbf{v}_2(q)\,\mathbf{v}_3(q)\Big]$ is the matrix containing the eigenvectors as columns.

%%%%%%%%%%%%%%%%%%%%%%%%%%%%%%%%%%%%%%%%%%%%%%%%%%%%%%%%%%%%%%%%%%%%%%%%%%%%%%%%%%%%%%%%%%%%%%%%%%%%%

    \begin{figure}[t]
	\centering
\includegraphics[scale=.3]{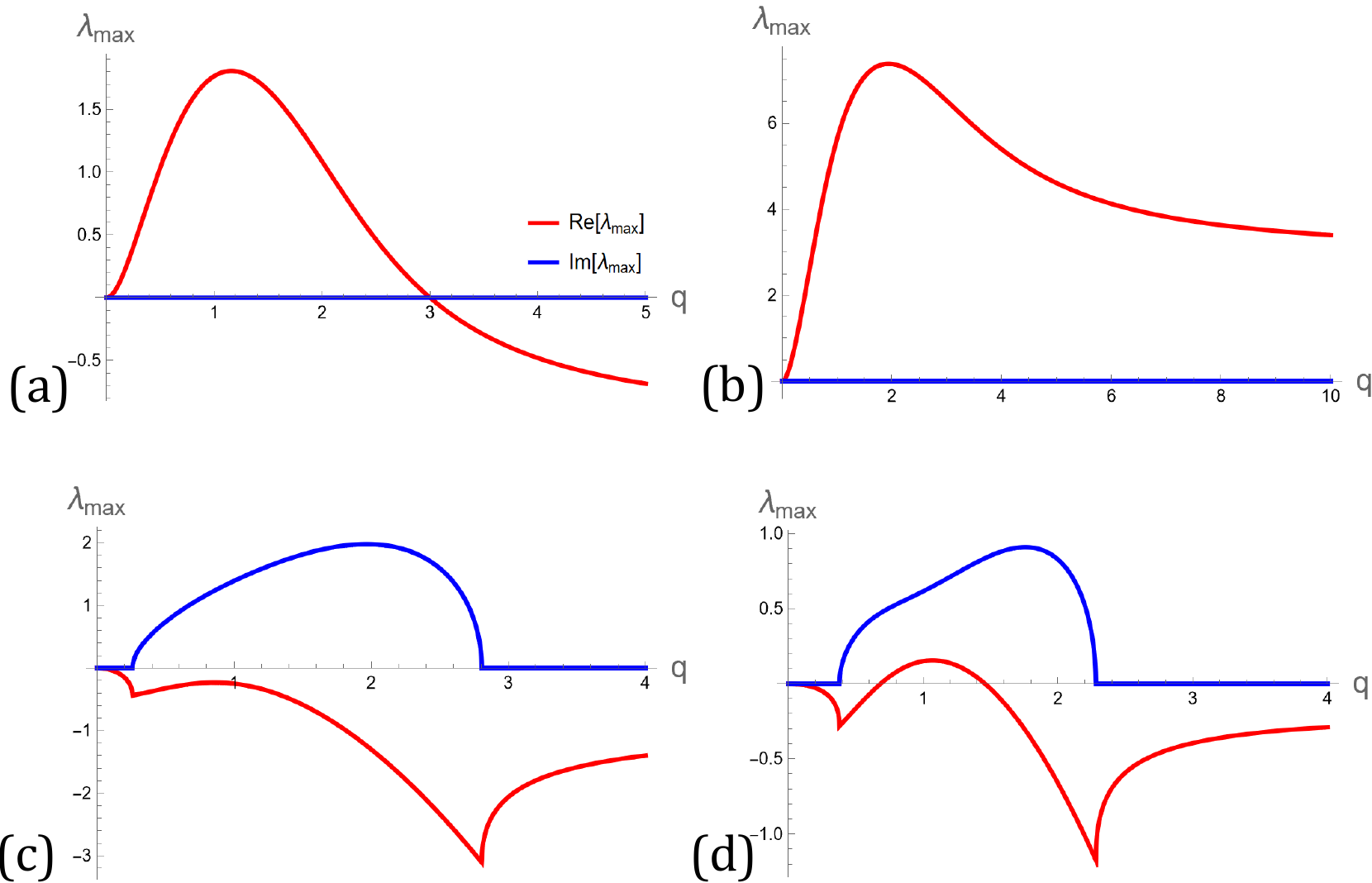}
\caption{Linear stability dispersion curves, with $C=1$, $A=1$, $D=1$, $k_1=1$, $k_2=0$, $\chi(\rho_a^0)=1$, $\chi'(\rho_a^0)=1$, showing (a) segregation instability, (b) elastic contractile instability, (c) stable wave and (d) unstable wave.}
\label{dispersionsi}
\end{figure}
%%%%%%%%%%%%%%%%%%%%%%%%%%%%%%%%%%%%%%%%%%%%%%%%%%%%%%%%%%%%%%%%%%%%%%%%%%%%%%%%%%%%%%%%%%%%%%%%%%%%%

%%%%%%%%%%%%%%%%%%%%%%%%%%%%%%%%%%%%%%%%%%%%%%%%%%%%%%%%%%%%%%%%%%%%%%%%%%%%%%%%%%%%%%%%%%%%%%%%%%%%%%%%%%%%%%%%%%%%%%%%%%%%%%%%%%%%%%%%%%%%%%%%%%%%%%%%%%%%%%%%%%%%%%%%%%%%%%%%%%%%%%%%%%%%%%%%%%%%%%%%%%%%%%%%%%%%%%%%%%%%%%%%%%%%%%%%%%%%%%%%%%%%%%%%%%%%%%%%%%%%%%%%

	\section{Linear Instabilities and Phases}\label{sect:b3}

	\begin{table}[H]
		\centering
		\begin{tabular}{|c|c|}
			\hline
			{\it (In)stability description} & {\it Definition} \\ 
			\hline
			\hline
			Mechanical stability of the elastomer & $\tilde{B}+\mu> 0$ \\
			\hline
			Mechanical instability of the elastomer & $\tilde{B}+\mu\le0$  \\
			\hline
			\hline
			\makecell{Segregation instability}  & $\tilde{B}+\mu>0$, $\lambda_b>0$, $\lambda\ge 0$  \\
			\hline
			\makecell{Monotonic stability} & $\tilde{B}+\mu>0$, $\lambda_b> 0$, $\lambda< 0$  \\
			\hline
			\hline
			\makecell{Damped oscillations} & $\tilde{B}+\mu> 0$, $\lambda_b<0$, $\lambda_a< 0$  \\
			\hline
			\makecell{Growing oscillations (unstable)} & $\tilde{B}+\mu> 0$, $\lambda_b<0$, $\lambda_a>0$  \\
			\hline
			\makecell{Sustained oscillations\\(swap, travelling waves)} & $\tilde{B}+\mu> 0$, $\lambda_b<0$, $\lambda_a= 0$  \\
			\hline
		\end{tabular}
		\caption{Definition of the instabilities and phases in the linear analysis.}
		\label{table:phases}.
	\end{table}
	Our first observation is that $\lambda_1$ is real and  $\lambda_1< 0$ for all $q$, since $k_1>0$ and $D>0$ by definition. Hence, all modes corresponding to the first eigenvalue $\lambda_1$ are asymptotically  stable (i.e., monotonically decaying). Secondly, we observe that  $Re[\lambda_3]\ge Re[\lambda_2]$ for all $q$, i.e., $\lambda_{max}=\lambda_3$. Hence, the largest eigenvalue  $\lambda_{max}=\frac{-\lambda_a+\sqrt{\lambda_b}}{2(1+q^2)}$ determines the asymptotic stability of the linearized system, i.e., the `linear' phases. We give the definitions of various phases for 1D case in Table \ref{table:phases}.
    
	Based on the definitions given in Table \ref{table:phases}, we can construct various linear stability phase diagrams, which, in principle, depend on the wave vector magnitude $q$. The phase diagrams presented in the main text are constructed assuming $q$ to be small, i.e., in the long wavelength limit.  
 %We Taylor expand $\lambda_{max}(q)$ and $\lambda_b(q)$: $\lambda_{max}(q)=P_0+P_2\,q^2+P_4\,q^4+\cdots$, $\lambda_{b}(q)=Q_0+Q_2\,q^2+Q_4\,q^4+\cdots$; then, for small $q$, sign of $\lambda_{max}(q)$ and $\lambda_b(q)$ will be determined by the signs of the lower order coefficients $P_{0,2,4}$, $Q_{0,2,4}$ etc.~which are dependent on various parameters of our system. Based on the signs of these coefficients, we construct $q$-independent phase diagrams.

	\subsection{Segregation} 
	
	When $\lambda_b>0$, then $\lambda_{2,3}$ are real, hence, we have the monotonic stable/unstable phases.  The  onset of segregation instability is dictated by the sign change of $\lambda_{max}$ from negative to positive.
	
	The fastest growing mode $q_{seg}$ is where $\partial_q\lambda_{max}=0$. With $\chi(\rho_a^0)=\chi'(\rho_a^0)=1$, $C=1$, $A=1$, $D=1$ and $k_1=1$, we find
\begin{equation}
	    q_{seg}=\frac{1}{\sqrt{2}}\sqrt{\frac{B+\mu-1-2\zeta_{\text{avg}}(2+k_3)-2\zeta_{\text{rel}}k_4}{G}}\, ,\label{eqnqseg}
	\end{equation}
	where
\begin{eqnarray}
    G&:=&B+\mu-1-6\zeta_{\text{avg}}(1+k_3)+2(B+\mu)k_3\zeta_{\text{avg}}\nonumber\\
    &&-4k_3\zeta_{\text{avg}}^2(2+k_3)-2\zeta_{\text{rel}}k_4(3-B-\mu)\nonumber\\
    &&-8\zeta_{\text{avg}}\zeta_{\text{rel}}k_4(1+k_3)-4\zeta_{\text{rel}}^2k_4^2.
\end{eqnarray}
	
	We observe that $\lambda_{max}=0$ when $\lambda_a=\sqrt{\lambda_b}$, which, using \eqref{lambda_b}, implies that
	
	\begin{equation}
		(\tilde{B}+\mu)\,D\,q^2+k_1\bigg(\tilde{B}+\mu-2\frac{\zeta_{\text{avg}}}{k_1}\bigg(\frac{C}{A}+\frac{k_3}{k_1}\bigg)-2\frac{\zeta_{\text{rel}}}{k_1}\frac{k_4}{k_1}\bigg)=0.
	\end{equation}
	The (positive) solution to this equation
	\begin{equation}
		q=\sqrt{\frac{k_1}{D(\tilde{B}+\mu)}}\sqrt{-\Bigg(\tilde{B}+\mu-2\frac{\zeta_{\text{avg}}}{k_1}\bigg(\frac{C}{A}+\frac{k_3}{k_1}\bigg)-2\frac{\zeta_{\text{rel}}}{k_1}\frac{k_4}{k_1}\Bigg)}
	\end{equation}
	is real and non-zero for $\tilde{B}_0>0$ if and only if
	\begin{equation}
		\tilde{B}+\mu-2\frac{\zeta_{\text{avg}}}{k_1}\bigg(\frac{C}{A}+\frac{k_3}{k_1}\bigg)-2\frac{\zeta_{\text{rel}}}{k_1}\frac{k_4}{k_1}< 0.
		\label{segrecond}
	\end{equation}

	Hence, if the condition \eqref{segrecond} is met, $\lambda_3$ becomes non-negative in the long wavelength limit, thus, triggering segregation instability. The characteristic width of the segregated regime is $\sim\frac{1}{q_{seg}}$.

    %%%%%%%%%%%%%%%%%%%%%%%%%%%%%
%%%%%%%%%%%%%%%%%%%%%%%%%%%%%%%%%%%%
%%%%%%%%%%%%%%%%%%%%%%%%%%%%%%%%%%%%%%%%%%%%%%%%%%%%%%%%%%%%%%%%%%%%%%%%%%%%%%%%%%%%%%%%%%%%%%%%%%%%%%%%%%%%%%%%

\section{Nonlinear Analysis of governing equations}
\label{sect:nonlin}

These scaling forms imply the following for the derivatives,
\begin{subequations}
	\begin{align}
		& \dot \rho=\frac{C_0}{(t_0-t)^{1+s}}\bigg[s\,R(\xi)+r\,\xi\,R'(\xi)\bigg],\\
        &\dot \phi=\frac{A_0}{(t_0-t)^{1+p}}\bigg[p\,\Phi(\xi)+r\,\xi\,\Phi'(\xi)\bigg],\\
		&\dot \epsilon=\frac{B_0}{(t_0-t)^{1+q}}\bigg[q\,E(\xi)+r\,\xi\,E'(\xi)\bigg],
  \end{align}
	\label{simtrans-dert}
\end{subequations}
and
\begin{subequations}
	\begin{align}
		&\partial_x \rho=\frac{C_0}{(t_0-t)^{s+r}}R'(\xi),~~\partial^2_{xx} \rho=\frac{C_0}{(t_0-t)^{s+2r}}R''(\xi),\\
		&\partial_x \phi=\frac{A_0}{(t_0-t)^{p+r}}\Phi'(\xi),~~\partial^2_{xx} \phi=\frac{A_0}{(t_0-t)^{p+2r}}\Phi''(\xi),\\
		&\partial_x \epsilon=\frac{B_0}{(t_0-t)^{q+r}}E'(\xi),~~\partial^2_{xx} \epsilon=\frac{B_0}{(t_0-t)^{q+2r}}E''(\xi),
	\end{align}
	\label{simtrans-derx}
\end{subequations}
where, $(\cdot)'$ denotes differentiation with respect to $\xi$.

We substitute the similarity forms \eqref{simtrans}, \eqref{simtrans-dert} and \eqref{simtrans-derx} into the system \eqref{dyn-1d-simple}, and ignore diffusion as it cannot produce any finite time singularity. %we will ignore the diffusion term in the $\rho$ and $\phi$-equations. 
The $\epsilon$-equation gives,
\vspace{-10mm}
\begin{widetext}
\begin{align}
&\frac{B_0}{(t_0-t)^{1+q}}\bigg(q\,E+r\,\xi\,E'\bigg) = 2\,\chi(\rho^0_a)\,\bigg(\zeta_{\text{avg}}\,\frac{C_0}{(t_0-t)^{s+2r}}R''+\zeta_{\text{rel}}\,\frac{A_0}{(t_0-t)^{p+2r}}\Phi''\bigg)\nonumber\\
&+\big(B-C^2\big)\frac{B_0}{(t_0-t)^{q+2r}}E''-2\,\chi'(\rho^0_a)\,C\bigg(\zeta_{\text{avg}}\,\frac{C_0}{(t_0-t)^{s}}R+\zeta_{\text{rel}}\,\frac{A_0}{(t_0-t)^{p}}\Phi\bigg)\,\frac{B_0}{(t_0-t)^{q+2r}}E''\nonumber\\
&-4\,\chi'(\rho^0_a)\,C\bigg(\zeta_{\text{avg}}\,\frac{C_0}{(t_0-t)^{s+r}}R'+\zeta_{\text{rel}}\,\frac{A_0}{(t_0-t)^{p+r}}\Phi'\bigg)\,\frac{B_0}{(t_0-t)^{q+r}}E'\nonumber\\
&-2\,\chi'(\rho^0_a)\,C\bigg(\zeta_{\text{avg}}\,\frac{C_0}{(t_0-t)^{s+2r}}R''+\zeta_{\text{rel}}\,\frac{A_0}{(t_0-t)^{p+2r}}\Phi''\bigg)\,\frac{B_0}{(t_0-t)^{q}}E\nonumber\\ &+2\chi''(\rho^0_a)\,C^2\,\bigg(\zeta_{\text{avg}}\,\frac{C_0}{(t_0-t)^{s+r}}R' +\zeta_{\text{rel}}\,\frac{A_0}{(t_0-t)^{p+r}}\Phi'\bigg)\,\frac{B_0^2}{(t_0-t)^{2q+r}}EE'\nonumber\\
&+\chi''(\rho^0_a)\,C^2\,\bigg(\zeta_{\text{avg}}\,\frac{C_0}{(t_0-t)^{s+2r}}R'' +\zeta_{\text{rel}}\,\frac{A_0}{(t_0-t)^{p+2r}}\Phi''\bigg)\,\frac{B_0^2}{(t_0-t)^{2q}}E^2\nonumber\\
&+2\chi''(\rho^0_a)\,C^2\,\bigg(\zeta_{\text{avg}}\,\frac{C_0}{(t_0-t)^{s}}R +\zeta_{\text{rel}}\,\frac{A_0}{(t_0-t)^{p}}\Phi\bigg)\,\frac{B_0^2}{(t_0-t)^{2q+2r}}\Big(EE''+E'^2\Big)\nonumber\\
&+2\chi''(\rho^0_a)\,C^2\,\bigg(\zeta_{\text{avg}}\,\frac{C_0}{(t_0-t)^{s+r}}R' +\zeta_{\text{rel}}\,\frac{A_0}{(t_0-t)^{p+r}}\Phi'\bigg)\,\frac{B_0^2}{(t_0-t)^{2q+r}}EE'.\label{epsilon-ode}
\end{align}
\end{widetext}

The $\rho$-equation gives,
\begin{widetext}
\begin{align}
&\frac{C_0}{(t_0-t)^{1+s}}\bigg(s\,R+r\,\xi\,R'\bigg)=-\frac{C_0}{(t_0-t)^{s+r}}R'\Bigg[2\,\chi(\rho^0_a)\,\bigg(\zeta_{\text{avg}}\,\frac{C_0}{(t_0-t)^{s+r}}R'+\zeta_{\text{rel}}\,\frac{A_0}{(t_0-t)^{p+r}}\Phi'\bigg)\nonumber\\
&+\big(B-C^2\big)\frac{B_0}{(t_0-t)^{q+r}}E'-2\,\chi'(\rho^0_a)\,C\bigg(\zeta_{\text{avg}}\,\frac{C_0}{(t_0-t)^{s}}R+\zeta_{\text{rel}}\,\frac{A_0}{(t_0-t)^{p}}\Phi\bigg)\,\frac{B_0}{(t_0-t)^{q+r}}E'\nonumber\\
&-2\,\chi'(\rho^0_a)\,C\bigg(\zeta_{\text{avg}}\,\frac{C_0}{(t_0-t)^{s+r}}R'+\zeta_{\text{rel}}\,\frac{A_0}{(t_0-t)^{p+r}}\Phi'\bigg)\,\frac{B_0}{(t_0-t)^{q}}E\nonumber\\ &+2\chi''(\rho^0_a)\,C^2\,\bigg(\zeta_{\text{avg}}\,\frac{C_0}{(t_0-t)^{s}}R +\zeta_{\text{rel}}\,\frac{A_0}{(t_0-t)^{p}}\Phi\bigg)\,\frac{B_0^2}{(t_0-t)^{2q+r}}EE'\nonumber\\
&+\chi''(\rho^0_a)\,C^2\,\bigg(\zeta_{\text{avg}}\,\frac{C_0}{(t_0-t)^{s+r}}R' +\zeta_{\text{rel}}\,\frac{A_0}{(t_0-t)^{p+r}}\Phi'\bigg)\,\frac{B_0^2}{(t_0-t)^{2q}}E^2
\Bigg]\nonumber\\
&-\frac{C_0}{(t_0-t)^{s}}R\times\Bigg[2\,\chi(\rho^0_a)\,\bigg(\zeta_{\text{avg}}\,\frac{C_0}{(t_0-t)^{s+2r}}R''+\zeta_{\text{rel}}\,\frac{A_0}{(t_0-t)^{p+2r}}\Phi''\bigg)\nonumber\\
&+\big(B-C^2\big)\frac{B_0}{(t_0-t)^{q+2r}}E''-2\,\chi'(\rho^0_a)\,C\bigg(\zeta_{\text{avg}}\,\frac{C_0}{(t_0-t)^{s}}R+\zeta_{\text{rel}}\,\frac{A_0}{(t_0-t)^{p}}\Phi\bigg)\frac{B_0}{(t_0-t)^{q+2r}}E''\nonumber\\
&-2\,\chi'(\rho^0_a)\,C\bigg(\zeta_{\text{avg}}\,\frac{C_0}{(t_0-t)^{s+r}}R'+\zeta_{\text{rel}}\,\frac{A_0}{(t_0-t)^{p+r}}\Phi'\bigg)\,\frac{B_0}{(t_0-t)^{q+r}}E'\nonumber\\
&-2\,\chi'(\rho^0_a)\,C\bigg(\zeta_{\text{avg}}\,\frac{C_0}{(t_0-t)^{s+r}}R'+\zeta_{\text{rel}}\,\frac{A_0}{(t_0-t)^{p+r}}\Phi'\bigg)\,\frac{B_0}{(t_0-t)^{q+r}}E'\nonumber\\
&-2\,\chi'(\rho^0_a)\,C\bigg(\zeta_{\text{avg}}\,\frac{C_0}{(t_0-t)^{s+2r}}R''+\zeta_{\text{rel}}\,\frac{A_0}{(t_0-t)^{p+2r}}\Phi''\bigg)\,\frac{B_0}{(t_0-t)^{q}}E\nonumber\\ &+2\chi''(\rho^0_a)\,C^2\,\bigg(\zeta_{\text{avg}}\,\frac{C_0}{(t_0-t)^{s+r}}R' +\zeta_{\text{rel}}\,\frac{A_0}{(t_0-t)^{p+r}}\Phi'\bigg)\,\frac{B_0^2}{(t_0-t)^{2q+r}}EE'\nonumber\\
&+\chi''(\rho^0_a)\,C^2\,\bigg(\zeta_{\text{avg}}\,\frac{C_0}{(t_0-t)^{s+2r}}R'' +\zeta_{\text{rel}}\,\frac{A_0}{(t_0-t)^{p+2r}}\Phi''\bigg)\,\frac{B_0^2}{(t_0-t)^{2q}}E^2\nonumber\\
&+2\chi''(\rho^0_a)\,C^2\,\bigg(\zeta_{\text{avg}}\,\frac{C_0}{(t_0-t)^{s}}R +\zeta_{\text{rel}}\,\frac{A_0}{(t_0-t)^{p}}\Phi\bigg)\,\frac{B_0^2}{(t_0-t)^{2q+2r}}\Big(EE''+E'^2\Big)\nonumber\\
&+2\chi''(\rho^0_a)\,C^2\,\bigg(\zeta_{\text{avg}}\,\frac{C_0}{(t_0-t)^{s+r}}R' +\zeta_{\text{rel}}\,\frac{A_0}{(t_0-t)^{p+r}}\Phi'\bigg)\,\frac{B_0^2}{(t_0-t)^{2q+r}}EE'\Bigg].\label{rho-ode}
\end{align}
\end{widetext}

The $\phi$-equation gives,
\begin{widetext}
\begin{align}
 &\frac{A_0}{(t_0-t)^{1+p}}\bigg(p\,\Phi+r\,\xi\,\Phi'\bigg)=-\frac{A_0}{(t_0-t)^{p+r}}\Phi'\times\Bigg[2\,\chi(\rho^0_a)\,\bigg(\zeta_{\text{avg}}\,\frac{C_0}{(t_0-t)^{s+r}}R'+\zeta_{\text{rel}}\,\frac{A_0}{(t_0-t)^{p+r}}\Phi'\bigg)\nonumber\\
 &+\big(B-C^2\big)\frac{B_0}{(t_0-t)^{q+r}}E'-2\,\chi'(\rho^0_a)\,C\bigg(\zeta_{\text{avg}}\,\frac{C_0}{(t_0-t)^{s}}R+\zeta_{\text{rel}}\,\frac{A_0}{(t_0-t)^{p}}\Phi\bigg)\,\frac{B_0}{(t_0-t)^{q+r}}E'\nonumber\\
&-2\,\chi'(\rho^0_a)\,C\bigg(\zeta_{\text{avg}}\,\frac{C_0}{(t_0-t)^{s+r}}R'+\zeta_{\text{rel}}\,\frac{A_0}{(t_0-t)^{p+r}}\Phi'\bigg)\,\frac{B_0}{(t_0-t)^{q}}E\nonumber\\ &+2\chi''(\rho^0_a)\,C^2\,\bigg(\zeta_{\text{avg}}\,\frac{C_0}{(t_0-t)^{s}}R +\zeta_{\text{rel}}\,\frac{A_0}{(t_0-t)^{p}}\Phi\bigg)\,\frac{B_0^2}{(t_0-t)^{2q+r}}EE'\nonumber\nonumber\\
&+\chi''(\rho^0_a)\,C^2\,\bigg(\zeta_{\text{avg}}\,\frac{C_0}{(t_0-t)^{s+r}}R' +\zeta_{\text{rel}}\,\frac{A_0}{(t_0-t)^{p+r}}\Phi'\bigg)\,\frac{B_0^2}{(t_0-t)^{2q}}E^2
 \Bigg]\nonumber\\
&-\frac{A_0}{(t_0-t)^{p}}\Phi\times\Bigg[2\,\chi(\rho^0_a)\,\bigg(\zeta_{\text{avg}}\,\frac{C_0}{(t_0-t)^{s+2r}}R''+\zeta_{\text{rel}}\,\frac{A_0}{(t_0-t)^{p+2r}}\Phi''\bigg)\nonumber\\
&+\big(B-C^2\big)\frac{B_0}{(t_0-t)^{q+2r}}E''-2\,\chi'(\rho^0_a)\,C\bigg(\zeta_{\text{avg}}\,\frac{C_0}{(t_0-t)^{s}}R+\zeta_{\text{rel}}\,\frac{A_0}{(t_0-t)^{p}}\Phi\bigg)\,\frac{B_0}{(t_0-t)^{q+2r}}E''\nonumber\\
&-4\,\chi'(\rho^0_a)\,C\bigg(\zeta_{\text{avg}}\,\frac{C_0}{(t_0-t)^{s+r}}R'+\zeta_{\text{rel}}\,\frac{A_0}{(t_0-t)^{p+r}}\Phi'\bigg)\,\frac{B_0}{(t_0-t)^{q+r}}E'\nonumber\\
&-2\,\chi'(\rho^0_a)\,C\bigg(\zeta_{\text{avg}}\,\frac{C_0}{(t_0-t)^{s+2r}}R''+\zeta_{\text{rel}}\,\frac{A_0}{(t_0-t)^{p+2r}}\Phi''\bigg)\,\frac{B_0}{(t_0-t)^{q}}E\nonumber\\ &+2\chi''(\rho^0_a)\,C^2\,\bigg(\zeta_{\text{avg}}\,\frac{C_0}{(t_0-t)^{s+r}}R' +\zeta_{\text{rel}}\,\frac{A_0}{(t_0-t)^{p+r}}\Phi'\bigg)\,\frac{B_0^2}{(t_0-t)^{2q+r}}EE'\nonumber\\
&+\chi''(\rho^0_a)\,C^2\,\bigg(\zeta_{\text{avg}}\,\frac{C_0}{(t_0-t)^{s+2r}}R'' +\zeta_{\text{rel}}\,\frac{A_0}{(t_0-t)^{p+2r}}\Phi''\bigg)\,\frac{B_0^2}{(t_0-t)^{2q}}E^2\nonumber\\
&+2\chi''(\rho^0_a)\,C^2\,\bigg(\zeta_{\text{avg}}\,\frac{C_0}{(t_0-t)^{s}}R +\zeta_{\text{rel}}\,\frac{A_0}{(t_0-t)^{p}}\Phi\bigg)\,\frac{B_0^2}{(t_0-t)^{2q+2r}}\Big(EE''+E'^2\Big)\nonumber\\
&+2\chi''(\rho^0_a)\,C^2\,\bigg(\zeta_{\text{avg}}\,\frac{C_0}{(t_0-t)^{s+r}}R' +\zeta_{\text{rel}}\,\frac{A_0}{(t_0-t)^{p+r}}\Phi'\bigg)\,\frac{B_0^2}{(t_0-t)^{2q+r}}EE'\Bigg].\label{phi-ode}
\end{align}
\end{widetext}
Then by balancing the powers of $(t_0-t)$ on the LHS and $\chi''$ terms on the RHS, equations \eqref{dombalance} in the main text are obtained.

 %%%%%%%%%%%%%%%%%%%%%%%%%%%%%%%%%%%%%%%%%%%%%%%%%%%%%%%%%%%%%%%%%%%%%%%%%%%%%%%%%%%%%%%%%%%%%%%%%%
%%%%%%%%%%%%%%%%%%%%%%%%%%%%%%%%%%%%%%%%%%%%%%%%%%%%%%%%%%%%%%%%%%%%%%%%%%%%%%%%%%%%%%%%%%%%%%%%%%%%%%%%%%%%%%%%%%%%%%%%%%%%%%%%%%%%%%%%%%%%%%%%%%%%%%%%%%%%%%%%%%%%%%%%%%%%%%%%%%%%%%%%%%%%%%%%%%%%%

%%%%%%%%%%%%%%%%%%%%%%%%%%%%%%%%%%%%%%%%%%%%%%%%%%%%%%%%%%%%%

\section{Kinematics and Balance Laws for 1D Singular Fields}\label{sect:singfield}

\subsection{Compatibility condition}

The bulk displacement field $u(x,t)$ is continuous at the singular point $x=x_0(t)$, i.e., 
\begin{equation}
    u^+(x_0^+(t),t)=u^-(x_0^-(t),t)
\end{equation}
where $x_0^{\pm}(t)$ represent the limiting points as one approaches the singularity from the right and the left, respectively, and similarly, $u^{\pm}$ represent the corresponding limiting values of the displacement. Taking time derivative of the above equality implies that
\begin{equation}
    \frac{\partial u^+}{\partial x}\frac{dx_0^+}{dt} +\frac{\partial u^+}{\partial t}=\frac{\partial u^-}{\partial x}\frac{dx_0^-}{dt} +\frac{\partial u^-}{\partial t},
\end{equation}
i.e.,
\begin{equation}
\llbracket\dot{{u}}\rrbracket +V\,\llbracket\epsilon\rrbracket ={0},\label{comp2}
\end{equation}
where $V:=\frac{dx_0^+}{dt}=\frac{dx_0^-}{dt}=\dot{x}_0$ denotes material velocity of the singular point. Thus, the spatial velocity of the singularity is then defined by  ${v}_S:=\langle\dot{{u}}\rangle+V\,\langle\epsilon\rangle$.

%%%%%%%%%%%%%%%%%%%%%%%%%%%%%%%%%%%%%%%%%%%%%%%%%%%%%%%%%%%%%%%%%%%%%%%%%%%%%%%%%%%%%

\subsection{Divergence Theorem}

Consider a piece-wise smooth bulk scalar field $f$ on an arbitrary closed interval $[a,b]\subset\mathbb{R}$ which suffer jump discontinuity $\llbracket f\rrbracket$ at a point $x_0\in[a,b]$. Let us partition the interval $[a,b]$ into subintervals $[a,x_0^-]$ and $[x_0^+,b]$ such that $[a,x_0^-]\cup[x_0^+,b]=[a,b]$ and $[a,x_0^-]\cap[x_0^+,b]=x_0$. Hence, $\llbracket f\rrbracket:=f(x_0^+)-f(x_0^-)$. Since $f$ is smooth within each of these subintervals, where the standard 1D divergence theorem (a.k.a.~fundamental theorem of calculus) applies:
\begin{subequations}
    \begin{align}
        &\int_{a}^{x_0^-}\partial_x f\,dx=f(x_0^-)-f(a),\\
        &\int_{x_0^+}^{b}\partial_x f\,dx=f(b)-f(x_0^+),
    \end{align}
\end{subequations}
Adding these two identities gives 1D divergence theorem for the piece-wise smooth field $f$ over $[a,b]$: 
\begin{equation}
\int_{a}^{b}\partial_x f\,dx=f(b)-f(a)-\llbracket f\rrbracket.
\label{divthm}
\end{equation}

%%%%%%%%%%%%%%%%%%%%%%%%%%%%%%%%%%%%%%%%%%%%%%%%%%%%%%%%%%%%%%%%%%%%%%%%%
\subsection{Transport Theorem}
Next, consider a moving singular point $x_0(t)\in[a,b]\equiv\Omega$, with material speed $V:=\dot{x}_0$. Consider a small $\varepsilon$-neighbourhood $\Omega_{\varepsilon}\subset\Omega$ of $x_0$, such that points within $\Omega_{\varepsilon}$ are parametrized by $x=x_0+\psi(t)$, with $\psi=0$ characterizing $x_0$, and $-\varepsilon<\psi(t)<\varepsilon$, $\forall t$, with a constant $\varepsilon>0$. Further, partition $\Omega_{\varepsilon}$ into subintervals $[-\varepsilon,\psi^-(t)]$ and $[\psi^+(t),\varepsilon]$, such that $[-\varepsilon,\psi^-(t)]\cup[\psi^+(t),\varepsilon]=[-\varepsilon,\varepsilon]$ and $[-\varepsilon,\psi^-(t)]\cap[\psi^+(t),\varepsilon]=\psi(t)$. Hence, $\dot{\psi^-}=\dot{\psi^+}=\dot{\psi}$, and, moreover, $\dot{\psi^-}=\dot{\psi^+}=V$ for $\psi=0$. In this proof, we have closely followed \cite{GuptaSteigmann2012}.

Then,  for a piece-wise smooth bulk scalar field $f$ on $[a,b]$ which suffer jump discontinuity $\llbracket f\rrbracket$ at $x_0(t)\in[a,b]$,
\begin{eqnarray}
    \frac{d}{dt}\int_{\Omega}f\,dx&=&\frac{d}{dt}\int_{\Omega\setminus\Omega_\varepsilon}f\,dx+\frac{d}{dt}\int_{\Omega_\varepsilon}f\,dx\nonumber\\
    &=&\int_{\Omega\setminus\Omega_\varepsilon}\dot{f}\,dx+\frac{d}{dt}\Big(\int_{-\varepsilon}^{\psi^{-}(t)}f\,dx+\int_{\psi^{+}(t)}^{\varepsilon}f\,dx\Big).\nonumber\\
    \label{appeq}
\end{eqnarray}

Using the Leibniz integral rule for smooth fields valid on each subintervals, we obtain
\begin{align}
    &\frac{d}{dt}\int_{-\varepsilon}^{\psi^-(t)}f\,dx=f(\psi^-(t))\,\dot{\psi^-}(t)+\int_{-\varepsilon}^{\psi^-(t)}\dot{f}\,dx,\\
    &\frac{d}{dt}\int_{\psi^+(t)}^{\varepsilon}f\,dx=-f(\psi^+(t))\,\dot{\psi^+}(t)+\int_{\psi^+(t)}^{\varepsilon}\dot{f}\,dx
\end{align}

Substituting these two integrals into \eqref{appeq} and then taking the limit $\varepsilon\to 0$, imply the following transport theorem for the 1D piece-wise smooth field $f$:
\begin{equation}
	\frac{d}{dt}\int_{a}^bf\,dx=\int_{a}^b\dot{f}\,dx-V\llbracket f\rrbracket.
    \label{transportthm}
\end{equation}

%%%%%%%%%%%%%%%%%%%%%%%%%%%%%%%%%%%%%%%
\subsection{Mass Balance}

Let the bulk and singular  mass densities be $\rho$ and $\rho_S\delta_{x_0}$, respectively, where $\delta_{x_0}$ denotes the Diract distribution supported at $x_0$, the piecewise smooth bulk  mass  flux be ${j}$ with jump discontinuity $\llbracket j\rrbracket$ at $x_0$, and the bulk and singular mass turnover (due to binding and unbinding) be $\Pi$ and $\Pi_S\delta_{x_0}$, respectively. Then, the equation for the global balance of mass reads
  \begin{equation}
	\frac{d}{dt}\int_{a}^b\big(\rho+\rho_S\delta_{x_0}\big)\,dx=\int_{a}^b(\Pi+\Pi_S\delta_{x_0})\,dx-\big(j(b)-j(a)\big).
\end{equation}
Using the theorems \eqref{divthm} and \eqref{transportthm}, this equation yields
\begin{equation}
	\int_{a}^b\dot{\rho}\,dx-V\llbracket\rho\rrbracket_{x_0}+\dot{\rho}_S=\int_{a}^b\Pi\,dx+\Pi_S
		-\int_{a}^b\partial_x j\,dx-\llbracket{j}\rrbracket.
\end{equation}
The arbitrariness of the domain $[a,b]$ and the point $x_0$ then yield the local laws
\begin{subequations}
	\begin{align}
		&\dot{\rho}=-\partial_x{j}+\Pi~~~\text{in}~[a,b]\setminus x_0,\\
		&\dot{\rho}_S-V\llbracket\rho\rrbracket=-\llbracket{j}\rrbracket+\Pi_S~~~\text{at}~x_0.
	\end{align}
\end{subequations}

%%%%%%%%%%%%%%%%%%%%%%%%%%%%%%%%%%%%%%%
\subsection{Force Balance}

Let the piecewise smooth bulk (scalar) stress field be ${\sigma}$. Then, the global linear momentum balance equation, with environmental viscosity and in absence of inertia, is
\begin{equation}
\int_{a}^b(\Gamma\dot{{u}}+\Gamma_S{v}_S\,\delta_{x_0})\,dx={\sigma}(b)-\sigma(a),
%+\int_{\partial\mathcal{S}}\boldsymbol{\sigma}_S,
\end{equation}
where the left hand side represents the total frictional force. Then, using the theorem \eqref{divthm}, and arbitrariness of the domain $[a,b]$ and the point $x_0$, we obtain the local laws
\begin{subequations}
	\begin{align}
		&\Gamma\dot{{u}}=\partial_x{\sigma}~~~\text{in}~[a,b]\setminus x_0,\\
		&\Gamma_S{v}_S=\llbracket{\sigma}\rrbracket~~~\text{at}~x_0.
	\end{align}
\end{subequations}

\begin{figure}[h]
\centering
\includegraphics[scale=.2]{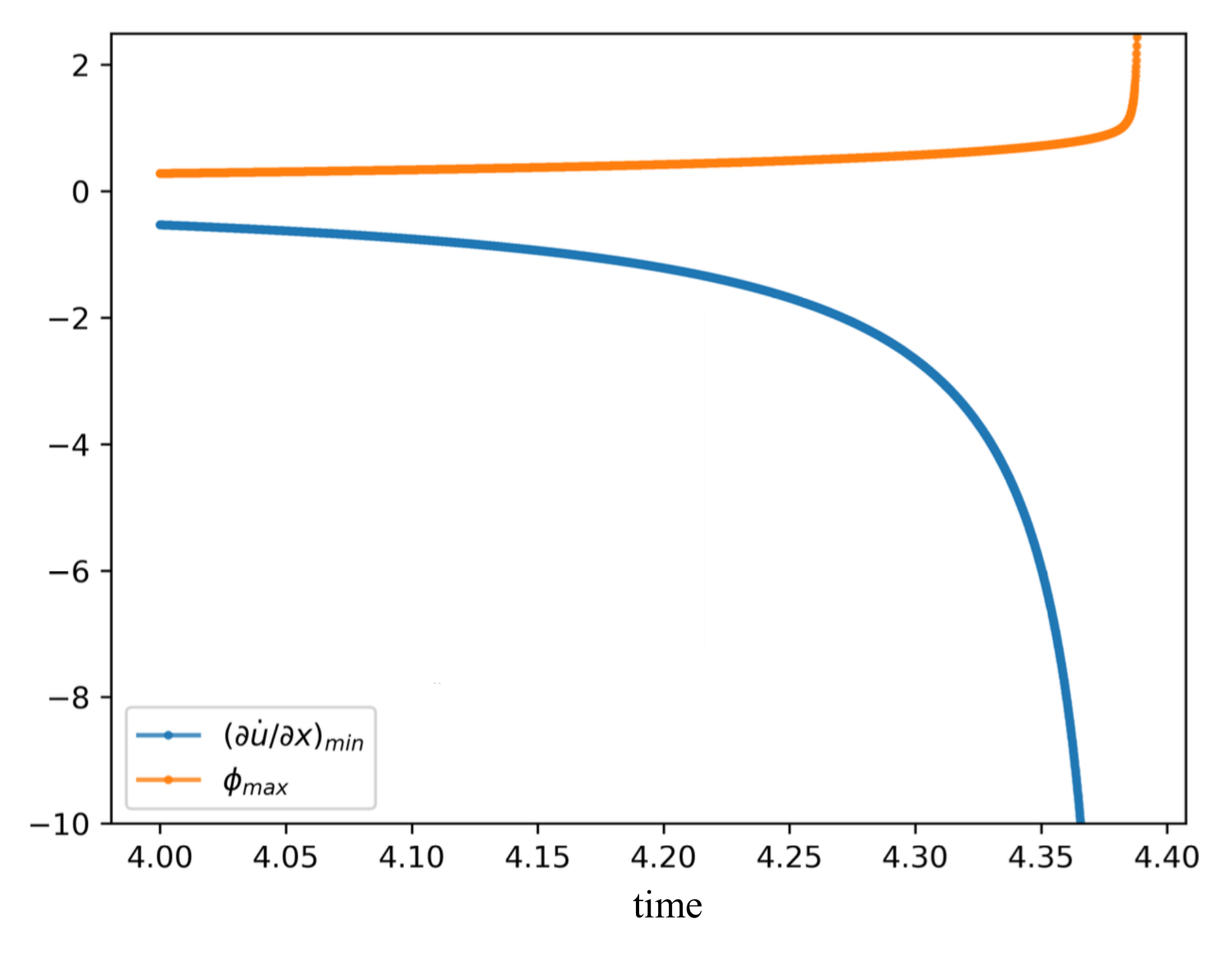}
\caption{Strain rate $\dot{\epsilon}$ blows up faster than the segregation field $\phi$, as the singularity is approached.
}
\label{caustic}
\end{figure}

\begin{figure}[h]
\centering
\includegraphics[scale=.77]{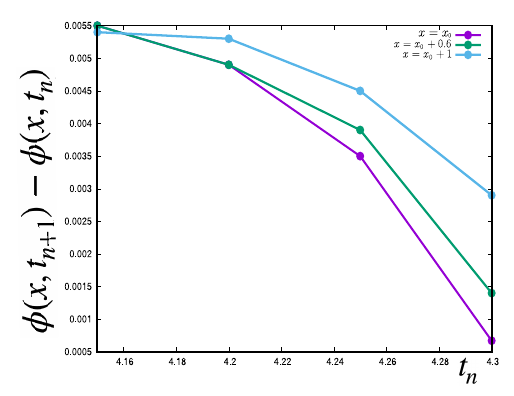}
\caption{Cauchy convergence of $\phi(x,t)$ in
time $t$ as the singularity is approached; points closer to the
blow-up location $x_0$ diverge faster.}
\label{fig:cauchyconvergence}
\end{figure}

%%%%%%%%%%%%%%%%%%%%%%%%%%%%%%%%%%%%%%%%%%%%%%%%%%%%%%%%%%%%%
%%%%%%%%%%%%%%%%%%%%%

%%%%%%%%%%%%%%%%%%%%%%%%%%%%%%%%%%%%%%%%%%%%%%%%%%%%%%%%%%%%%%%%%%%%%%%%%%%%%%%%%%%%%%%%%%%%%%%%%%%%%%%%%%%%%%%%

\section{Movie Captions}\label{sect:movie}

\subsection{Segregation of a binary mixture of stresslets with density peak co-localization } 

\noindent
\href{https://drive.google.com/file/d/14CVxT7SFx-76mTdeD2PDLCU-ONjg75jI/view?usp=sharing}{{\bf {\blue Movie 1.}}}
This movie shows segregation, followed by the formation of singularities in the profiles of $\rho$, $\phi$ and $\epsilon$ of the stresslets, where the density peaks co-localize. Initial conditions are
homogeneous unstrained state with noise. Both stresslets are of catch bond type. Parameters set at
$B=6$, $C=1$, $D=1$, $k_1=1$, $k_2=0$, $k^b_{\text{rel}}=0$, $\zeta_{avg}=2$, $\zeta_{rel}=2$, $\alpha_1=4.5$,$\alpha_2=0.5$, $\chi(\rho_a^0)=1$, $\chi'(\rho_a^0)=1$, $\chi''(\rho_a^0)=0.5$, $\chi'''(\rho_a^0)=-10$, $\eta= 10$, $K=0.5$, $L=32$.
\vspace{10mm}
%%%%%%%%%%%%%%%%%%%%%%%%%%%%%%%%%%%

%%%%%%%%%%%%%%%%%%%%%%%%%%%%%%%%%%%
\subsection{Segregation of a binary mixture of stresslets with density peak separation} 

\noindent
\href{https://drive.google.com/file/d/16a1KWvMt95U8zEvvgcskpv9YLXXqApWs/view?usp=sharing}{{\bf {\blue Movie 2.}}} This movie shows segregation, followed by the formation of singularities in the profiles of $\rho$, $\phi$ and $\epsilon$ of the stresslets, where the density peaks separate. Initial conditions are
homogeneous unstrained state with noise.  One stresslet is of a catch bond type and the other is of a slip bond type. Parameters are set at
$B=6$, $C=1$, $D=1$, $k_1=1$, $k_2=0$, $k^b_{\text{rel}}=0$, $\zeta_{avg}=2$, $\zeta_{rel}=2$, $\alpha_1=0.5$,$\alpha_2=-3.5$, $\chi(\rho_a^0)=1$, $\chi'(\rho_a^0)=1$, $\chi''(\rho_a^0)=0.5$, $\chi'''(\rho_a^0)=-10$, $\eta= 10$, $K=0.5$.

\vspace{10mm}
%%%%%%%%%%%%%%%%%%%%%%%%%%%%%%%%%%%

%%%%%%%%%%%%%%%%%%%%%%%%%%%%%%%%%%%
\subsection{Dynamics of singularity formation} 

\noindent
\href{https://drive.google.com/file/d/1nAkaJR_G6udob4C0n7vA6DdXK_b3O7D4/view?usp=sharing}{{\bf {\blue Movie 3.}}}
This movie (left panel) shows the approach to a finite time elastic singularity, and (right panel) shows its  physical resolution by a steric term ($B_3 \neq 0$) . Parameters are set at $B=6$, $C=1$, $D=1$, $k_1=1$, $k_2=0$, $k^b_{\text{rel}}=0$, $\zeta_{avg}=2$, $\zeta_{rel}=2$, $\alpha_1=4$,$\alpha_2=0$, $\chi(\rho_a^0)=1$, $\chi'(\rho_a^0)=1$, $\chi''(\rho_a^0)=0.5$, $\chi'''(\rho_a^0)=0 $ (for left panel), $\chi'''(\rho_a^0)=-10$ (for right panel), $\eta= 10$, $K=0.1$, $L=10$.

\vspace{10mm}
%%%%%%%%%%%%%%%%%%%%%%%%%%%%%%%%%%%

%%%%%%%%%%%%%%%%%%%%%%%%%%%%%%%%%%%
\subsection{Coarsening Dynamics: catch bond with turnover} 

\noindent
\href{https://drive.google.com/file/d/1ci_D42aP20kXC45w0JwtzV5Nisizz1hS/view?usp=sharing}{{\bf {\blue Movie 4.}}}
This movie shows the coarsening dynamics of elastic singularities, in a system where both stresslets have catch bonds. At first, the singularities merge quickly following which the coarsening slows down, showing signs of dynamical arrest.  Parameters are set at $B=5$, $C=1$, $D=1$, $k_1=1$, $k_2=0$, $k^b_{\text{rel}}=0$, $\zeta_{avg}=1.4$, $\zeta_{rel}=1$, $\alpha_1=3$,$\alpha_2=1$, $\chi(\rho_a^0)=1$, $\chi'(\rho_a^0)=1$, $\chi''(\rho_a^0)=0.5$, $\chi'''(\rho_a^0)=-15$, $\eta= 0.5$, $K=0.01$, $L=600$.

\vspace{10mm}
%%%%%%%%%%%%%%%%%%%%%%%%%%%%%%%%%%%

\subsection{Coarsening Dynamics: catch bond without turnover} 

\noindent
\href{https://drive.google.com/file/d/1XYkDBGoiNu1z5DrmqNfOgcsHBNfQXiD2/view?usp=sharing}{{\bf {\blue Movie 5.}}}
The configurations generated in the previous numerical simulation of coarsening dynamics of the species with catch bonds at $t=100$, is further evolved without turnover. The singularities broaden and merge faster in the absence of turnover and even undergo splitting on occasion. Parameters are set at $B=5$, $C=1$, $D=1.5$, $k_1=1$, $k_2=0$, $k^b_{\text{rel}}=0$, $\zeta_{avg}=1.4$, $\zeta_{rel}=1$, $\chi(\rho_a^0)=1$, $\chi'(\rho_a^0)=1$, $\chi''(\rho_a^0)=0.5$, $\chi'''(\rho_a^0)=-15$, $\eta= 25$, $K=10$, $L=600$.

\vspace{10mm}
%%%%%%%%%%%%%%%%%%%%%%%%%%%%%%%%%%%

%%%%%%%%%%%%%%%%%%%%%%%%%%%%%%%%%%%
\subsection{Coarsening Dynamics: slip bond with turnover} 

\noindent
\href{https://drive.google.com/file/d/1ezNY2PXOxmdJnOaqTryy-zk9IpxBE7jo/view?usp=sharing}{{\bf {\blue Movie 6.}}}
This movie shows the coarsening dynamics of the singularities where one stresslet has a catch bond and the other, a slip bond. Parameters are set at $B=5$, $C=1$, $D=1$, $k_1=1$, $k_2=0$, $k^b_{\text{rel}}=0$, $\zeta_{avg}=1.4$, $\zeta_{rel}=1$, $\alpha_1=0.5$,$\alpha_2=-1.5$, $\chi(\rho_a^0)=1$, $\chi'(\rho_a^0)=1$, $\chi''(\rho_a^0)=0.5$, $\chi'''(\rho_a^0)=-10$, $\eta= 0.5$, $K=0.01$, $L=600$.

\vspace{10mm}
%%%%%%%%%%%%%%%%%%%%%%%%%%%%%%%%%%%

The pseudospectral code and the analysis notebook used to create the Movies 1-6 can be accessed \href{https://github.com/saptarshid999/active_force_patterning}{{\bf {\blue here}}}.

\bibliographystyle{apsrev4-1}

%%%%%%%%%%%%%%%%%%%%%%%%%%%%%%%%%%%%%%%%%%%%%%%%%%%%%%%%%%%%%%%%

\end{document}